\documentclass[aps,rmp,amsmath,amsfonts,floatfix,superscriptaddress,twocolumn,showpacs]{revtex4}

\usepackage{graphicx}
\usepackage{epstopdf}
\input{epsf.sty}

\font\msytw=msbm9 scaled\magstep1

\let\a=\alpha \let\b=\beta  \let\g=\gamma  \let\d=\delta 
  \let\h=\eta  \let\k=\kappa \let\l=\lambda
\let\m=\mu    \let\n=\nu    \let\x=\xi     \let\p=\pi   
 \let\t=\tau   \let\f=\varphi \let\c=\chi
   
\let\G=\Gamma \let\D=\Delta  \let\Th=\Theta 
    \let\Si=\Sigma     
 \let\ee=\epsilon \let\r=\rho \let\th=\theta
\let\io=\infty
\def\ie{{i.e. }}\def\eg{{e.g. }}

\def\PP{{\cal P}} 
\def\FF{{\cal F}} 
\def\NN{{\cal N}} 
  
\def\DD{{\cal D}}\def\GG{{\cal G}} \def\SS{{\cal S}}

 \def\xx{{\bf x}} \def\yy{{\bf y}}

\def\erf{\text{erf}}

\def\to{\rightarrow}
\def\la{\left\langle}
\def\ra{\right\rangle}

\def\RRR{\hbox{\msytw R}}

\newcommand{\beq}{\begin{equation}}
\newcommand{\eeq}{\end{equation}}
\newcommand{\wh}{\widehat}
\newcommand{\wt}{\widetilde}
\newcommand{\Tr}{\text{Tr}}

\def\fICP{\varphi_{GCP}}
\def\gG{g_{G}}

\begin{document}

\title{
Mean field theory of hard sphere glasses and jamming
}

\author{Giorgio Parisi}
\email{giorgio.parisi@roma1.infn.it}
\affiliation{Dipartimento di Fisica, INFM-CNR SMC, INFN,
Universit\`a di Roma ``La Sapienza'',
P.le A. Moro 2, 00185 Roma, Italy
}
\author{Francesco Zamponi}
\homepage{http://www.lpt.ens.fr/~zamponi}
\affiliation{Laboratoire de Physique Th\'eorique, 
\'Ecole Normale Sup\'erieure,
24 Rue Lhomond, 75231 Paris Cedex 05, France
}


\begin{abstract}
Hard spheres are ubiquitous in condensed matter: they have been used as models for liquids, crystals,
colloidal systems, granular systems, and powders. Packings of hard spheres are of even wider interest,
as they are related to important problems in information theory, such as digitalization of
signals, error correcting codes, and optimization problems. In three dimensions the densest packing 
of identical hard spheres has been proven to be the FCC lattice, and it is conjectured that the closest
packing is ordered  (a regular lattice, e.g, a crystal) in low enough dimension. 
Still, amorphous packings have attracted a lot of
interest, because for polydisperse colloids and granular materials the 
crystalline state is not obtained in experiments for kinetic reasons.
We review here a theory of amorphous packings, and more generally glassy states, of hard spheres 
that is based on the replica method: this theory  gives predictions
on the structure and thermodynamics of these states.
In dimensions between two and six these predictions can be successfully compared
with numerical simulations. We will also discuss the limit of large dimension where an exact solution
is possible. 

Some of the results we present here have been already published, but others are original:
in particular we improved the discussion of the large dimension limit and we obtained new
results on the correlation function and the contact force distribution in three dimensions.
We also try here to clarify the main assumptions that are beyond our theory and in particular
the relation between our static computation and the dynamical procedures used to construct
amorphous packings. 

There remain many weak points in our theory that should be
better investigated. We hope that this paper can be useful to 
present the state of the art of the method and to stimulate new research in this field.
\end{abstract}


\maketitle

\tableofcontents

\section{Introduction}

The study of amorphous states of hard spheres is relevant for a large class
of physical systems, including liquids, glasses, colloidal dispersions, granular
matter, powders, porous media. Therefore,
after pioneering works done in the sixties~\cite{BM60,BM62,Sc62,SK69,MC65,MC66,Fi70},
a huge amount of precise numerical and
experimental data is now available, see \eg
\cite{CJ93,Be83,Be72,Ma74,Po79,LS90,To95,RT96,Sp98,OLLN02,OSLN03,PV86,Torquato,DTS05,SLN06,SDST06,SHESS07,AF07,JSSSSA08,MSLB07,Krauth}.
Moreover, the sphere packing problem is related to many mathematical problems
and arises in the context 
of signal digitalization and of error correcting codes, and it has been investigated
in detail by the information theory community \cite{ConwaySloane,Rogers}.
Nevertheless, a satisfactory characterization of the amorphous states of a system of
identical hard spheres is not yet available and the definition of amorphous close
packed states is still matter of debate \cite{TTD00,OLLN02,KL07}.
From the rigorous point of view, for space dimension $d>3$ 
only some not very restrictive bounds have been obtained,
and in particular it is still unclear whether
the densest packings for $d\to \io$ are amorphous or crystalline
(see \cite{SlWeb} for a list of all known densest packings up to $d=128$).

Dense amorphous packings of hard spheres are usually produced according to some specific
dynamical protocol. Typically one starts from an initial random configuration of the spheres,
obtained \eg by throwing them into a container, and then shake, tap, or agitate in some way
the spheres until a jammed structure is found \cite{SK69,Be72,PV86,Torquato,SGS05,DB06,DMB05,AD06,PNC07,JSSSSA08,MSLB07}.
In numerical simulations, amorphous packings are produced by inflating the particles while
avoiding superposition either by molecular dynamics \cite{LS90,DTS05,SDST06} or by using
soft particles and minimizing the energy \cite{CJ93,OLLN02,OSLN03,SLN06,SHESS07}.
As a matter of fact, most of these procedures, if crystallization is avoided, 
lead to a final packing fraction
close to $0.64$ in $d=3$ and to $0.84$ in $d=2$. These values of density,
that are approximately $10\%$ smaller than the values of the ordered close packing,
have been called ``random close packing density''.

Unfortunately, the algorithms (or procedures) that are used to create such packings are
complicated {\it dynamical non-equilibrium} procedures. Obtaining analytical results for the
properties of the final states requires an analytical solution of such complicated dynamical
processes, that is very difficult even in the simplest theoretical models~\cite{KK07,TS06}.
The aim of this paper is then to identify a class of amorphous packings that might be described
using {\it equilibrium} statistical mechanics, that is, in a {\it static} framework.
These packings will be defined as the infinite pressure limit of glassy states of hard spheres:
such glassy states, if dense enough, are well defined metastable states with very long
life times, and should be then correctly described by equilibrium statistical mechanics.
The idea of studying amorphous packings as the infinite pressure limit of a metastable state has
been already discussed in the literature \cite{AC04,BM01,PTCC03,RBMM04,TCFNC04,PZ05,Za07,KK07,KL07} and is 
appealing because it converts a difficult dynamical problem into a much simpler equilibrium problem.

Our approach to study glassy states will be based on the so-called
Random First Order Transition (RFOT) theory of glasses, whose theoretical foundations
were posed in a series of papers by Kirkpatrick, Thirumalai and 
Wolynes~\cite{KT87,KW87}, see \cite{Ca09} for a detailed and recent review.
In this theory the glass transition of particle systems 
is assumed to be in the universality class of 
the 1-step Replica Symmetry Breaking (1RSB) transition that happens in some mean-field 
exactly solvable spin glass models~\cite{MPV87,GM84}. 
Under this assumption, the glassy states of realistic finite-dimensional systems
can be studied analytically, {\it within some approximation},
using equilibrium statistical mechanics
by means of density functional theory~\cite{DV99,CKDKS05,KW87,KM03,SSW85,SW84,YYO07}
and of the replica trick \cite{Mo95,MP96}. In particular, the replica method seems to give
good quantitative estimates of the glass
transition temperature (or density) and of the equation of state of the glass 
for Lennard-Jones systems \cite{MP99,MP99b,MP00,CMPV99} and
hard spheres \cite{CFP98,CFP98b,PZ05,PZ06a}.  

Beside the large amount of numerical and experimental data available in the literature,
there is an important advantage in working with hard spheres with respect to Lennard-Jones--like potentials.
``Ground states'' of hard spheres are obtained in the infinite pressure limit, and correspond
then to sphere packings that have interesting geometrical and topological properties, that can
be investigated by looking at the network of contacts and at contact forces. This is very
interesting because one can identify a set of geometrical observables that can be computed
within the theory and directly compared with simulation and experiments. In this paper we will
show that this allows a very precise test of the theory. 
Moreover, a more direct geometrical approach is possible; it has been largely 
exploited, see \eg~\cite{ASS05,AM07,MFC08,As05,ROTG99,AMA08,DCST07,DTS05,SDST06,KTS02,LAEMS06},
and led to some important successes in the characterization of sphere packings. 
This is not possible for Lennard-Jones--like particles
since the (zero temperature) ground states of the system do not have special geometrical properties,
and indeed their structure is quite similar to typical liquid configurations.
One might consider instead a soft potential that vanishes outside a finite radius; then
the zero-energy ground states correspond to hard sphere configurations. In this way one obtains
a system that displays the same geometrical properties of hard spheres at zero temperature and energy, but at
the same time becomes soft at finite temperature. This has been largely exploited~\cite{OSLN03,OLLN02,BW08,BW09,SO08}
to obtain important informations on the properties of amorphous hard sphere packings.
Therefore in this paper we will focus on the hard sphere case and we will discuss in details 
the limits of the theory and how it compares with numerical and experimental results.
We will show that despite the strong idealizations involved in the theory, the agreement with
numerical data is surprisingly good. Moreover, we will be able to study in full detail the limit
of large space dimension, and obtain the asymptotic value of the density of amorphous packings.

Remarkably, a class of mean-field hard sphere models have been recently formulated, 
for which the RFOT scenario is exact~\cite{BM01,PTCC03,RBMM04,TCFNC04,SBT05,MKK08,KTZ08}.
These models allowed to test the methods used here confirming that they are reliable,
at least at the mean-field level. In particular, \cite{MKK08} formulated a model in this class
whose phase diagram is exactly the same as the one we will discuss below for finite dimensional
hard spheres.

It is worth to stress that in experiments on granular systems and powders
the role of friction is very important~\cite{SGS05,DB06,DMB05,AD06,LDBB08,PNC07,SHESS07,SHS07},
for instance in determining the existence of loose packings~\cite{OL90,JSSSSA08,SWM08}.
Friction complicates a lot the
theoretical analysis of the packing problem, since the system is intrinsically
out of equilibrium and standard equilibrium statistical mechanics is in principle
useless. Nevertheless, since the pioneering work of Edwards~\cite{EO89,Ed98},
statistical mechanics
ideas have been used to describe frictional systems, leading to remarkable
results~\cite{Gold}.
The comparison with experimental results is made difficult by the fact that in most
experiments samples are polydisperse, often with a quite large range of particle
sizes as in the case of many granulars.
 
For reasons of space, 
this paper will be focused on our approach that we wish to discuss in full detail;
therefore in the following we will consider mainly the statistical properties of
a system of frictionless spheres since our
method is based on equilibrium statistical mechanics. 
We will not discuss in detail neither the geometrical properties (unless needed to compare numerical
data with our results) of
amorphous packings, nor how their properties are influenced by the presence of friction.
While in principle polydispersity can be
included in our theory, 
here we limit ourselves to discuss the simple case of monodisperse systems, and only 
at the end we consider the case of binary mixtures.
The literature on hard spheres in general (and on the role of friction and geometry
in particular) is immense and covering
it here would require a considerable effort: the reader is therefore referred to the original
literature and to the existing excellent books and 
reviews, \eg~\cite{Al98,LN01,As05,Gold,ConwaySloane,Rogers,Torquato,Krauth}.
Similarly, many excellent general books and reviews on the physics of glasses 
and the glass transition
exist, see \eg \cite{Ca09,BK05,DS01,EAN96,Do01,LN07} just to quote a few of them.

\subsection{Organization of the paper (and how to read it)}

This paper is organized as follows. 
Section \ref{sec:phasediagram} is devoted to a general discussion of the ideas that
lead to the connection between packings that are produced by dynamic protocols and
infinite pressure glassy states.
In section \ref{sec:method} the replica method is introduced in a general context and
it is explained how it can be used to compute the phase diagram of glassy states.
In section \ref{sec:HNC} a first implementation of the method, that works far from jamming,
is discussed. Sections \ref{sec:effectivepotentials}, \ref{sec:larged}, \ref{sec:smallcage},
\ref{sec:beyond} contain the core of the paper: the method is implemented in a way that
works up to jamming and most of the results are presented. In particular section \ref{sec:larged}
contains the discussion of the $d\to\io$ limit.
Many technical parts of the paper are in the Appendix.

We tried where possible to make the different parts of the paper independent. There are different ways,
of increasing difficulty, to read the paper:
\begin{itemize}
\item
The discussion of section \ref{sec:phasediagram} is self-contained.
We suggest to the reader not interested in technical details to read only section~\ref{sec:phasediagram},
then skip the computations and look directly to the figures and tables that contain most of our results, and
finally jump to the conclusions. Each figure where results are presented contains a reference
to the section where the corresponding calculations are discussed.
\item
The general discussion of the replica method in section~\ref{sec:method}
is as self-contained as possible. This might be read together with Appendix~\ref{sec:metastability} if
one wants to obtain more insight into the method without going into the technical details of the computations.
\item
The technical 
sections \ref{sec:HNC}, \ref{sec:effectivepotentials}, \ref{sec:larged}, \ref{sec:smallcage},
\ref{sec:beyond} are all independent one from the other; to understand one of them in detail
one needs only to read section~\ref{sec:method} and the Appendices.
\item Finally, in \ref{sec:binary} we present some results on the extension of the theory to binary
mixtures, that are based on the method of section~\ref{sec:smallcage}.
\end{itemize}

It is important to stress that we decided to present different implementations of the method
separately, together with the corresponding results. Clearly, another possible choice could be
to collect all the results together in a separate section.
Our choice has been made to stress that each approximation scheme gives slightly 
different results; although the qualitative picture stays the same, it is difficult to
compare quantitatively different approximations. Moreover, each approximation has some advantages
and disadvantages that we tried to discuss in detail; in particular, some observables can be computed
within some approximation and not within others.
We leave to the reader the task to compare the different methods, and choosing the best one
according to his personal taste.

\subsection{Notations}

It is useful to give here some definitions that will be widely used in the following.
We will consider a system of hard hyperspheres in $d$ dimension with diameter $D$.
We will denote by 
\beq\begin{split}
&V_d(D) =\frac{\pi^{d/2}}{\G(1+d/2)} D^d \ , \\
&\Si_d(D) = d \, D^{-1} \, V_d(D)  =\frac{2 \pi^{d/2}}{\G(d/2)} D^{d-1} \ ,
\end{split}\eeq
respectively the volume and surface of a sphere of 
radius $D$.
It is also convenient to define
\beq\begin{split}
&V_d = V_d(1) = \frac{\pi^{d/2}}{\G(1+d/2)} \ , \\
&\Omega_d = \Si_d(1) = \frac{2 \pi^{d/2}}{\G(d/2)} \ .
\end{split}\eeq
Note that $\Omega_d$ is the $d$-dimensional solid angle.

Often, if there is no ambiguity, we will use the symbols $x, \ y, \ \cdots$ to
denote vectors in $\RRR^d$. If there is an ambiguity, we will use the letter
$r$ and in this case we will use $\vec r$ for a $d$-dimensional vector and $r=|\vec r|$ for its modulus.
Correspondingly $\int d\vec r F(\vec r)$ will indicate an integral over $\vec r \in \RRR^d$ while
$\int_a^b dr F(r)$ will indicate an integral over the real $r$. 
We will use the notation $\d(|\vec r|-D)$ for the distribution defined by
$\int d\vec r F(|\vec r|) \d(|\vec r|-D) = \Si_d(D) F(D)$, while 
$\d(r-D)$ is defined by
$\int_0^\io dr \, F(r) \d(r-D) = F(D)$. They are related by
$\d(|\vec r|-D) = \Omega_d \d(r-D)$.

For a generic potential $\phi(r)$ we define $b(r) = e^{-\phi(r)}$ and the Mayer
function $f(r)=b(r)-1$. For the specific case of Hard Spheres we have $b(r) =
\c(r) = \th(|r|-D)$.

Defining $\r = N/V$ the number density of the spheres, we introduce as usual the
{\it packing fraction} $\f=\r V_d(D/2)$, \ie the fraction of volume covered by the spheres.
In the following, when talking about ``density'', we will usually refer to the packing
fraction.

\begin{figure*}
\centering
\includegraphics[width=8cm]{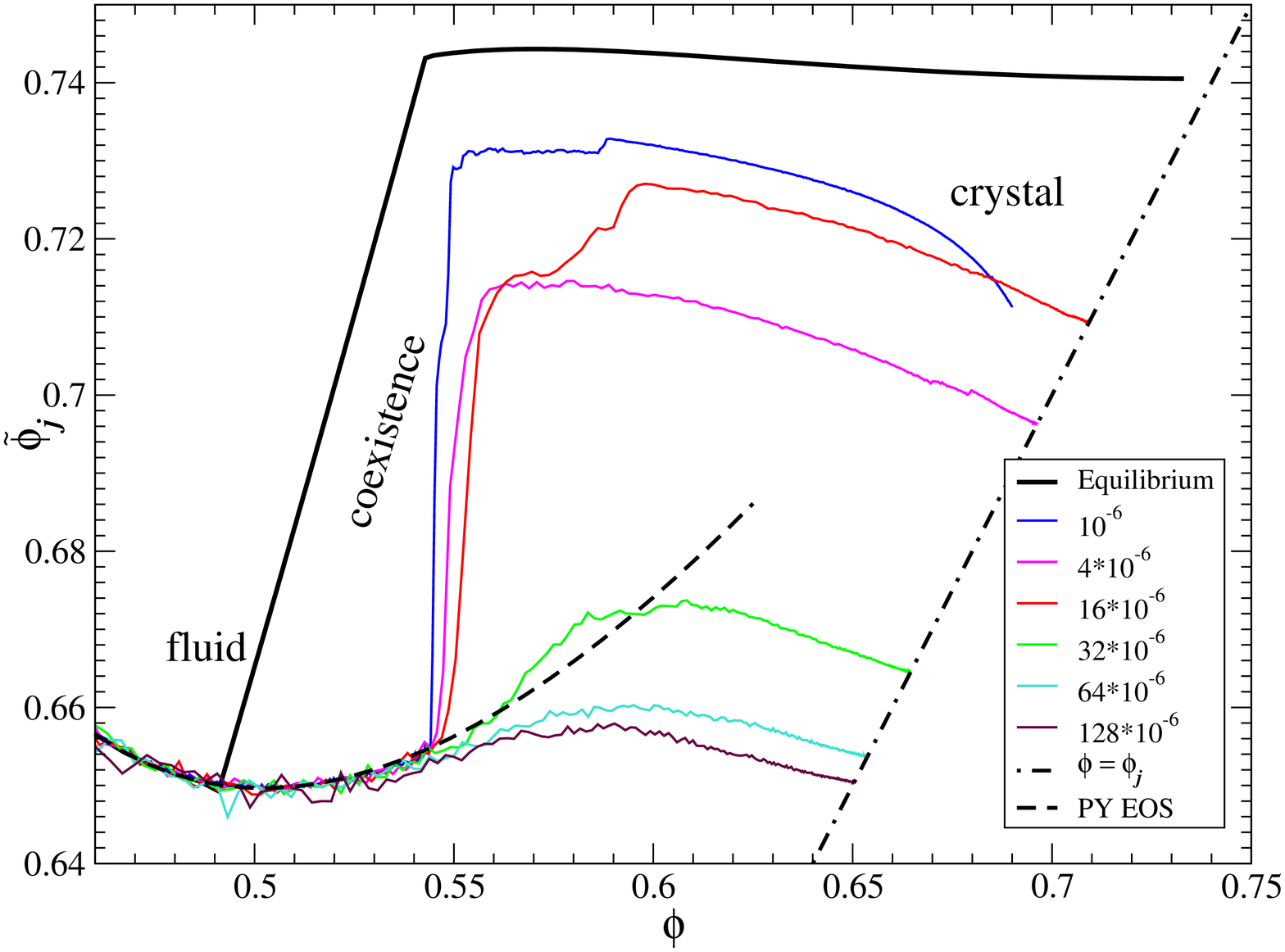}
\includegraphics[width=8cm]{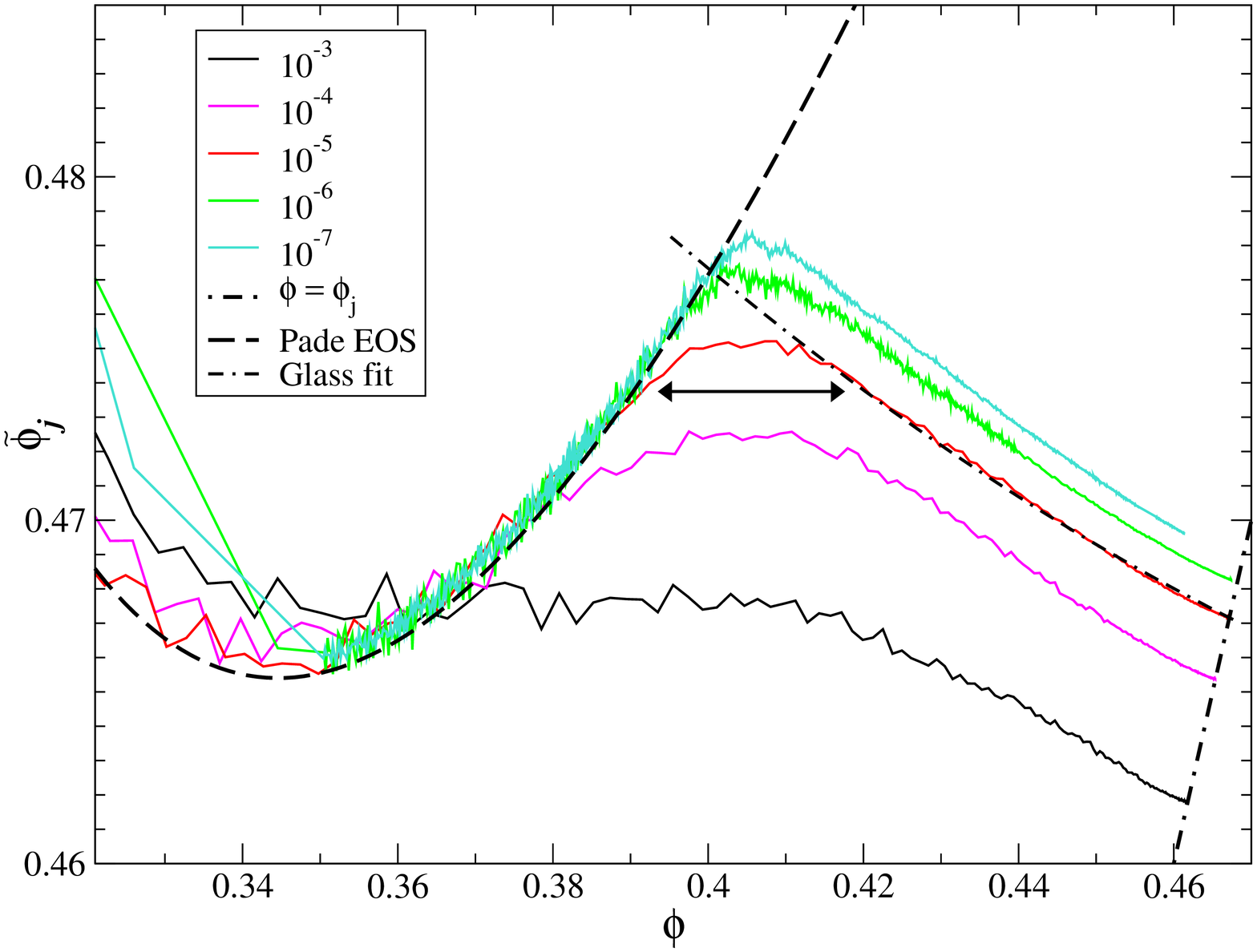}
\vskip5pt
\caption{(From~\cite{SDST06}) Evolution of the pressure during compression at rate $\g$ in
$d=3$ (left) and $d=4$ (right). The density $\f$ is increased at rate $\g$ and the reduced pressure 
$p(\f) = \b P / \r$ is measured during the process. See \cite{SDST06} for details. 
The quantity $\wt \f_j(\f) = \frac{p(\f) \f}{p(\f)-d}$ is plotted as a function of $\f$. If the
system jams at density $\f_j$, $p \to \io$ and $\wt \f_j \to \f_j$. Thus the final jamming density
is the point where $\wt \f_j(\f)$ intersects the dot-dashed line $\wt \f_j =\f$.
(Left) The dotted line is the liquid (Percus-Yevick) equation of state. The curves at high $\g$
follow the liquid branch at low density; when they leave it, 
the pressure increases faster and diverges at $\f_j$.
The curves for lower $\g$ show first a drop in the pressure, which signals crystallization.
(Right) All the curves follow the liquid equation of state 
(obtained from Eq.(9) of \cite{BW05}) 
and leave it at a density
that depends on $\g$. 
In this case no crystallization is observed. 
For $\g=10^{-5}$ the dot-dashed line is a fit to the high-density part of the pressure (glass branch).
The arrow marks the region where the pressure crosses over from the liquid to the glass branch.
}
\label{fig:Skoge}
\end{figure*}

\section{A class of amorphous packings: infinite pressure glassy states}
\label{sec:phasediagram}

In this section we will define a class of ``amorphous packings'' of hard spheres
that can be studied within the framework of equilibrium statistical mechanics.
We start by discussing some algorithms that are commonly used to construct amorphous packings.
Then we argue that the final states reached by these algorithms are well defined metastable
states whose properties can be investigated by
a ``static'' computation (\ie without any knowledge of the dynamical process that generated
the packings). This point is very delicate and is receiving a lot of attention in the context
of optimization problems~\cite{KK07}, where it has not yet been solved. The discussion that 
follows is tentative and the problem surely deserves further investigation.

\subsection{Algorithms to construct amorphous packings}
\label{sec:algorithms}

The usual way to construct amorphous packings in experiments or numerical simulations is to compress
the system according to some protocol. In early experiments particles were thrown 
randomly in a box which was then shaken~\cite{SK69}, or were deposited
randomly around a small seed cluster~\cite{Be72}. 
In numerical simulations a common protocol \cite{LS90} 
is to slowly increase, at a given rate $\g$, 
the particle diameter during a molecular dynamics run; it has been recently
used extensively in three~\cite{DTS05} and higher~\cite{SDST06} dimensions to produce amorphous
packings of $N \sim 10^4$ spheres.
Another possibility is to increase the diameter of the spheres until two of them overlap, then
eliminate the overlap by following a gradient descent using some potential vanishing outside the radius of the 
particle~\cite{XBO05,GBO06}; or, alternatively, to start from a random configuration and minimize
the energy at fixed density, repeating the procedure while increasing the density until it becomes impossible to find
a zero-energy final configuration~\cite{OLLN02,OSLN03}. These two procedures give very similar results~\cite{SO08}. 
Other similar algorithms have been proposed and analyzed in \cite{JT85,CJ93,LAEMS06}.

Based on standard concentration arguments, it is believed that, 
in the limit $N \to \io$, the density of the final state
is independent on the randomness built in the algorithm\footnote{Similar
  results have been shown for some classes of algorithms to solve
  optimization problems, see \eg~\cite{SM03,BHV03}.
} (\eg the initial
configuration). This has been numerically verified for the soft-potential
algorithms~\cite{OLLN02,OSLN03,XBO05,GBO06}; the final density is very close to 0.64 and has been called 
{\it J-point}. 
It is a fact that for all algorithms that have been devised to construct
amorphous packings of monodisperse frictionless spheres, for $N\to\io$
the final density of the system converges to
a value of $\f$ close to 0.64 in three dimensions.
It has been proposed to call the value
$\f=0.64$ {\it random close packing density} and different more precise definitions of
this concept have been proposed in the literature.
However the precise numerical value of the latter quantity is found to depend on the 
details of the experimental protocol, and this led some authors to criticize
the notion of random close packing \cite{TTD00}.

\begin{figure*}
\centering
\includegraphics[width=8cm]{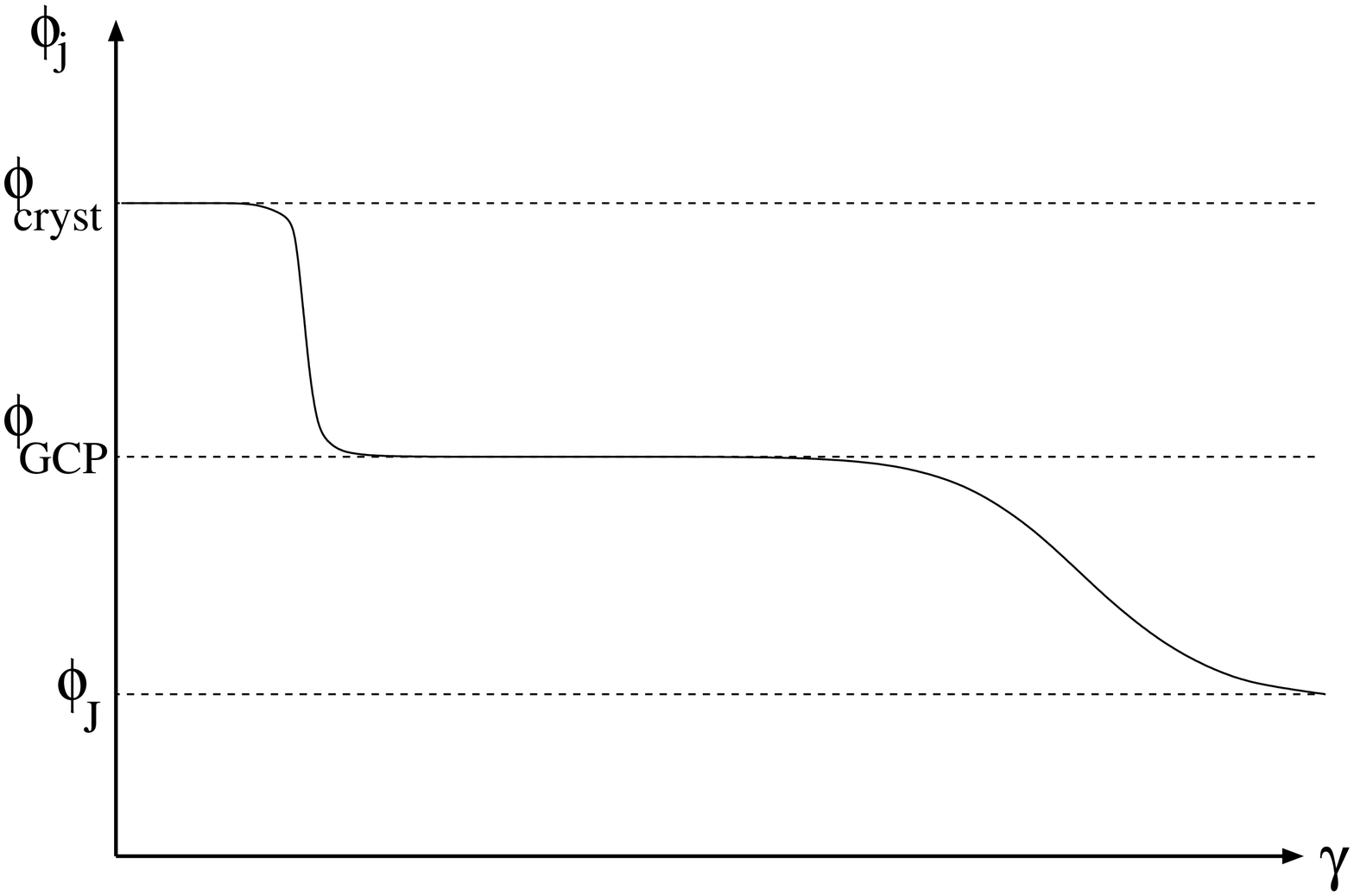}
\includegraphics[width=8cm]{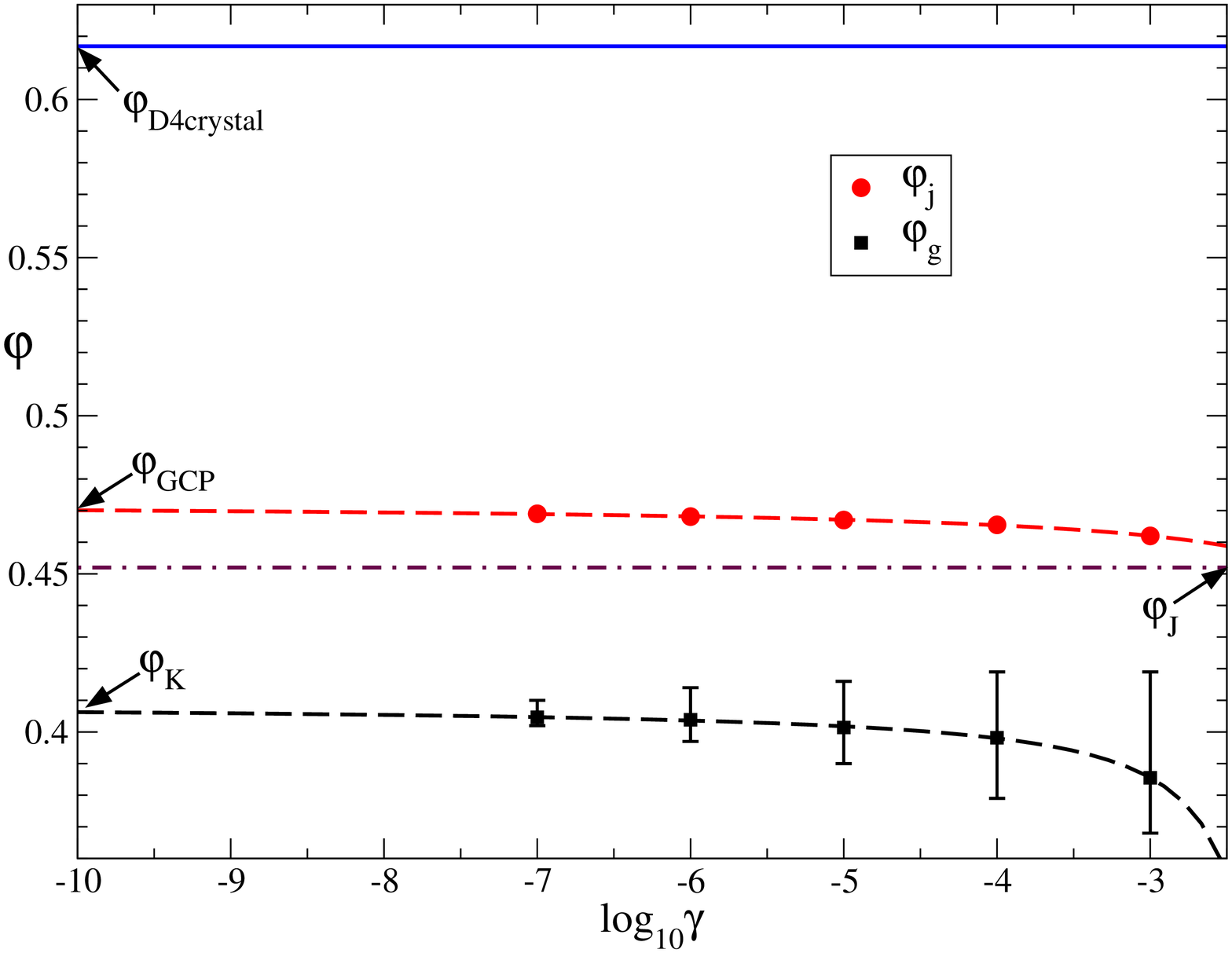}
\caption{(Left) A schematic plot of the jamming density $\f_j(\g)$ 
as a function of
compression rate $\g$ for the Lubachevsky-Stillinger algorithm.
At quite high rates, the algorithm should be similar to the 
$J$-point procedure and converge to packings of density 
$\f_j(\g \text{ large}) \sim \f_J$.
On decreasing $\g$ the
jamming density decreases (compare with figure~\ref{fig:Skoge}) and reaches a {\it plateau}
at $\f_j=\fICP$.
For smaller $\g$ the system is able to crystallize and $\f_j(0)=\f_{crystal}$. 
In $d=3$, crystallization is fast and the plateau is not observed, while in $d>3$ crystallization
is so slow that it is not observed at all in numerical simulations \cite{SDST06}.
(Right) The same plot for hyperspheres in $d=4$. The full line is the close packing density 
achieved by the D4 crystal, 
$\f_{D4} = \p^2/16$~\cite{ConwaySloane}. Circles are values of $\f_j(\g)$
(the point where the curves intersect the dashed line in right panel of
figure~\ref{fig:Skoge}). The dot-dashed line is the value of the J-point 
density~\cite{SO08}.
In addition, we report the estimates of the glass transition density
$\f_g(\g)$ (squares); error bars mark the amplitude of the crossover region and correspond
to the arrow in right panel of figure~\ref{fig:Skoge}.
Dashed lines are fits to $\f_j(\g) = 0.473 + 0.023/(\log_{10}(\g)+0.85)$ and
$\f_g(\g) = 0.409 + 0.02/(\log_{10}(\g)+2.13)$.
}
\label{fig:phi_gamma}
\end{figure*}

To illustrate this difficulty, let us concentrate on data obtained using
the Lubachevsky-Stillinger procedure at different rates $\g$~\cite{SDST06} and reproduced
in figure~\ref{fig:Skoge}. In this algorithm, during the compression, the (hard) spheres evolve
according to molecular dynamics at some temperature; the value of the temperature is irrelevant
and fixes the unit of time~\cite{SDST06}. One can measure the (kinetic) reduced pressure $p(\f) = \b P/\r$,
where $\b= 1/T$, during the evolution. 
This quantity is reported in figure~\ref{fig:Skoge} for different values of 
$\g$.
One observes that, on compressing the low
density liquid at a constant rate $\g$, the pressure of the system follows the 
equilibrium pressure of the liquid up to some density $\f_g(\g)$
(often called {\it glass transition density}) around which the pressure starts
to increase faster than in equilibrium, and diverges on approaching
a value of density $\f_j(\g)$, which is called {\it jamming density}.
At this point the algorithm stops because the system cannot be compressed anymore: most
of the spheres are in contact with their 
neighbors\footnote{At low compression rates, 
crystallization is observed in $d=3$, but it seems to be strongly suppressed 
by kinetic effects in dimension $d>3$ so we will neglect it for the moment.}.
Values of $\f_j(\gamma)$ have been accurately 
measured in \cite{SDST06} as a function of $\gamma$. On the contrary,
$\f_g(\gamma)$ is
not precisely determined as long as $\g > 0$: the glass transition is smeared
and happens in a crossover region $[\f^-_g(\g),\f^+_g(\g)]$ 
(marked by an arrow in the right panel
of figure~\ref{fig:Skoge}). However, the amplitude
of the crossover interval seems to decrease\footnote{See 
\cite{MGS06} for a recent theoretical discussion of
these effects.} for $\g \to 0$, see figure~\ref{fig:Skoge}.

To characterize the typical configurations reached by the algorithm
at $\f = \f_j(\g)$, one could try to solve the dynamics of the algorithm and compute
for instance the correlation function $g(r)$ as a function of ``time''. This has been
done for much simpler algorithms, such as the {\it Ghost Random Sequential Addition (GRSA)}
algorithm~\cite{TS06}, which however are unable to reach interesting 
densities\footnote{For instance in $d=3$ the limiting density for the algorithm is
$\f=0.125$.}. More efficient algorithms such as the {\it Random Sequential Addition (RSA)},
where one attempts to add a sphere randomly and accepts the move only if there are no overlaps,
already cannot be analyzed analytically and one has to resort to numerical 
investigation~\cite{TTVV00,TUS06}.

For this reason, in order to compute analytically the properties of jammed configurations,
one would like to make use of a ``static'' computation. 
This can be justified as follows:
if we plot the jamming
density $\f_j(\g)$ as a function of $\g$, we obtain the plot reported
in right panel of figure~\ref{fig:phi_gamma}.
For $\g\to 0$, the algorithm is equivalent to an equilibrium compression
and the pressure should follow the equilibrium equation of state; then the final state will
be the most dense state, which is a crystal at least if the dimension is not very large.
However, for small but non-zero $\g$, crystallization is not observed and
the data of~\cite{SDST06} suggest 
the existence of a {\it plateau} at some value of density, that we will call $\fICP$ for
reasons that will be clear in the following.
This is a hint of the existence of
a long-lived metastable state; if this is the case, one can compute its properties
by mean of a thermodynamic computation by restricting the partition function to the region
of amorphous configurations. In section~\ref{sec:eq_phase_diag} we discuss this issue in more detail.

It seems that many different
algorithms that produce packings starting from a random configuration and without allowing the
particles to relax much~\cite{OSLN03,OLLN02,XBO05,GBO06} lead
to very close values of density (the J-point).
Therefore for large $\g$ (but still small enough to allow
for some minimal relaxation of the particles) the final density of the Lubachevsky-Stillinger
algorithm should be close to the one
reached by the J-point procedure, leading to 
the schematic plot reported in left panel of 
figure~\ref{fig:phi_gamma}.
This is what indeed happens in a simple mean-field model~\cite{CK93},
and seems confirmed by the data for spheres in $d=4$ reported
in the right panel of figure~\ref{fig:phi_gamma}.
However, the validity of this statement is very debated 
for more general mean-field models~\cite{KK07}. 
We will come back to this point in the following.

Before concluding this section, let us stress that here we focused only on the behavior of
algorithms that are currently used to construct amorphous packings; these algorithms typically
perform local moves in phase space. Many other algorithms to simulate hard spheres have been
invented. Smarter algorithms can be designed, that are able to sample
the equilibrium measure at higher density~\cite{Krauth,DK95,GP01}.

\subsection{The ``equilibrium'' phase diagram}
\label{sec:eq_phase_diag}

The equilibrium phase diagram of hard spheres in $\RRR^3$ is sketched in Fig.~\ref{fig:dialiquido}, 
where
the pressure is reported as a function of the packing fraction. At low density
the system is in a liquid phase (as defined \eg by the low density virial expansion); the maximum
possible density is realized by the FCC lattice, as conjectured by Kepler and
rigorously proven by Hales~\cite{Ha05}. A first order phase 
transition between the liquid phase and the FCC crystal phase is found by numerical simulations~\cite{AW57,WJ57}
and in experiments on colloidal systems \cite{PV86,PRCZCDO96}.
This is the equation of state that the system will follow if compressed at {\it really infinitesimal}
rate. 
We wish instead to focus on a small but finite rate in such a way to follow the liquid branch
of the equation of state inside its metastability region.

\subsubsection{What is the fate of the liquid above freezing?}

\begin{figure*}
\includegraphics[width=8.5cm]{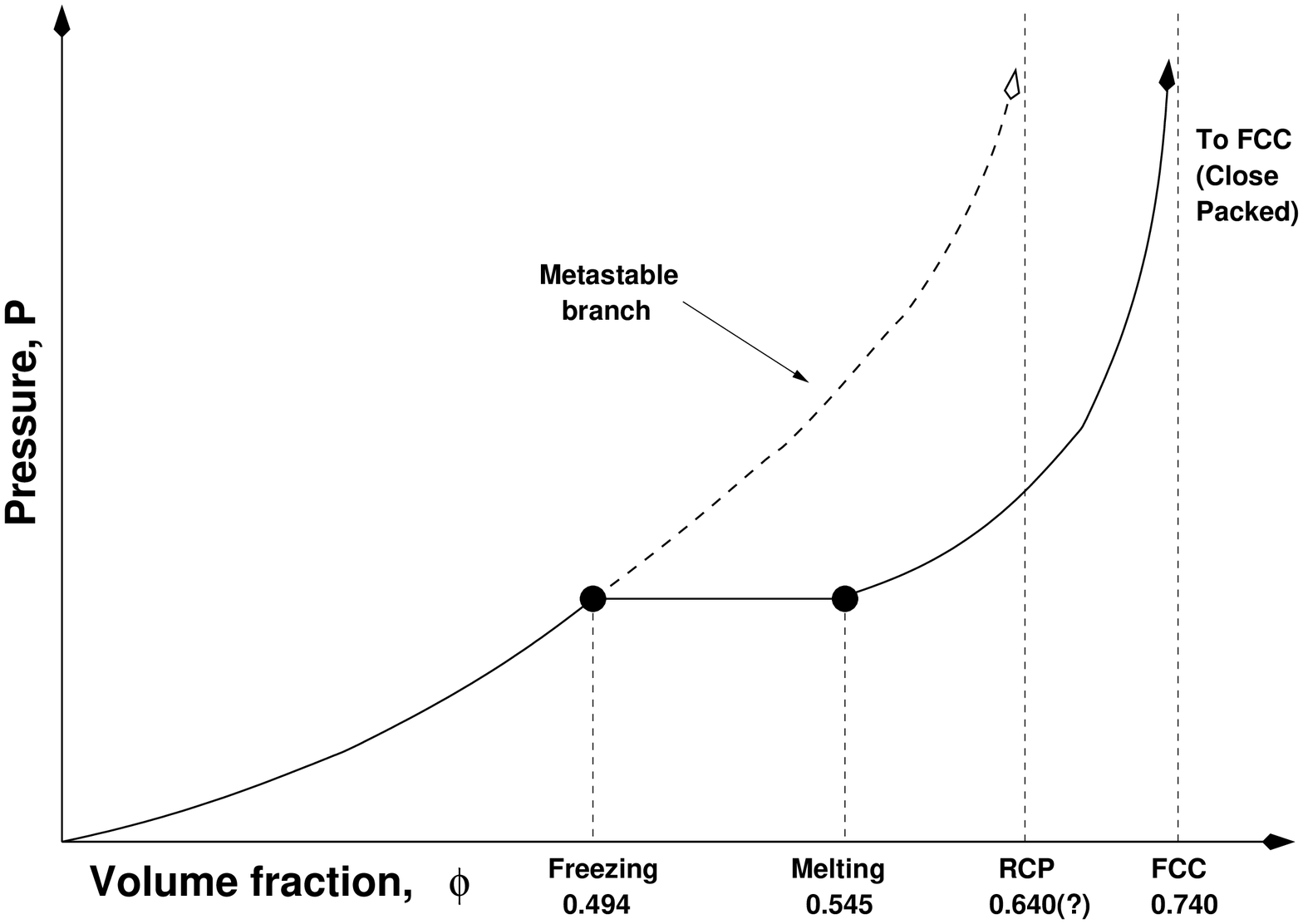}
\includegraphics[width=8.5cm]{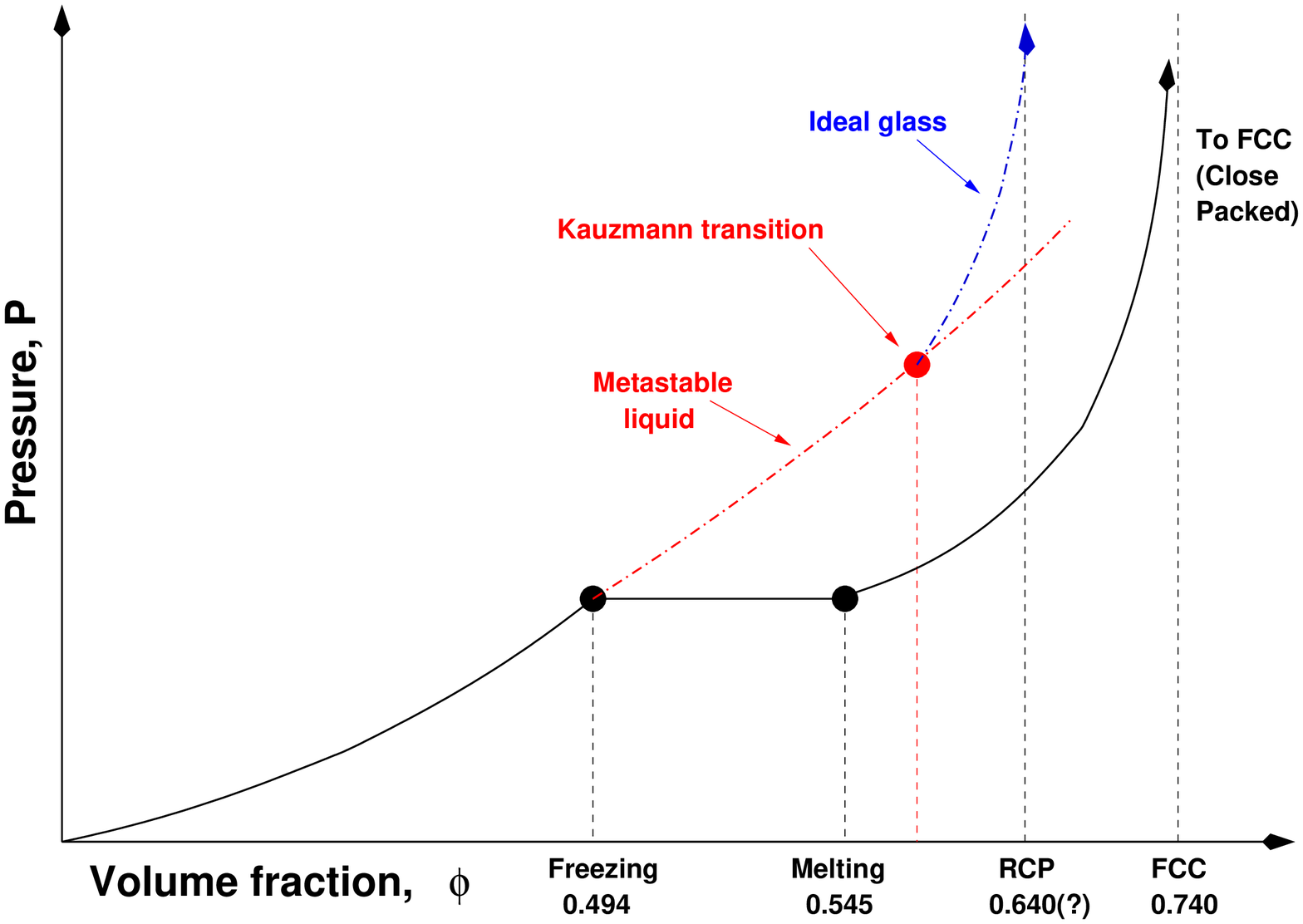}
\caption{Schematic phase diagram of hard spheres in $\RRR^3$. (Left)
  Continuation of the liquid equation of state in the metastable
  region. (Right) Expected behavior in presence of a thermodynamic glass transition.}
\label{fig:dialiquido}
\end{figure*}

The first idea to define amorphous states of hard spheres at high density is to assume
that the liquid phase can be continued above the freezing density $\f_f$ and to look at its properties
at large density~\cite{AC04,KL07}. 
For instance one can choose a functional form that represents well 
the equation of state of the liquid below $\f_f$ (\eg the Carnahan-Starling or Percus-Yevick equation of 
state~\cite{Hansen}) and assume that it describes the liquid phase also above $\f_f$.
On increasing the density the pressure of the liquid increases as the
average distance between particles decreases: one may expect that it 
diverges at a point where the particles get in contact with their neighbors,
and the system
cannot be further compressed. Then one might identify this point with the
random close packing density, see left panel of Fig.~\ref{fig:dialiquido}.

The first objection that has been raised against 
this proposal is that an intrinsic stability limit of the liquid 
(a spinodal point) might exist at a density above $\f_f$ due to thermodynamic or kinetic
reasons. It is probable that a thermodynamic spinodal does not exist,
because any reasonable continuation
of the liquid equation of state does not predict such an instability (manifested \eg by an infinite
compressibility). A kinetic spinodal, related to the existence of the crystal
\cite{CAL05}, could instead exist, at least in monodisperse systems. This
would imply the impossibility of reaching amorphous jammed states if the
compression rate is not very high.
We will assume in the following that the metastable liquid can be compressed
as slow as wished avoiding crystallization. This is not a very good
assumption for monodisperse three-dimensional spheres but seems to be more close 
to reality for larger dimension~\cite{SDST06}, see figure~\ref{fig:Skoge}, or for
suitably chosen binary mixtures. 
This point requires a better investigation, \eg following the analysis of \cite{CAL05},
and we will not discuss it further here.

A second objection is that in presence of a first order phase transition the
continuation of one phase into its metastable region is not well defined due to the appearance
of essential singularities at $\f_f$. Many possible continuations of the
low density equation of state above $\f_f$ are possible: the properties
of the ``liquid'' above $\f_f$ depend on the history of the sample, much as it happens for
the hysteresis of a ferromagnetic system. 
Nevertheless, the ambiguity is expected to be exponentially small in the distance from $\f_f$,
and as the distance between $\f_f$ and the maximum density $\f_{FCC}$ is not so large, one might
expect to obtain a meaningful result in this way. And indeed different possible
continuation of the liquid equation of state (\eg Carnahan-Starling, Percus-Yevick, Hypernetted Chain)
differ by less than $10\%$ in the dense region 
around $\f \sim 0.64$ in $d=3$. Moreover, the ambiguity becomes smaller and smaller on
increasing the dimension. 

Still the extrapolation to infinite pressure can give very different results;
in~\cite{KL07} it is shown that one can reasonably fit the
pressure in the metastable region by a free volume equation of state, and obtain a divergence of
the pressure at $\f_{RCP}\sim 0.64$ in $d=3$. On the other hand,
all possible analytic continuations of the liquid equation of state, based on resummations
of the virial series, predict a divergence of the pressure
at unphysical large values of $\f$: \eg the CS equation predicts a divergence in $\f = 1$ which is clearly
wrong since it is larger than the FCC value and moreover implies that the available volume is 
completely covered by the spheres. Moreover, the $g(r)$ computed using these resummations do not show
the characteristic features observed in amorphous jammed packings.
Therefore, from a theoretical point of view it is very difficult to justify the validity of the
phenomenological free volume fit discussed in \cite{KL07} on the basis of an analytical continuation
of the liquid equation of state.
Clearly we cannot exclude that a more refined resummation of the virial series will give a continuation
of the liquid branch with a divergence in $0.64$ and a $g(r)$ similar to the one of amorphous packings;
but such theory has not been found yet. This motivates the study of different proposals, such
as the existence of a phase transition in the liquid branch.
This idea is also supported by accurate integral equations based on improvements
of the PY equation that seem to indicate the existence of a phase transition at 
high density~\cite{RL03}.

\subsubsection{The ideal glass transition}
\label{sec:idealglass}

In this section we discuss the existence of a thermodynamic glass transition in
the metastable liquid branch of the phase diagram~\cite{SW84,WA81,CFP98,Sp94,Sp98}.
Such a transition can be expected for various reasons: it is predicted by mean-field 
models~\cite{BM01,PTCC03,RBMM04}, and is suggested by the results of~\cite{RL03}.
Moreover, from a dynamical point of view, 
an ergodicity breaking transition is predicted by Mode-Coupling theory~\cite{BGS84,MU93,Go99,SBT05},
and a recently published set of accurate experimental and numerical data~\cite{BEPPSBC08,BW08,BW09}
on three dimensional spheres
strongly indicates a divergence of the equilibrium relaxation time in the metastable liquid phase
at a density $\f_0$ at which the pressure is still finite. This supports the existence of
a phase transition at $\f_0$ in three dimensions. In two dimensions the situation seems quite 
different~\cite{SK00,DTS06,Ta07}.

Still, the existence of an ideal glass transition in a finite-dimensional
system is a matter of intense debate, see \eg~\cite{XW01a,XW01b,BB04,SK00,DTS06,Ta07,BR04,Ca09}.
To avoid entering in this delicate discussion, for the moment, we will take a more ``pragmatic'' point of view.
We will assume that a thermodynamic glass transition exists and investigate its
consequences. It will become clear in a moment that for the comparison with numerical and
experimental data the existence of a true thermodynamic transition is not a key issue, since in real life
one is always stuck in metastable glassy states.

So let us assume again that the liquid phase can be continued above $\f_f$ and neglect the
(small) ambiguity in its definition due to its intrinsic metastability with respect 
to crystallization. We assume that at a density $\f_K$ a {\it thermodynamic glass transition} 
(sometimes called {\it ideal glass transition}) happens\footnote{The density $\f_K$ has
been associated with the name of Kauzmann, who first observed that in some molecular glasses,
extrapolating the liquid entropy at low temperature, one would obtain at some point an entropy smaller
then that of the crystal~\cite{Ka48}. Based on the intuitive notion that a liquid should have more states than a crystal,
hence more entropy, he concluded that some kind of transition should happen preventing the liquid entropy
to cross the crystal one. However, in the case of hard spheres, it is well known that the crystal entropy becomes bigger
than that of the liquid at the freezing/melting transition. Hence the entropy of the liquid
is smaller than that of the crystal in the whole metastable liquid branch. At $\f_K$ the {\it complexity} 
(that will be introduced below) vanishes; but this quantity has nothing to do with the difference between liquid and crystal
entropies in the case of hard spheres.
}.
The transition is signaled by a jump in the compressibility of the system. A
simple qualitative argument
to explain this is the following: in the dense liquid phase particles vibrate on a fast time
scale in the cages made by their neighbors, while on a much larger time large scale
cooperative relaxation processes happen ({\it structural relaxation}). 
If we change the density by $\D \f$ the pressure
will instantaneously increase by a $\D P_0$; very rapidly the average size of the cages will
decrease a little due to the increase in density and the pressure will relax to a value
$\D P_f < \D P_0$. Then, on the time scale of structural relaxation, the structure will change
to follow the change in density and the pressure will relax further to a value $\D P_\io < \D P_f$.
We assume that at the glass transition the latter relaxation is frozen and 
the corresponding time scale becomes infinite: in other words, as usually done in the glass
transition literature~\cite{EAN96},
we identify $\f_K$ with the density $\f_0$ defined in~\cite{BEPPSBC08,BW08,BW09}.
Thus, in the glass phase the increase in pressure following a change in density will
be larger than in the liquid phase, leading to a smaller compressibility $K = \f^{-1} (\D \f/\D P)$.

The schematic phase diagram that we expect in presence of a glass transition is in 
right panel of Fig.~\ref{fig:dialiquido}.
The existence of a glass transition can cure the behavior
of the pressure of the liquid, that seems to diverge at a density bigger than the FCC density.
The pressure of the glass diverges at a ``Glass Close Packing'' density $\fICP$
and we could be tempted to identify
$\fICP$ with the RCP density. 
In next section we discuss a further complication: the existence of 
a very large number of glassy states, in addition to the ideal (thermodynamic) glass, 
with different densities.
Before concluding this section, we wish to stress again that, although recent very accurate numerical
data~\cite{BW09} support the existence of a glass transition of the type discussed in this
section for Hard Spheres, this remains one of the most debated problems in the community.

\subsubsection{Many glassy states: the ``mean-field'' phase diagram}
\label{sec:MFphasediagram}

The glass transition, in the standard picture coming from
the analysis of mean field models, is related to the appearance of 
many {\it metastable glassy states}\footnote{There are here two ``types'' of metastability: the first
is the metastability of the whole amorphous branch with respect to crystallization. We are now
discussing a second type of metastability, \ie the existence of glassy states that have lower
density (or smaller entropy) with respect to the glassy state with maximal density (the ideal one) and consequently
they are metastable with respect to it.
In order to avoid confusion, we will indicate that some of the glassy states are metastable with
respect to the ideal glass only when needed. Otherwise, we will use ``glassy states'' to indicate
both metastable glasses and the ideal glass. In any case these states will have a small, but finite probability of decaying into
the ideal state; this probability should go exponentially to zero when their density approaches the density of the ideal glassy state.
}
in addition to the ideal glass one. 
These states appear in the liquid above some density $\f_d$ and can be defined for instance as minima of a suitable density
functional~\cite{TAP77,MPV87,DV99,CKDKS05,KW87} (see Appendix~\ref{sec:metastability} for a more detailed discussion).
In the interval of densities $\f_d \leq \f \leq \f_K$, 
particles vibrate around these locally stable structures, that are visited subsequently
on the scale of the structural relaxation.
If we ``artificially'' freeze the structural relaxation
and compress the system, the pressure increases faster than if the structure is allowed to relax
for the same reason as above: the system is forced to reduce the size of the cages to respond
to a change in density. The pressure diverges at the point where the particles get in contact
with their neighbors and the average size of the cages is zero.
In this way, {\it to each configuration of the liquid at a given density} $\f \in [\f_d,\f_K]$
one can associate a jammed configuration at a density $\f_j(\f)$ that is obtained by compressing
this configuration fast enough to avoid structural relaxation. A ``glass state'' can be roughly
thought as a set of configurations leading to the same jammed configuration after a fast 
compression\footnote{In
  systems with soft potentials a similar procedure has been proposed in
  \cite{SW82,SW85} in order to
  associate to each configuration of energy $e$ a mechanically stable
  configuration or ``inherent structure'' of lower energy $e_{IS}$:
  starting from the reference configuration one quenches the system
  at zero temperature and finds a minimum of the potential, which is the
  corresponding inherent structure. We are describing the same procedure, with
  energy replaced by (inverse) density and temperature by (inverse) pressure.
  See \cite{Sp98} for details.
}~\cite{Sp98}.

In the Lubachevsky-Stillinger protocol discussed before, 
(see Fig.~\ref{fig:Skoge}) one
chooses {\it a priori} a compression rate $\g$ and the system can equilibrate only
up to a density $\f_g(\g)$ where the relaxation time becomes of the order of the compression rate.
At this density the structure cannot follow anymore the compression and basically the system
responds to the compression by reducing the cages up to the jamming density $\f_j(\g)$.
Calling $\f_{th}$ the value of the jamming density
of the states that first appear at $\f_d$,
it follows that (in the mean field picture) we can produce {\it jammed}
configurations in a whole range of densities $\f_{th} \leq \f \leq \fICP$~\cite{Za07,KK07}.
The notation $\fICP$ (Glass Close Packing density) is appropriate since $\fICP$ is the highest 
possible density of infinite pressure glassy states.

\begin{figure*}[t]
\centering
\includegraphics[width=8cm]{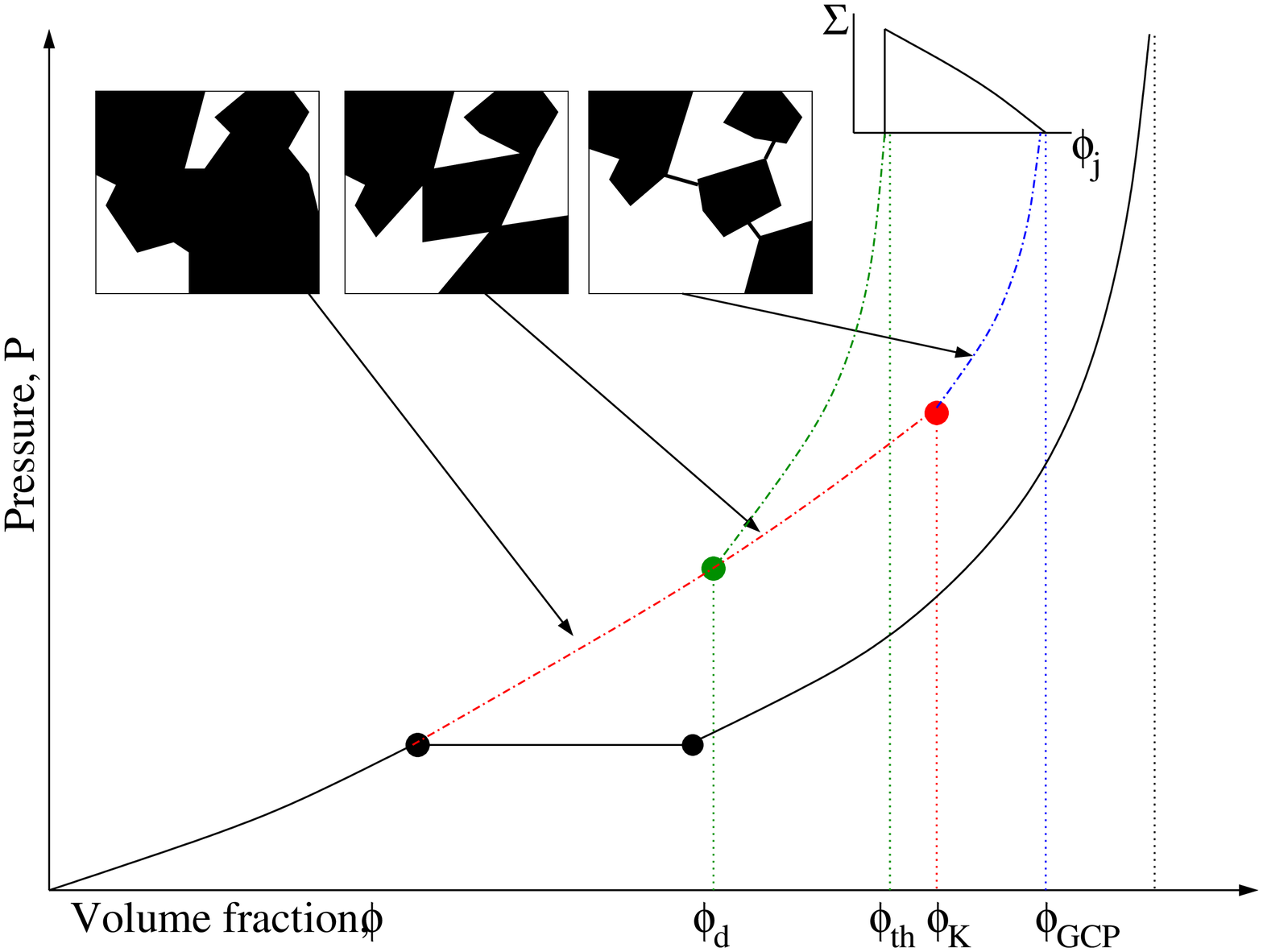}
\includegraphics[width=8cm]{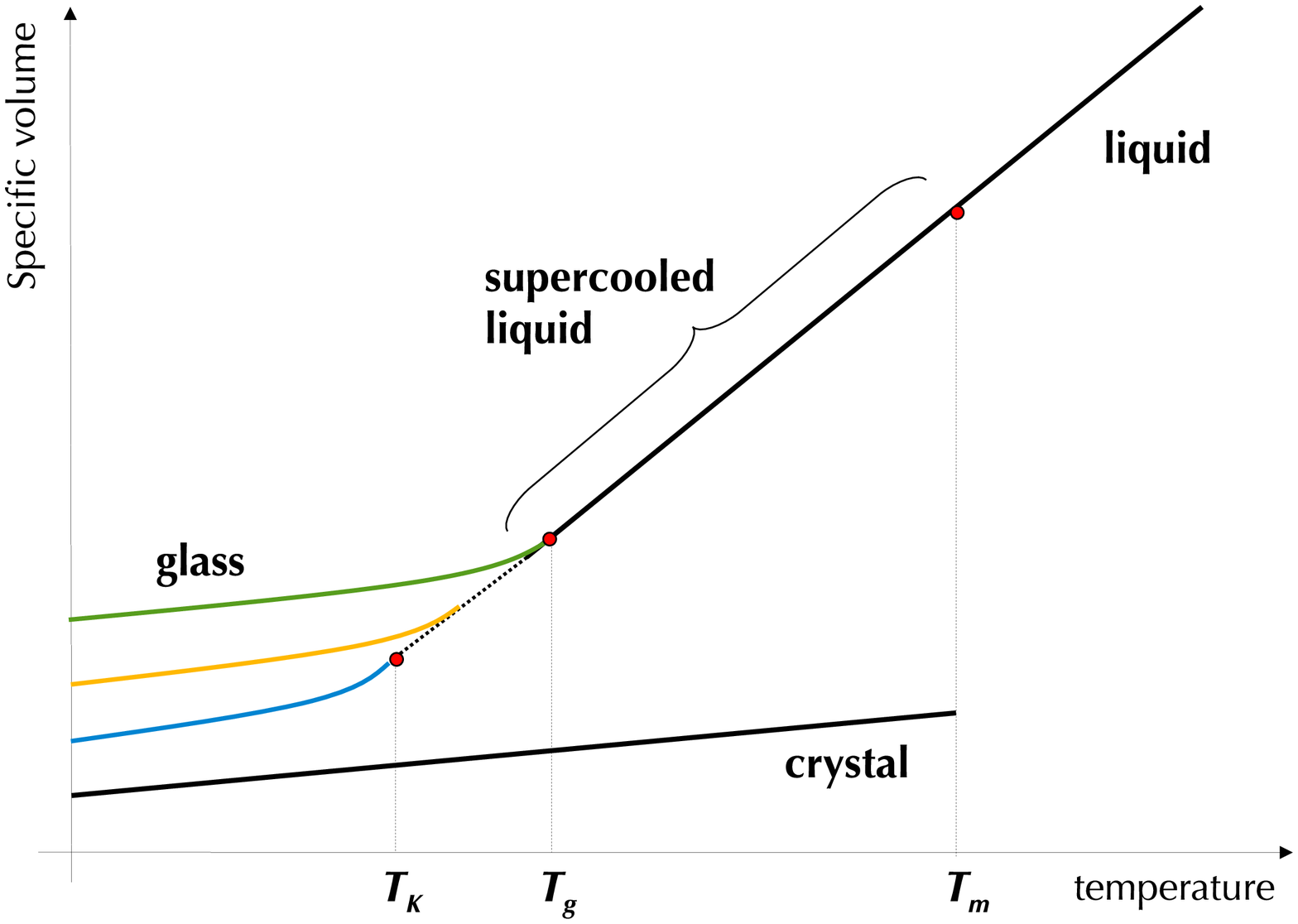}
\caption{Schematic mean-field phase diagram of hard spheres in $\RRR^3$,
see text for a detailed description. 
(Left) $(P,\f)$ diagram:
The full black line represents the
equilibrium phase diagram with the liquid-solid transition 
(see figure~\ref{fig:dialiquido}).
The metastable liquid is made by a single state below $\f_d$, while
above this density it is the superposition of many glassy states. If the
system is stuck in one of these states and compressed, it follows one of
the glass branches of the phase diagram. At $\f_K$ the system reaches
the most dense states, and if further compressed enters the ideal glass state.
The pressure of the latter diverges at $\fICP$. In the inset, the complexity,
\ie the logarithm of the number of glassy states, is plotted as function of
the jamming density $\f_j$. The boxes show a schematic picture of the ($dN$-dimensional) 
phase space of the system: black configurations are allowed, white ones are forbidden by the
hard-core constraint. In the supercooled liquid phase the allowed configurations form a connected
domain; however, on approaching $\f_d$ the connections between different metastable regions 
become smaller and smaller. Above $\f_K$, they disappear in the thermodynamic limit and glassy
states are well defined.
(Right) For comparison, the standard glassy phase diagram for a soft potential 
(\eg Lennard-Jones), specific volume vs
temperature at fixed pressure,
is reported~\cite{EAN96,DS01,Ca09}. The similarity is evident if one identifies
$v \to 1/\f$ and $T\to 1/P$ (\ie one should reflect the left diagram on a diagonal line
joining the upper left corner and the lower right corner).
}
\label{fig:diatot}
\end{figure*}

Note that, as the structural relaxation time scale
is expected to diverge on approaching $\f_K$~\cite{BEPPSBC08,BW08,BW09}, at some point it will
fall beyond any experimentally accessible time scale and necessarily the system will be frozen
into a glassy state which is not the ideal glass. 
The {\it ideal} glass states are unobservable in
practice: one will observe instead a {\it nonequilibrium} glass transition to a state that, again,
will depend on the experimental protocol. This is very familiar in the structural glass literature,
and in fact the diagram in figure~\ref{fig:diatot} (left panel) is analogous to the usual
specific volume vs temperature plot (right panel in figure~\ref{fig:diatot})
that characterizes structural glasses,
if one identifies $\f$ with the inverse of specific volume and $1/P$ with temperature.

To summarize, the phase diagram inspired by mean-field models 
is reported in Fig.~\ref{fig:diatot} and in table~\ref{tab:replica}. 
It is characterized by the presence of the
following phases and transitions:
\begin{itemize}
\item At equilibrium, a low density-liquid phase and a 
high-density crystal phase separated by a first-order transition, with corresponding
freezing density $\f_f$ and melting density $\f_m$ (black full line and black dots in
Fig.~\ref{fig:diatot}).
\item A metastable dense liquid phase; above some density $\f_d$ this phase is made by a
collection of glassy states corresponding to locally stable configurations around which
the system vibrates for a long time. Thus, $\f_d$ is defined, in a static framework, as the
density at which glassy states 
first appear\footnote{At the mean-field level this point 
corresponds to the Mode-Coupling transition $\f_{MCT}$} and for this reason is often
called {\it clustering transition density}\footnote{Glassy states are 
called {\it clusters} in optimization problems}.
\item The liquid exists up to a density $\f_K$ where an {\it ideal glass transition} happens.
The ideal glass transition is signaled by a jump in the compressibility and by a divergence
of the equilibrium relaxation time of the liquid\footnote{Mode-Coupling Theory would predict
a divergence of the relaxation time at the smaller density $\f_{MCT}\sim\f_d$; however, in a finite
dimensional system there are very strong arguments indicating that the liquid can be equilibrated
up to $\f_K$, thanks to activated processes that allow to jump over the barriers separating metastable
glassy states, see Appendix~\ref{sec:metastability}.
}~\cite{BEPPSBC08,BW08}.
\item For each density $\f \in [\f_d,\f_K]$, a different group of glassy states
dominate the partition function. These can
be followed by compressing very fast, and each group is characterized by a jamming density $\f_j(\f)$.
We call $\fICP = \f_j(\f_K)$ the jamming density of the ideal glass state, and 
$\f_{th} = \f_j(\f_d)$ the jamming density of the less dense glassy states.
\end{itemize}
The number of glassy states corresponding to jamming densities 
$\f_j \in [\f_{th},\fICP]$ grows {\it exponentially}
with the size of the system, $\NN(\f_j) = \exp[N\Si(\f_j)]$. The function $\Si(\f_j)$,
usually called {\it complexity} or {\it configurational entropy}\footnote{
Although sometimes used as synonyms, in other cases these words denote different concepts:
in fact, especially in the experimental literature the word ``configurational entropy'' denotes
the difference between the entropy of a system in its liquid and crystalline phases. This difference
is often used as an estimate of the complexity. Already in the case of soft potentials, this gives
rise to a number of interpretation problems; see \eg \cite{BK05,Ca09} for a more detailed discussion.
In the particular case of hard spheres, the situation is even worse since energy is irrelevant and 
the liquid-crystal transition is completely driven by entropy; hence the crystal entropy is bigger
than the liquid one above the melting density.  
It follows that the configurational entropy defined as the difference between liquid and crystal 
entropy has nothing to do with the complexity; in particular it will be negative at all densities
above the melting density. In this paper, we will always use the word ``complexity'' in order to
avoid confusion.
}, has the
shape reported in the inset of Fig.~\ref{fig:diatot}: it jumps at a positive value at
$\f_{th}$ and vanishes at $\fICP$ (therefore, the number of {\it ideal glasses} is {\it not}
exponential in the system size).
A more precise definition of this
equilibrium complexity is presented
in Appendix~\ref{sec:metastability}.

\subsubsection{Remarks on the static phase diagram}

The striking similarity of the phase diagram in Fig.~\ref{fig:diatot}
with the numerical results\footnote{Recall that in Fig.~\ref{fig:Skoge} the pressure has been
rescaled.} of \cite{SDST06}, see Fig.~\ref{fig:Skoge}, makes it
a good starting point to understand the physics of jammed amorphous states.
However one should keep in mind the following important remarks:
\begin{enumerate}
\item This phase diagram is an idealization that discards many difficulties in the definition 
of amorphous packings,
mainly related to {\it metastability}, either of metastable glass states with
respect to the ideal glass state, or of the ensemble of glassy states with respect to the crystal. 
The ambiguity in the definition of the liquid equation of
state due to its metastability \cite{KL07} affects also the glass: the glass equation
of state is {\it theoretically} not well defined, with an ambiguity of
the order of $10\%$, depending on the equation of state one chooses to describe
the liquid.
Note that these difficulties might be not so important in some cases where the nucleation time 
of the crystal is very large, \eg high-dimensional spheres or binary mixtures, as discussed above.
In particular, in the limit
$d\to\io$ we believe that the theory should be exact.
\item In a finite dimensional system, pure equilibrium states
cannot exist in exponential number. 
In fact, the states corresponding to $\f_{j} < \fICP$
can be stable only on a finite length scale and for a finite time in finite
dimensional systems as discussed originally in \cite{KW87,KT87}, in more detail
in \cite{KW87b,KTW89}, and more recently in \cite{MP00}.
Therefore the notion of {\it complexity} makes sense only on a finite time and
length scale.
The problem of evaluating this length and time scales has been
recently reformulated in term of a nucleation problem~\cite{XW01a,XW01b,BB04}
and is probably the most active subject of research in the glass transition
community. The subject is difficult and we cannot discuss it in detail here.
The interested reader should look to the original literature~\cite{XW01a,XW01b,BB04,Fr05,MS06,CGV07}.
In the rest of this paper we will neglect this difficulty and assume that mean
field states are stable on every length scale;
however we will try to give a more precise definition of states in
Appendix~\ref{sec:metastability},
where we explicitly show the origin of the
 difficulties in  finite dimension and explain why our theory might be a
 reasonable approximation of the real situation.
\begin{table*}[t]
\centering
\begin{tabular}{|c|c|c|c|}
\hline
Density & Definition & Type & Value in $d=3$ \\
\hline
$\f_{MRJ}$ & Maximally random jammed configurations \cite{TTD00} & Geometrical & $\sim 0.64$ \\
\hline
$\f_d$ & The liquid state splits in an exponential number of states & MF, Static & $\sim 0.58$ \\
$\f_K$ & Ideal glass phase transition - jump in compressibility & Static & $\sim 0.62$ \\
$\f_{th}$ & Divergence of the pressure of the less dense states  & MF, Static & $\sim 0.64$ \\
$\fICP$ & Divergence of the pressure of the ideal glass & Static & $\sim 0.68$ \\
\hline
$\f_{MCT}$ & Mode-Coupling transition \cite{MU93} & MF, Dynamic & $\sim 0.58$ \\
$\f_g$ & Glass transition density; depends on the compression rate & Dynamic & $0.58 \div 0.62$ \\
$\f_0$ & Divergence of the equilibrium relaxation time & Dynamic & $\sim 0.62$ \\
$\f_{J}$ & J-point: final state of the algorithm of~\cite{OLLN02} & Dynamic & $\sim 0.64$ \\
\hline
$\f_{RCP}$ & No general agreement on the definition & ?? & $\sim 0.64$ \\
\hline
\end{tabular}
\caption{
Summary of the relevant densities defined in the text.
The values given in $d=3$ (for a monodisperse system) are indicative,
more detailed values will be given in the following. The label ``MF'' indicates
that the corresponding concept can be defined in mean-field theory, but lacks
an unambiguous operative definition in finite dimension.
}
\label{tab:densities}
\end{table*}
\item The structure of the states close to $\f_{th}$ might be more complicated than described here.
In mean field models
processes such as state
crossing, temperature (density) chaos, birth and death of states, etc. are known to
happen. This complicates the analysis; unfortunately
at present we cannot say much more on this important issue.
More insight should come from the study of mean-field hard sphere models such as
the ones discussed in \cite{MKK08,BM01,KTZ08}.
\item For finite dimensional systems, the transition to a metastable glass state is smeared
and becomes a crossover, as seen in figure~\ref{fig:Skoge}, see also~\cite{EAN96,MGS06}. Only the ideal
glass transition has a chance to survive as a real phase transition. For this reason the free volume
fit of \cite{KL07}, that describes the metastable liquid branch without assuming a phase transition,
is not incompatible with the point of view presented here.
\item When looking to actual configurations, it might be very difficult to distinguish between
a configuration representative of a pure glass state and one representative of a mixture of
glass and crystal~\cite{TTD00}. Thus if one really wants to look at {\it single configurations of finite
systems}, the definition
of amorphous states discussed above is not very useful, and a definition based on order metrics
may be more suitable~\cite{TTD00}. This leads to the concept of maximally random jammed packings (MRJ),
defined as follows: one chooses a function $\psi$ that measures the order in some way, with
$\psi = 1$ corresponding to most ordered and $\psi = 0$ to most disordered configurations. Then one
defines $\f_{MRJ}$ as the density of the jammed configurations that minimize $\psi$, 
see~\cite{TTD00} for details. 
This definition may be suitable when one studies single configurations, and it has
been shown numerically that a value $\f_{MRJ} \sim 0.64$ is obtained for many different order metrics.
It is therefore important to stress that the phase diagram above refers to 
{\it thermodynamic states} and not to configurations.
Note that it is likely that the local order in the configurations typical of the infinite pressure
states at density $\f_j$ depends on $\f_j$, \ie $\la\psi\ra(\f_j)$ is not constant; this does
not necessarily mean that the packings at higher $\f_j$ cannot be considered as amorphous, since
$\la \psi \ra(\f)$ increases with density also in the liquid state, which is definitely an
amorphous state at all densities.
\end{enumerate}

Before turning to the technical part, it is useful, having in mind this phase diagram,
to come back to the behavior of algorithms that are used to construct jammed amorphous packings,
and to review some basic features that are observed in the structure of such packings.

\subsection{On the protocol dependence of the random close packing density}

Up to now we discussed
a specific algorithm: a molecular dynamics simulation during which the system is compressed 
at a given rate, the Lubachevsky-Stillinger algorithm.
In a more general setting, we can consider an algorithm or experimental protocol attempting to produce
jammed amorphous configurations of an hard sphere system.
The algorithm stops when the system is jammed: the final
density is a random variable depending on the initial data and possibly
on some randomness built in the algorithm itself. 
To investigate the
probability $\PP(\f_j)$ of reaching a final density $\f_j$, let us make the
assumption that {\it when the system is jammed, each glassy state contains
  only one configuration}\footnote{If this assumption is false one
  should change the definition of complexity but the following argument would
  remain true. See Appendix~\ref{sec:metastability} for a more detailed
  discussion.}.
Then the number $\NN(\f_j)$ of jammed states with density $\f_j$ is given by
$\NN(\f_j)\sim\exp[N \Si(\f_j)]$ where $\Si(\f_j)$ is the complexity
introduced above.

Let us define also the probability $\PP_a(\f_j)$
that the algorithm finds {\it one particular configuration} with density
$\f_j$. This probability is related to the ``basin of attraction'' of configurations,
\ie to the number of trajectories of the algorithm that lead to the final configuration.
Again, on the basis of concentration arguments, we expect the scaling
$\PP_a(\f_j) \sim \exp[N s_a(\f_j)]$.

Note that
$\NN(\f_j)$ depends only on geometrical
properties of the configuration space of hard spheres, while $\PP_a(\f_j)$ encodes the properties
of the algorithm~\cite{XBO05,KK07}.
The resulting $\PP(\f_j) = \NN(\f_j) \PP_a(\f_j) \sim \exp[ N
(\Si(\f_j)+s_a(\f_j))]$ 
will be strongly peaked around a given value $\f_j^a$,
the maximum of $\Si(\f_j)+s_a(\f_j)$ \cite{KK07}.
The important point is that $\f^a_j$
will depend on the particular algorithm through the function
$s_a(\f_j)$, giving rise to the protocol dependence observed in experiments and numerical 
simulations~\cite{XBO05,CBS09}.

Therefore it is highly non trivial to associate the phase diagram of Fig.~\ref{fig:Skoge}
to the behavior of a given algorithm or experimental protocol. The same problem
has been extensively discussed in the context of optimization problems, where a similar
phase diagram is found, without reaching yet many conclusive statements.
We can only make some remarks and conjectures, some of them already discussed in the
introduction to this section:
\begin{enumerate}
\item In the static computation, infinite pressure states are found only in the interval $[\f_{th},\fICP]$. 
One cannot exclude the existence of states with very small volume in phase space that are
not seen in the static computation. However, due to the smallness of their volume, to observe
them one should design a very specific algorithm. Thus we can assume that reasonable procedures,
that allow for some exploration of phase space, will always find states with jamming density
$\f_j \in [\f_{th},\fICP]$.
\item In the case of equilibrium dynamics (\eg a molecular dynamics simulation
with infinitely slow compression), the system is expected to equilibrate at each
density. Thus, neglecting crystallization, we expect the system to follow the liquid
branch up to $\f_K$ and then the ideal glass branch up to $\fICP$.
\item The Mode-Coupling theory instead predicts a divergence of the relaxation time
at $\f_{MCT} < \f_K$. However in the context of structural glasses it is well known that
the Mode-Coupling transition is avoided and becomes a crossover to a nonequilibrium glass
state at a compression-rate dependent $\f_g$. At the mean field level, the Mode-Coupling
transition corresponds to the point $\f_d$ where the liquid state breaks into an exponential
number of disconnected states.
\item The equilibration is particularly slow above $\f_{MCT}\sim \f_d$, being due
to {\it activated processes}~\cite{XW01a,XW01b,BB04,Fr05}
\footnote{In the constant-pressure ensemble the times are proportional to $\exp(P \Delta V)$, where $\Delta V$ is
the volume barrier to go from one state to another state}.
For moderately high rates, the system will leave the liquid branch very close
to $\f_d$ and will jam at $\f_{th}$. Thus the procedure of O'Hern et al.~\cite{OLLN02},
that correspond to a very fast compression, should produce packings around $\f_{th}$.
At the mean field level one would have $\f_J \sim \f_{th}$. However one should keep in mind
(see Appendix~\ref{sec:metastability}) that states close to $\f_{th}$ are particularly unstable
in finite dimension and therefore $\f_{th}$ is an ill-defined concept in this case.
\item Remarkably, the value $\fICP$ defined as the point where $\Si(\f_j)$ vanishes is a
property of the system that does not depend on the algorithm (at least if one
accepts the idealizations discussed in the previous subsections, \ie if one
neglects the existence of the crystal).
\item The behavior of more complex algorithms should be discussed on a case-by-case basis.
Thus, if one defines the random close packing density as the final density for a given
protocol, its value can be everywhere in $[\f_{th},\fICP]$ (and maybe outside this interval,
for very particular protocols). Depending on the personal taste, one would like to identify
$\f_{RCP}$ with $\fICP$, if one is thinking to slow quasi-equilibrium compressions, or with
$\f_{th}\sim \f_J$, if one is thinking to fast quenches.
\end{enumerate}
In table~\ref{tab:densities} we summarize the different densities defined up to now.

\begin{table}[t]
\centering
\begin{tabular}{|c|c|c|}
\hline
Point & Densities & Replica parameter \\
\hline
Dynamic transition & $\f_d$, $\f_{MCT}$ & $m=1$ \\
Static transition & $\f_K$, $\f_0$ & $m=1$ \\
\hline
Glass close packing & $\fICP$ & $m=0$ \\
Threshold states & $\f_{th}$, $\f_{J}$(?), $\f_{MRJ}$(?) & $m=0$ \\
\hline
\end{tabular}
\caption{
Special points of the phase diagram obtained within the replica computation, see figure~\ref{fig:diatot}.
For each special point, we list the densities defined in table~\ref{tab:densities} 
associated to it.
Note that $\f_g$ and $\f_{RCP}$ depend on the protocol used,
hence they are not associated to any special point in the replica computation.
The identification of $\f_{J}$ and $\f_{MRJ}$ with $\f_{th}$ is tentative and 
probably not stricly true; still we expect these points to be quite close (see text
for details).
}
\label{tab:replica}
\end{table}

\subsection{Structural properties of amorphous states}
\label{sec:gofr}

Despite the difficulty in determining the final jamming density for a given experimental
protocol, it turns out that many structural properties of the final state, manifested for instance
in the pair correlation function $g(r)$, are roughly independent
of the protocol. This suggests that at least some of these properties are common
to all amorphous packings in the range $\f_j \in [\f_{th},\fICP]$, and are therefore
characteristic of ``random close packed'' structures. We will be able
to compute the correlation function $g(r)$ for these states; therefore it is interesting to
review the main features that are observed in numerical simulations.

Detailed numerical and experimental results for $g(r)$ at jamming have been reported 
in~\cite{CJ93,ASS05,OLLN02,OSLN03,DTS05,DTS05b,SLN06,SDST06}. Depending on the procedure one approaches
the jamming point $\f_j$ from below, $\d \f = \f_j-\f > 0$ \cite{DTS05,DTS05b,SDST06}
for hard particles, or from above, $\d \f = \f- \f_j > 0$, for soft 
particles~\cite{OLLN02,OSLN03,SLN06}.
The main features that are observed are:
\begin{enumerate}
\item A delta peak close to $r=D$, due to particles in contact. One has $g(D) \sim \d\f^{-1}$
and the width of the peak $\propto \d\f$. The delta peak has
the scaling form $g(r)/g(D) = f(\l)$, where $\l \propto \frac{r-D}{D} \d\f^{-1}$, if $\f \to \f_j^-$ for
hard particles~\cite{DTS05}, with $f(t)$ a scaling function independent on $\d\f$.
The area of this peak gives the average number of particles in contact with a reference one,
which is found to be $z=2d$ once rattlers (particles without contacts) are removed;
this property is called {\it isostaticity}.
\item A square root singularity, $g(r) \sim (r-D)^{-\a}$ with $\a = 0.5$, close to $r=D$. This
singularity is integrable and does not contribute to the number of contacts. The value of the
exponent is debated: some claim that it is equal to $0.5$ irrespective of the value of $\f_j$,
the procedure, etc. \cite{SLN06}. 
However some dependence on the procedure has been claimed in \cite{DTS05},
where a value $\a \sim 0.4$ has been found.
\item A dip around $r/D \sim 1.2$ with respect to the liquid is observed. This is due to
the particles of the first shell which are pushed toward contact ($r=D$) for $\d\f \to 0$.
\item A split second peak in $r/D=\sqrt{3}$~\cite{SLN06,DTS05}. 
It is not clear what is the exact behavior of $g(r)$
close to this peak~\cite{SLN06}, in particular if $g(r)$ is divergent or has only divergent slope
for $r/D\to\sqrt{3}^-$. Even if $g(r)$ is not divergent for $r/D \to \sqrt{3}$, 
it has at least a jump in $\sqrt{3}$.
\item A similar behavior is observed in $r/D=2$. Both features have been interpreted in \cite{CJ93,SLN06}
as coming from the network of contacts. In fact, $r/D=2$ is the maximum distance at which two particles
sharing one neighbor can be found, while $r/D=\sqrt{3}$ is the maximum distance for two particles
sharing two neighbors.
\item Long range correlations ($h(r) \sim 1/r^4$ and $S(k) \sim |k|$ for large $r$ and small $k$) 
are found for hard particles
\cite{DTS05b} and seem to be present also in the soft particles case \cite{SLN06}.
This implies also that $S(0) = 0$, \ie the packings are incompressible. 
\item 
A particularly intriguing property of jammed amorphous configurations is the presence of an excess
of {\it soft modes}, \ie vibrational modes with very small frequency. This has been shown numerically
in~\cite{OSLN03,SLN05}. 
In~\cite{WNW05,WSNW05,Wyart} it has been argued that this excess of
soft modes is related to the isostaticity property of the network of contact; moreover, a diverging length
scale has been associated to these modes. It has also been proposed that the square-root singularity
of $g(r)$ is related to these modes~\cite{Wyart}. This set of results seems to suggest a ``critical'' 
nature
of jammed amorphous packings that is currently receiving a lot of interest in the 
community~\cite{MSLB07,OT07,ZSN08,HOS07}.
\end{enumerate}
One of the main aims of the following discussion is to show that the class of packings obtained
as infinite pressure glassy states share at least some of these
features, that are common to all disordered jammed packings produced with different
protocols. We will be able to compute partially the $g(r)$ of these states and show that it is 
consistent with the properties described above.
This fact supports
the main assumption of this paper, that the final states reached by typical algorithms belong
to the class of infinite pressure glassy states.

\section{The method}
\label{sec:method}

Assuming the phase diagram in fig.~\ref{fig:diatot}, 
and neglecting the ambiguities associated to the existence
of the crystal, the properties of the glass phase can be computed
using a replica method inspired by mean field models. 
The method has been described in great detail in a number of papers, see in
particular \cite{Mo95,MP96,Me99,MP99,MP00}; therefore we will briefly sketch it here,
but the reader who is interested in the details and has no previous knowledge
of the replica method should refer to the original papers for a more complete 
presentation.

An alternative route to compute the properties of glassy states is to use density
functional theory~\cite{DV99,CKDKS05,KW87,KM03,SSW85,YYO07}. In principle the two methods
should be equivalent (see~\cite{CC05} for a very pedagogical discussion in the
case of mean field models), still it seems that 
the replica method gives more accurate
quantitative results and for this reason we will focus 
on this method in the following.

\subsection{The replica method}

The phase diagram in figure~\ref{fig:diatot} is characterized by the existence
of many glassy states at densities above $\f_d$.
At constant density, states are characterized by
their {\it vibrational (or internal) entropy} $s$, defined as the entropy of the system 
constrained to be in this state without relaxing toward different states. The derivative
of the internal entropy with respect to density is the pressure of the states, that
is plotted schematically in figure~\ref{fig:diatot}.
Taking a constant $\f$ slice of the phase diagram in figure~\ref{fig:diatot}, 
one will meet different states, depending on the pressure (or equivalently
on the entropy). The number of states of entropy $s$ 
at a given density $\f$ is by definition $\NN(s) = \exp N \Si(s,\f)$.

\begin{figure}[t]
\centering
\includegraphics[width=8cm]{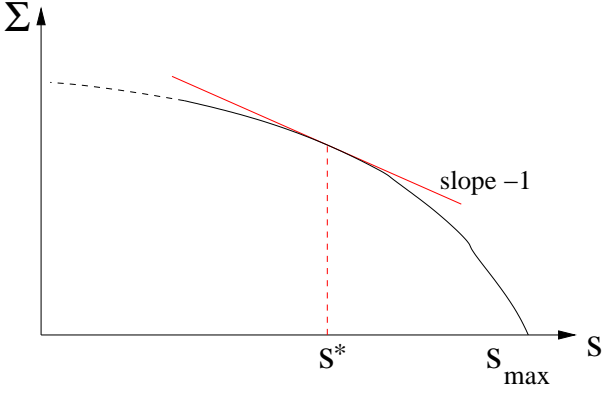}
\caption{
A schematic representation of $\Si(s,\f)$ at fixed $\f$. The behavior at small
$s$ depends strongly on the model and on the density (see \cite{KMRSZ07} for a more
detailed discussion). On increasing $s$, $\Si(s,\f)$ decreases and ultimately vanishes
at a value $s_{max}(\f)$. The value $s^*$ is defined by $d\Si/ds = -1$. For $\f < \f_K$,
$s^* < s_{max}$, while for $\f \geq \f_K$ there is no solution and $s^* = s_{max}$.
}
\label{fig:sigma_s}
\end{figure}

\subsubsection{The ideal glass transition}

The complexity 
$\Si(s,\f)$ (sketched in Fig.~\ref{fig:sigma_s}) is a concave function of $s$; 
it is reasonable to assume $\Si(s,\f)$ to be
a decreasing function of $s$, because at fixed density states of higher entropy correspond to more 
compact structures (in order
to have more free volume) and should be more rare. Moreover $\Si(s,\f)$ should continuously
vanish at some value $s_{max}(\f)$ corresponding to the entropy of the best amorphous 
structures at this density\footnote{As already discussed one can construct 
denser structures by allowing a small amount of local order: we are assuming
to be able to avoid this in some well-defined way, that will be discussed in the following.}.
The partition function of hard spheres at density $\f$ 
is just the total number of allowed configurations
at that density. In the thermodynamic limit, each relevant configuration
belongs only to one state; and $\exp ( N s_\a)$ is
the number of configurations belonging to the state $\a$.
Therefore one can write the partition function $Z$ in the following way:
\beq\begin{split}
\label{Zm1}
Z &= e^{N S(\f)} \sim \sum_\a e^{ N s_\a} 
= \int_{s_{min}(\f)}^{s_{max}(\f)}ds \, e^{N [\Si(s,\f)+s]} \\
&\sim  e^{N [\Si(s^*,\f)+s^*]} \ ,
\end{split}\eeq
where in the last line we performed a saddle-point approximation of the integral;
$s^*(\f)$ is the point where $\Si(s,\f)+s$ assumes its maximum in the interval $[s_{min},s_{max}]$. 
Indeed more compact structures have higher vibrational entropy (free volume) and lower complexity 
(their number), and the partition function is dominated by the best compromise,
as expressed by the saddle point evaluation of (\ref{Zm1}).
For densities $\f_d < \f < \f_K$, the saddle point $s^*$ falls inside the interval $[s_{min},s_{max}]$.
The ideal glass transition is met at the density $\f_K$ such that $s^*(\f_K) = s_{max}(\f_K)$, or equivalently
$\Si(s^*,\f_K)=0$. Above this density, the integral (\ref{Zm1}) is always dominated by the upper limit
of integration, therefore we have
\beq\label{SdiF}
S(\f) = \begin{cases}
\Si(s^*,\f) + s^* \ \ \ \ \f < \f_K \ , \\
s_{max}(\f) \ \ \ \ \ \ \ \ \ \ \f > \f_K \ ,
\end{cases}
\eeq
It is easy to see that (if $\Si(s,\f)$ is a smooth function)
at $\f_K$ a phase transition
happens, and is manifested by a discontinuity in the second derivative of $S(\f)$.
Under our assumptions, close to $s_{max}(\f)$ 
we have $\Si(s,\f) = \Si_1(\f) (s-s_{max}(\f)) + \frac12\Si_2(\f) (s-s_{max}(\f))^2 + \cdots$.
$\f_K$ is defined by $\Si_1(\f_K)=-1$ and
close to $\f_K$ we have $s^*(\f) \sim s_{max}(\f)- (1+\Si_1(\f))/\Si_2(\f)$. The derivative
of $S(\f)$ for $\f \to \f_K^-$ is given by 
$S'(\f) = \frac{\partial \Si}{\partial \f}(s^*,\f) \sim -\Si_1(\f) s_{max}'(\f) \sim s_{max}'(\f)$
which coincides with the derivative for $\f\to\f_K^+$. Therefore the pressure is continuous at $\f_K$,
while it is easy to show that the compressibility jumps at $\f_K$, as discussed in 
section~\ref{sec:idealglass}.

\begin{figure*} \centering
\includegraphics[width=8cm]{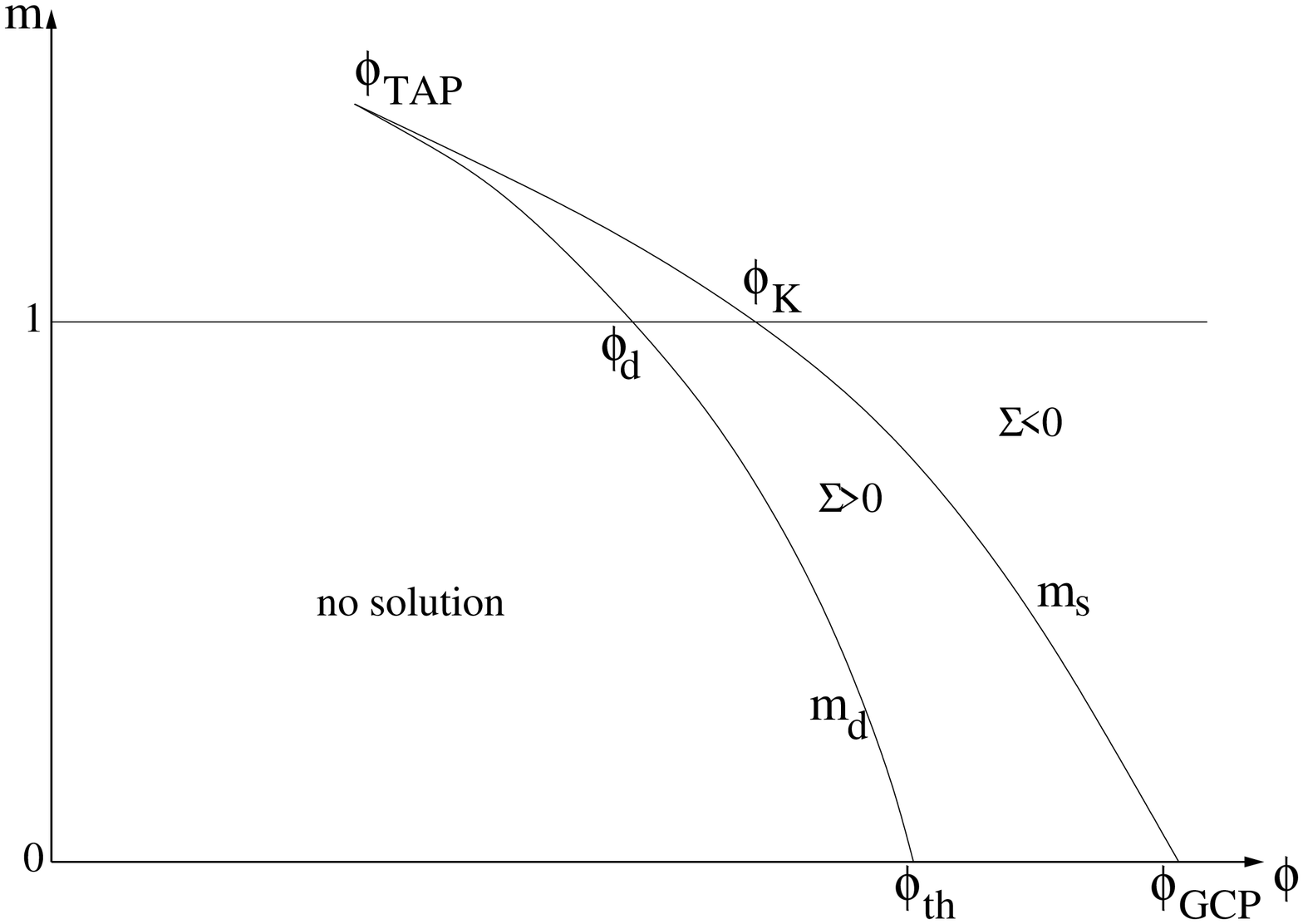}
\includegraphics[width=8cm]{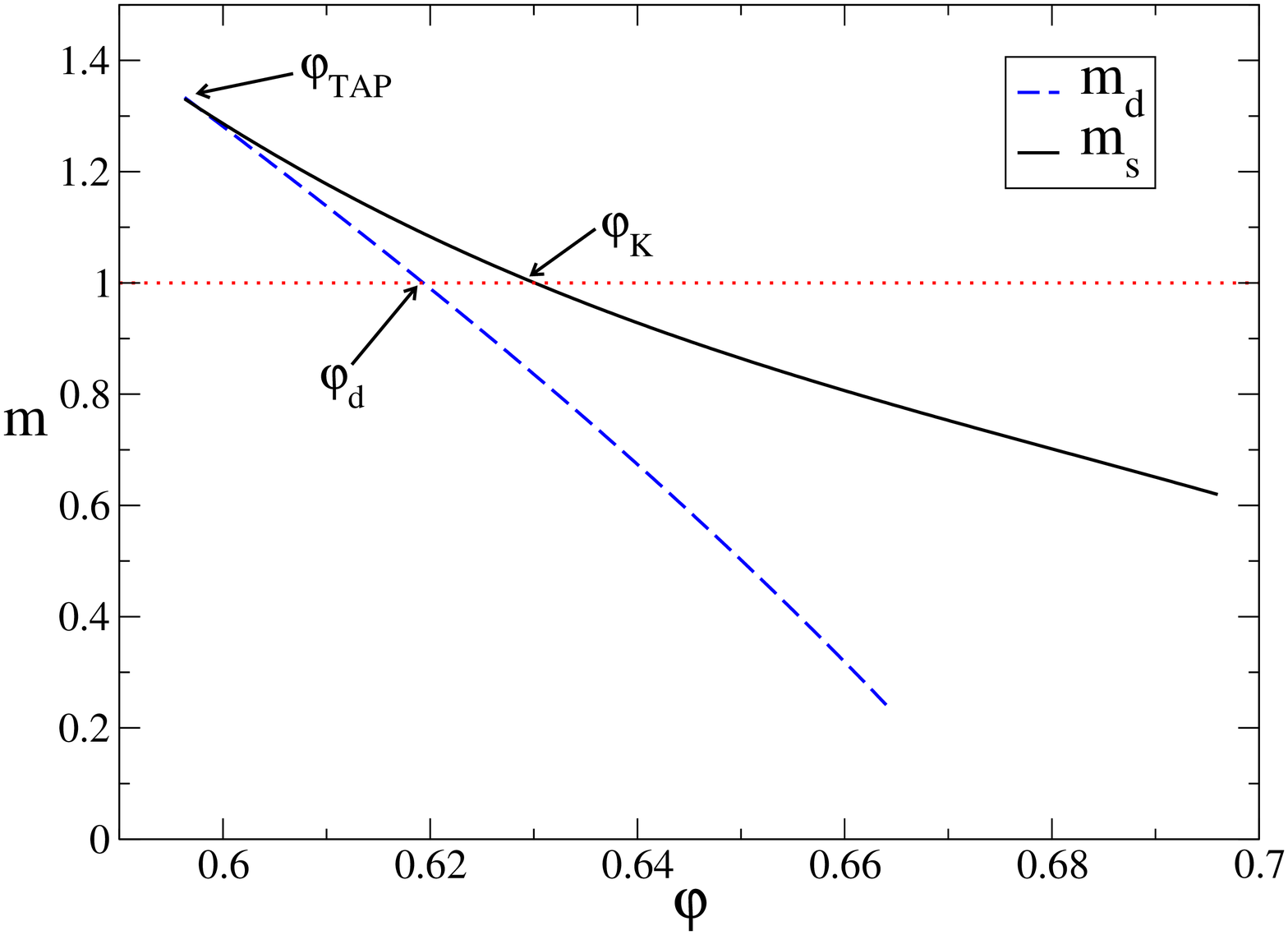}
\caption{
(Left) Schematic $(m,\f)$ diagram: above the clustering line $m_d(\f)$ a non
trivial solution for the inter-replica correlation is found. This solution gives
a positive complexity in the region enclosed between the lines $m_d$ and $m_s$, therefore
in this region glassy states are present. The
line $m_s(\f)$ is defined by the condition $\Si(m,\f)=0$ and corresponds to the
ideal glass state. The intersections of the line $m_d$ ($m_s$) with $m=1$ and $m=0$ define
$\f_d$ ($\f_K$) and $\f_{th}$ ($\fICP$), respectively. 
(Right) Phase diagram for $d=3$, as obtained by solving
the replicated HNC equations (section~\ref{sec:HNC}). 
Note that the static line has an unreasonable behavior
at small $m$, 
and that the densities $\f_d$ and $\f_{th}$ seem too big if compared with the accepted
$\f_{MCT}\sim 0.58$ and $\f_{J} \sim 0.64$ values. 
}
\label{fig:mstar}
\end{figure*}

\subsubsection{The replicated partition function and the $(m,\f)$ phase diagram}

The basic idea of the replica approach \cite{Mo95,MP99}
is to introduce in (\ref{Zm1}) a parameter $m$ conjugated to the internal entropy of the states.
One can think to $m$ as a control parameter that at fixed density allows to
select a given group of states.

In practice this can be done by considering $m$ copies of the original system, constrained to 
be in the same state by
a small attractive coupling (the practical implementation will be discussed in the following). 
The partition function of the replicated system is then
\beq
\label{Zm}
\begin{split}
Z_m &= e^{N \SS(m,\f)} \sim \sum_\a e^{N m s_\a} \\ &= \int_{s_{min}(\f)}^{s_{max}(\f)}ds \, 
e^{N [\Si(s,\f)+m s]} 
\sim  e^{N [\Si(s^*,\f)+ m s^*]} \ ,
\end{split}\eeq
where now $s^*(m,\f)$ is such that $\SS(m,s)=m s + \Si(s,\f)$ is minimum.
The introduction of the coupled replicas has exactly the effect of giving a weight $m$ to
the vibrational entropy in (\ref{Zm}).  Only for integer $m$ the quantity $Z_m$ has an explicit definition
(and it can be evaluated by direct numerical simulations),
but if $m$ is allowed to assume real values\footnote{This must be done by analytical continuation
once the partition function for integer $m$ has been computed using some approximation. In principle,
the analytic continuation might not be well defined, but it has been verified explicitely in mean
field models that the procedure gives the correct results; see \cite{Mo95,MP99,MPV87,Ta03,Me99,FT06} for a more
detailed general discussion of the replica method.
}, 
the complexity can be estimated from the knowledge
of the function $\SS(m,\f)=m s^*(m,\f) + \Si(s^*(m,\f),\f)$. 
Indeed, it is easy to show that
\beq
\label{mcomplexity}
\begin{split}
s^*(m,\f) &= \frac{\partial \, \SS(m,\f)}{\partial m} \ , \\
\Si(m,\f) &= \Si(s^*(m,\f),\f) =- m^2 \frac{\partial \,[ m^{-1} \SS(m,\f)]}{\partial m} \\ &= 
\SS(m,\f) - m  s^*(m,\f)  \ .
\end{split}
\eeq
The function $\Si(s,\f)$ can be reconstructed from the parametric plot of $s^*(m,\f)$ and $\Si(m,\f)$.

For each density, one can define a point $m_s(\f)$ 
as the solution\footnote{Note that the condition
$\Si(m,\f)=0$ is equivalent to $\frac{\partial (\SS/m)}{\partial m} = 0$, which correspond 
to the usual optimization of the free energy with respect to $m$ in the 1RSB computations.}
of $\Si(m,\f)=0$, see left panel of 
figure~\ref{fig:mstar}.
On the line $m_s(\f)$, from (\ref{Zm}) we see that $\SS(m,\f) = m s_{max}(\f)$, then
\beq
S_{glass}(\f) \equiv s_{max}(\f) = \frac{\SS(m_s(\f),\f)}{m_s(\f)} \ .
\eeq
This simple prescription allows to compute the entropy of the glass once
$\SS(m,\f)$ is known\footnote{A very important remark~\cite{MP99} is that the
  entropy of the replicated liquid $\SS(m,\f)$ is analytic as long as $m <
  m_s(\f)$. In fact, the introduction of $m$ shifts the phase transition that
  happens for $\f=\f_K$ at $m=1$ to higher values of density for $m < 1$ (see
  figure~\ref{fig:mstar}). Therefore $\SS(m,\f)$ can be computed, in the whole
  interesting glassy region, by analytic continuation of the low density
  (replicated) liquid entropy.}

Generically, the entropy $\SS(m,\f)$ turns out to be the maximum of a functional of some order parameter,
that typically represent the inter-replica correlation functions, and is
obtained by maximizing the functional
(explicit examples will be given in the following).
For each density, the inter-replica correlations are non-zero only above some
value $m_d(\f)$, and in this case the replicas are in the same state. In this region we
obtain non-trivial values of $\Si(m,\f)$, thus $m_d(\f)$ corresponds to the minimum value $s_{min}(\f)$
below which there are no states and $\Si=0$.
We will call $m_d(\f)$ {\it clustering} line, because above this line the space of configuration
is disconnected in many clusters corresponding to the glassy states.
The two lines $m_d(\f)$ and $m_s(\f)$ in the plane $(m,\f)$ define a phase diagram which is schematically reported
in figure~\ref{fig:mstar} and mirrors the $(P,\f)$ phase diagram; in fact $m$ is conjugated to the
entropy, which is related to the pressure; the two phase diagrams
are related by a Legendre transform.
The lines $m_s$ and $m_d$ touch at some value $\f_{TAP}$ below which there are no states
except the liquid one. Above $\f_{TAP}$, states are found for $m_d(\f) \leq m \leq m_s(\f)$.
When the line $m_d(\f)$ reach the line $m=1$, states begin to be present in the liquid phase:
this correspond to the point $\f_d$. When the static line $m_s$ crosses $m=1$, the liquid ceases
to exist and the ideal glass transition is met.

\subsubsection{The equation of state of metastable glassy states}
\label{sec:metastable}

The replica method allows to compute, for a given density, the function $\Si(s,\f)$.
But to access the equation of state of the metastable glassy states we would like to follow the
evolution of each state at different densities.
This is in general a very complicated problem already at the mean-field 
level.
Therefore in order to be able to perform the computation, 
we make a very strong additional assumption
on the phase space structure of the model: namely, that each state is labeled uniquely 
by its maximum possible density or {\it jamming density} $\f_j$.
This is the maximum value of density for this given
structure, corresponding to infinite pressure, where particles are in contact with their neighbors.
We assume that if one starts from the jammed structure at $\f_j$ and slowly decrease the density, 
the particles are allowed to vibrate slightly around the original structure but the state maintains its
identity until it merges with the liquid.
In other words, we assume that there are no bifurcations of states, and states can disappear 
only at $\f_j$.

\begin{figure*} \centering
\includegraphics[width=5cm]{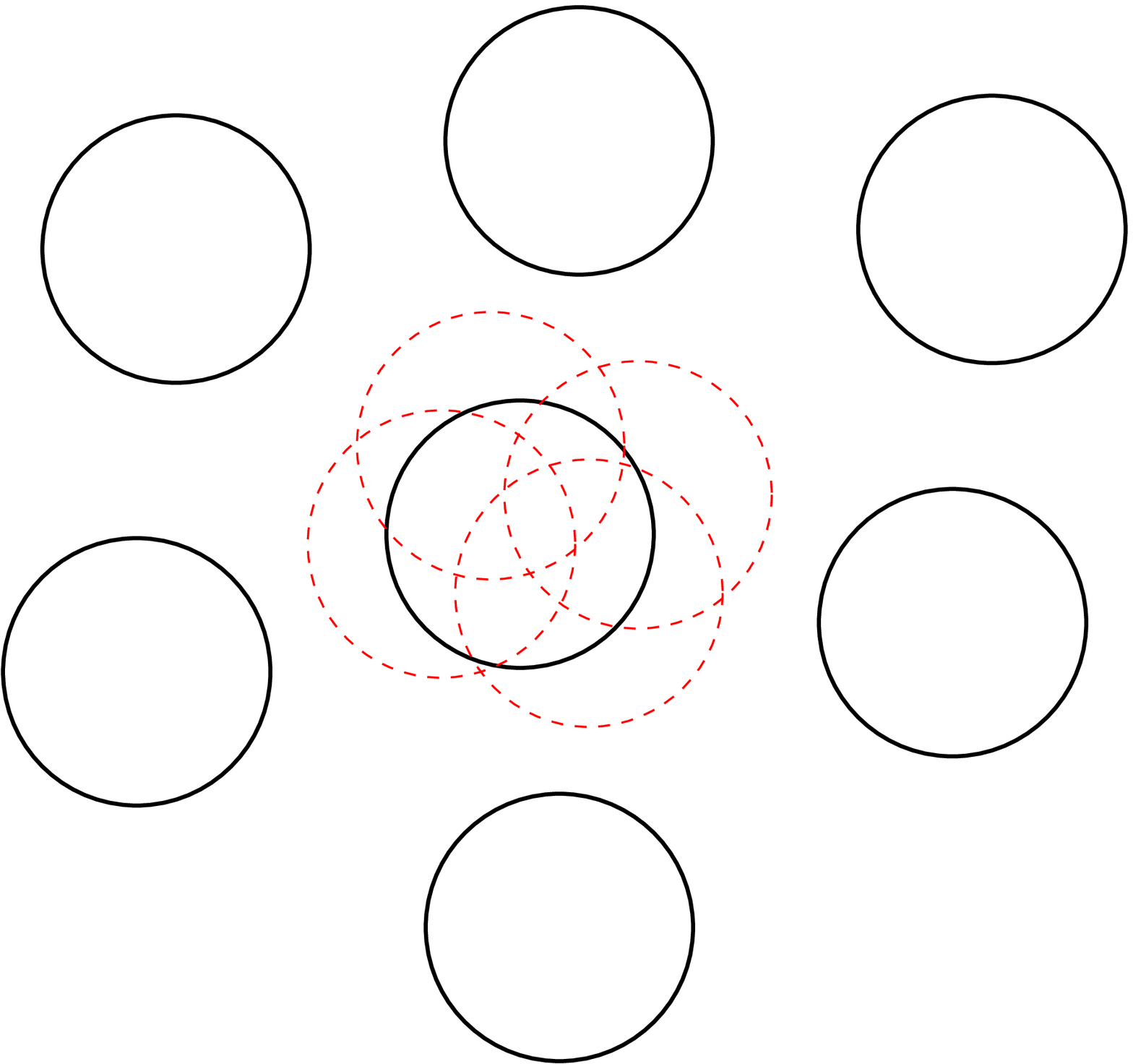}\hskip2cm
\includegraphics[width=8cm]{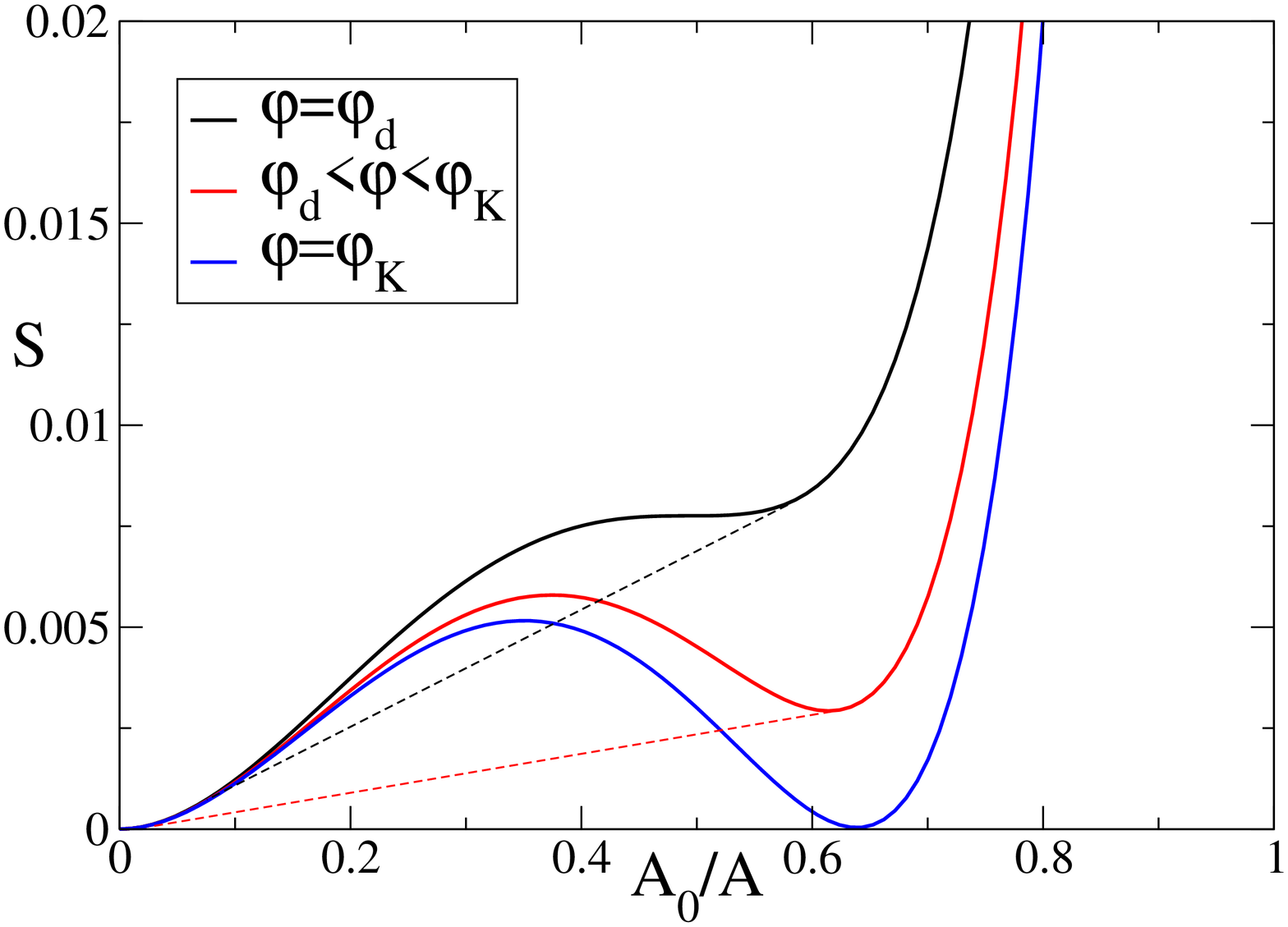}
\caption{(Left) A molecule of the replicated liquid: each full (black) sphere
  of the original liquid is replicated $m$ times (dashed red spheres), and the
  $m$ copies vibrate around the reference one. (Right) 
Replicated entropy as a function of the order parameter $A/A_0$, where $A_0$ is some
reference value and $m<1$. The full line is the mean field curve, while the dashed line
  take into account the finite dimensional nature of the system (see Appendix~\ref{sec:metastability}). 
}
\label{fig:sfere}
\end{figure*}

If one starts in a jammed state at density $\f_j$ and slightly decompress the system, 
the state acquires a finite entropy $s(\f,\f_j)$. 
We can invert this relation to get $\f_j(s,\f)$.
If there are no bifurcations, and we denote the logarithm of the number of structures 
with jamming density $\f_j$ as $\Si_j(\f_j)$, we have the simple relation
\beq
\Si(s,\f) = \Si_j(\f_j(s,\f)) \ .
\eeq
This is a consequence of our assumption\footnote{For this reason it is often called 
``isocomplexity'' assumption in the spin glass literature. It holds for the simplest spin
glass, the spherical $p$-spin glass with a single value of $p$~\cite{CK93}, 
but for more general models (\eg spherical spin glasses with interactions involving sets
of $p$ and $q$ spins with $p\neq q$, or Ising spin glasses) its exact validity is 
debated~\cite{BFP97,MR04,KTZ08}; 
still there is a general agreement that isocomplexity is approximately correct.},
because if states cannot bifurcate or die, their number remains constant;
this implies that we can label the states equivalently by their jamming density $\f_j$ 
or by their complexity $\Si_j$.

Then the procedure to extract $s(\f,\f_j)$ from
Eq.~(\ref{mcomplexity}) is the following: we fix a value $\Si_j$ and solve
\beq\label{Sijsolve}
\Si(m,\f) = \Si_j \ ,
\eeq
to get $m(\f,\Si_j)$. Then we have $s(\f,\Si_j) = s^*(m(\f,\Si_j),\f)$. 
As we will see below, in the limit $m\to 0$ we find $s^*(m,\f)\to -\io$, \ie the pressure diverges.
The jamming limit then corresponds to $m\to 0$, and the relation between $\f_j$ and $\Si_j$ is
simply
\beq
\Si_j(\f_j) = \lim_{m\to 0} \Si(m,\f_j) \ .
\eeq
Inverting this equation we obtain $\f_j(\Si_j)$, that we can substitute in $s(\f,\Si_j)$ to obtain
$s(\f,\f_j)$.

To summarize, under our assumption, 
in the $(m,\f)$ plane of figure~\ref{fig:diatot}, we can draw many lines, each defined 
by $\Si(m,\f) = \Si_j$. Each line identifies a group of states that share the same
jamming density $\f_j$, which is the point where the line crosses $m=0$, and the same
complexity $\Si_j$. For a given line, we can compute the internal entropy of the corresponding states
as $s(\f,\f_j) = s^*(m(\f,\f_j),\f)$, and differentiating this with respect to $\f$ we get the pressure
of the states, as drawn in the right panel of figure~\ref{fig:diatot}.
Clearly if we choose $\Si_j=0$ we recover the ideal glass state as 
discussed in the previous subsection.

\subsection{The molecular liquid}
\label{sec:pictorial}

The $m$ copies are assumed to be in the same state.  This can be implemented by constraining
each particle of a given replica to be
close to a particle of each of the other $m-1$ replicas. The liquid is made of {\it
molecules} of $m$ atoms, each belonging to a different replica of the original system, or in other
words the atoms of different replicas stay in the same cage.  The replica method allow us to define
and compute the properties of the cages in a purely equilibrium framework, in spite of the fact that
the cages have been defined originally in a dynamic framework\footnote{Note that it has been pointed
out that, already at the MCT level, ``cages'' are in fact extended objects, in the sense that single
atoms can always hop out of their local cage, and only groups of many atoms can be really blocked: 
strictly speaking only when the number of atoms goes to infinity the group is blocked.
Depending on the approximation, we will be able or not to take into account this effect.}.
The problem is then to compute the
free energy of a molecular liquid where each molecule is made of $m$ atoms.  The $m$ atoms are kept
close one to each other by a small inter-replica coupling that is switched off at the end of the
calculation, while each atom interacts with all the other atoms of the same replica via the original
pair potential.  

Note that the free energy of the replicated liquid is assumed to be analytic in the whole region
$m < m_s(\f)$, as the only phase transition happens at $m_s(\f)$ when the complexity vanishes.
Thus we can use approximations valid in the liquid phase to compute the free energy up to $m_s(\f)$.
This is enough, because the free energy is continuous on the transition line, and therefore the free
energy of the glass can be computed by approaching the line $m_s(\f)$ from below, as discussed above. 
Note also that we are interested in the regime of fairly high densities where we expect the cages
to be small and the motion of atoms inside a state to be quite localized. Thus, if all the replicas
are in the same state, inter-replica correlations remains strong when we switch off the coupling.
Our task is then to compute the entropy of the replicated liquid in a small cage regime, when atoms
in a molecule are quite close to each other.

\subsubsection{The partition function}
\label{sec:ZrepLeg}

We start from the grancanonical partition function of the replicated system of molecules of coordinates
$\bar x = (x_1,\cdots,x_m)$ in a volume $V$, and we put an harmonic attraction between particles in a molecule.
Particles belonging to the same replica interact via the hard sphere potential.
The replicated partition function is
\beq\label{Zstart}\begin{split}
Z_m(\ee) &= \sum_{N=0}^\io z^N \int_V \frac{d^N x_1 \cdots d^N x_m}{N!} \prod_{i<j}\prod_a \chi(x_{ai}-x_{aj}) 
\\ &\times
\prod_i \exp\left({-\frac\ee{m} \sum_{a<b} (x_{ai}-x_{bi})^2}\right) \\
&= \sum_{N=0}^\io \int_V \frac{d^N \bar x}{N!} \prod_i z(\bar x_i) \prod_{i<j}\bar\chi(\bar x_{i},\bar x_{j}) 
\end{split}
\eeq
where $\chi(x-y) = \th(|x-y|-D)$, 
$\bar \chi(\bar x,\bar y) = \prod_a \chi(x_a - y_a)$, and
\beq\label{zmolgauss}
z(\bar x) = z \exp\left({-\frac\ee{m} \sum_{a<b} (x_{a}-x_{b})^2}\right) \ .
\eeq
It is clear that the derivative of $\SS(m,\f;\ee) \equiv \la N \ra^{-1}\ln Z_m(\ee)$ 
with respect to $\ee$ gives the average cage radius, which is defined as the average
distance of atoms in two replicas $a\neq b$:
\begin{equation}\begin{split}
\label{cagedef}
A &\equiv \frac1{2Nd} \left\langle \sum_i (x_{ai} - x_{bi})^2 \right\rangle \\ &=
\frac1{m(m-1) d}\left\langle \sum_{a<b}(x_a-x_b)^2 \right\rangle
\\ &= - \frac{1}{(m-1)d} \frac{d\SS(m,\f;\ee)}{d\ee} \ .
\end{split}\end{equation}
Note that this cage radius could also be measured as the large time limit of
the mean square displacement of a single system~\cite{AF07}.
We are interested in the limit of zero coupling between the replicas. If there
is only one thermodynamic state, the liquid, then the replicas will
decorrelate for $\ee \to 0$ and $A(0)=\io$. On the contrary, if the are many
stable states, an infinitesimal coupling will be enough to send the replicas
into the same state; this will produce a correlation between the replicas and
$A$ will be of the order of the cage radius inside one state.
This phenomenon can be better understood in term of the Legendre transform of
$\SS(m,\f;\ee)$ with respect to $\ee$, which is a function of $A$ defined 
by\footnote{Due to a global $m-1$ factor, in the following discussion
minima and maxima are exchanged for $m<1$.}
\beq\label{LegA}
\SS(m,\f;A) = \min_{\ee} \left[ \SS(m,\f;\ee) + (m-1) d A \ee \right] \ ,
\eeq
and is schematically drawn in the right panel of figure~\ref{fig:sfere}. 
As $(m-1)d \ee = \frac{d\SS(m,\f;A)}{dA}$, the stationary points of $\SS(m,\f;A)$ 
correspond to zero coupling and in fact $\SS(m,\f) = \max_A \SS(m,\f;A)$.
In the liquid phase, $\SS(m,\f;A)$ 
will have a single maximum in
$A=\io$, while when glassy states are present, $\SS(m,\f;A)$ will have a
secondary maximum at finite $A$. 
See \cite{MP00} for a much more detailed discussion of the Legendre transform
with respect to $\ee$.

We will discuss in the following some approximation schemes to compute $Z_m(\ee)$. For the moment,
let us try to give a ``pictorial'' representation of this method.
A way to visualize the partition function $Z_m(\ee)$, Eq.~(\ref{Zstart}), is
that the original system is represented by the
reference coordinates $x_{1i}$ of the first replica; 
around each reference particle, $m-1$ other particles vibrate on a small scale $\sim 1/\sqrt{\ee}$. 
They are represented in Fig.~\ref{fig:sfere} as dashed spheres. The interaction is such that dashed
spheres belonging to the same replica cannot overlap. 

Alternatively, the system
can be thought as a ``molecular liquid'' in which each molecule is built by
the $m$ replicated particles vibrating around their center of mass $X = m^{-1}
\sum_a x_a$.
The entropy of the replicated system is given in Eq.~(\ref{mcomplexity})
by $\SS(m,\f) = \Si(s^*(m,\f),\f) + m s^*(m,\f)$.
In this picture, the entropy of the molecular system, $\SS(m,\f)$, 
is given by the entropy of the centers of mass, 
$\Si(m,\f)$, plus the contribution due to the vibrations of the $m$ particles in a molecule around the 
center of mass, which is $m$ times the vibrational entropy $s^*(m,\f)$, 
\ie the volume of a typical cage of the centers of mass system.
Note that for generic $m$, $s^*$
slightly depends on $m$ since the $m$ replicas are not independent. 

This representation of $Z_m(\ee)$ shows that a consistent way to
remove the crystal state in this computation is to describe the system of the centers of mass
by a low density virial expansion, as it will be done in the following (see Appendix~\ref{app:linkexp}).
In other words we are assuming that
the centers of mass $X$ represent structures that are {\it typical of the liquid state}. This is our
operational definition of amorphous states and for this reason in our computation partially crystallized
states do not appear.

\subsubsection{Correlation functions}
\label{sec:correlations}

The replicated liquid is characterized by the density $\r_a(x)=\la \sum_i \d(x_{ia}-x) \ra$ 
of each replica and
by the correlation function
\beq
\r_{ab}(x,y) = \left\langle \sum_{ij} \d(x_{ai}-x) \d(x_{bj}-y) \right\rangle \ . 
\eeq
where the sum is over all $ij$ is $a\neq b$ and over
$i\neq j$ if $a=b$.
If there is symmetry between the replicas, $\r_a(x) = \r$ and 
the two relevant correlations are the
intra-replica correlation $g(x,y) = \r_{aa}(x,y)/\r^2$ and the 
inter-replica correlation $\wt g(x,y) = \r_{a\neq b}(x,y)/\r^2$.
In the following we will show that $g(x,y)$ has a quite different
shape in the liquid and in the glass at high pressure. Hence,
we will denote by $\gG (x,y)$ the pair
distribution function in the glass phase, while we keep the notation
$g(x,y)$ for the one of the liquid. To avoid confusion, we 
stress that $g(x,y)$ and $\gG (x,y)$
refer to {\it the same observable in different phases}, while
$g(x,y)$ and $\wt g(x,y)$ are {\it different observables}.

If there are no correlations between different replicas, $\wt g = 1$. This happens for $m < m_d(\f)$, where no
glassy states are present. In the region $m \geq m_d(\f)$, there are glassy
states. Each state $\a$ is characterized by a frozen density 
profile $\r_\a(x)$ and by its correlation function $\r_\a(x,y)$.
The density profile can be computed, in principle, as the minimum
of a suitable density functional $F[\r(x)]$~\cite{DV99,CKDKS05,KW87,KM03,SSW85,YYO07}.

We are interested in averages over the states. Due to translational invariance,
$\overline{\r_\a(x)}=\r$, and for the intra-replica correlation we
have simply
\beq
g(x-y) =\r^{-2} \overline{\r_\a(x,y)} \ ,
\eeq
where the overline denotes average over the states.

We assumed that the coupling between different replicas is able to force all them to be in
the same state. Apart from that, in the limit of vanishing coupling no other correlations
are induced. Thus
\beq\label{eq:tgxydef}
\wt g(x-y) =\r^{-2} \overline{ \r_\a(x) \r_\a(y) } \ .
\eeq

\subsubsection{Nonergodicity factor}

Eq.~(\ref{eq:tgxydef}) allows to relate the inter-replica correlations to the so-called
{\it nonergodicity factor} of Mode-Coupling theory. We sketch here the connection
but the reader is referred to~\cite{MU93,Go99,BGS84} for more details.
In Mode-Coupling theory it is usual to work
in Fourier space; the Fourier-transformed density reads
\beq
\r_\a(q) = \int dx e^{iq x} \r_\a(x) =  \la \sum_i e^{i q x_i} \ra_\a \ ,
\eeq
where $\la \bullet \ra_\a$ denotes thermal average in state $\a$. The main object
of Mode-Coupling theory is the coherent normalized scattering function
\beq
F(q,t) = \frac{1}{N S(q)} \la \sum_{ij} e^{i q [x_i(t) - x_j(0) ] } \ra \ ,
\eeq
where $S(q) = N^{-1} \sum_{ij} e^{i q (x_i - x_j ) } = 1 + \r h(q)$ is the structure factor, 
which is related to the Fourier transform of $h(r) = g(r)-1$~\cite{Hansen},
and $x_i(t)$ is the 
position of particle $i$ at time $t$ assuming that the position at time $t=0$
has been extracted from the Gibbs distribution (Mode-Coupling theory is a theory of the equilibrium 
dynamics of glasses). By definition, $F(q,0)=1$; in the glass phase $F(q,t)$ develops a {\it plateau} that
becomes infinite at the Mode-Coupling transition $\f_{MCT}$. Hence for $\f > \f_{MCT}$ the
long-time limit of $F(q,t)$ is non zero; this is called nonergodicity factor $f_q$:
\beq
f_q \equiv \lim_{t \to \io} F(q,t) \ .
\eeq
In the replica interpretation, the fact that $F(q,t)$ does not vanish in the long time limit
(or in other word, that density fluctuations cannot relax completely)
signals the appearance of metastable states. Even for $t\to \io$, 
the system cannot escape from the
metastable state in which it was at time $t=0$. However, it can decorrelate {\it inside
the state}. Therefore, in the long time limit one has
\beq\begin{split}
& \la \sum_{ij} e^{i q [x_i(t) - x_j(0) ] } \ra_\a \sim  \la \sum_{i} e^{i q x_i(t)} \ra_\a \la \sum_j e^{-i q x_j(0) } \ra_\a \\
&= \r_\a(q) \r_\a(-q) \ ,
\end{split}\eeq
and, taking into account that the initial condition (hence the initial state $\a$) has been extracted from the
Gibbs distribution, one finally obtains
\beq\label{eq_fqdef}
f_q = \frac{\overline{ \r_\a(q) \r_\a(-q) } }{N S(q)} = \frac{\r \wt h(q)}{S(q)} \ ,
\eeq
where $\wt h(q)$ is the Fourier transform of $\wt h(r) = \wt g(r) - 1$.

\section{The replicated liquid: HNC equations}
\label{sec:HNC}

The simplest way to compute the property of the replicated liquid is the following.
We consider the system of the $m$ replicas as a $m$-component mixture, which is then
described by the number density $\r_a$ of type $a$ particles, and by the correlation function
$g_{ab}(x,y)$ which is the probability of finding a particle of type $b$ in $y$ given
that there is a particle of type $a$ in $x$.

\begin{figure} \centering
\includegraphics[width=8cm]{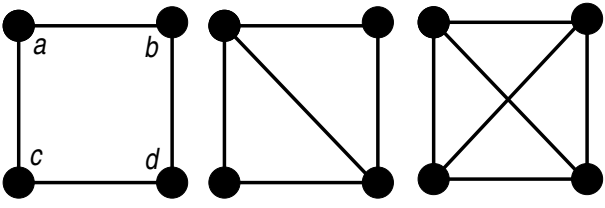}
\caption{The three diagrams contributing to order four in
density to the replicated free energy functional. Each vertex
is associated to a density $\r_a$, while each line is associated
to a $h_{ab}(x_i,x_j)=g_{ab}(x_i,x_j)-1$ factor; $a,b=1,\cdots,m$ are replica indeces.
The value of the diagram is obtained by integrating over the $x_i$
and summing over the replica indeces. Only the leftmost diagram is kept in the
HNC approximation~\cite{Hansen}; the other two are two-lines irreducible.
}
\label{fig:diag_HNC}
\end{figure}

\subsection{Replicated HNC equations}

The entropy of the replicated liquid can be expressed as a functional of these
quantities~\cite{Hansen,MH61,DM64}:
\begin{widetext}
\beq\label{completeS}
\begin{split}
\SS[\r_a,g_{ab}(x,y)] &= -\frac12 \sum_{ab} \int dx dy \r_a \r_b 
[ g_{ab}(x,y) \ln g_{ab}(x,y) -g_{ab}(x,y) + g_{ab}(x,y) \b \phi_{ab}(x,y) +1 ] \\
&- V \sum_a \r_a [ \ln \r_a -1] - \frac12 \sum_{n\geq 3} \frac{(-1)^n}n
\Tr [\r h]^n - \{\text{two-lines irreducible diagrams}\} \ ,
\end{split}\eeq
and the $g_{ab}(x,y)$ have to be determined by maximizing the entropy. In the
expression above we assume that the inter-replica coupling has already been
sent to zero and look for non-trivial solutions for the inter-replica correlation.
As an example, 
the diagrams contributing at order four in density are given in figure~\ref{fig:diag_HNC}.

The simplest approximation that allows to obtain a treatable functional amounts to
neglect two-lines irreducible diagrams; in this way one obtains the HNC equations for $g_{ab}$:
\beq
\ln g_{ab}(x,y) + \b \phi_{ab}(x,y) = h_{ab}(x,y) - c_{ab}(x,y) \ ,
\eeq
where $c$ is defined by the OZ relation
\beq
h_{ab}(x,y) = c_{ab}(x,y) + \sum_c \int dz h_{ac}(x,z) \r_c c_{cb}(z,y) \ .
\eeq

The interaction potential is $\phi_{ab}(x,y)=\phi(x-y) \d_{ab}$, where $\phi(x)$ is the hard
core potential.
Using translational invariance, and a replica symmetric structure, we have $\r_a = \r$, 
$g_{aa}(x,y)=g(x-y)$ and $g_{a\neq b}(x,y) = \wt g(x-y)$.
The replicated entropy becomes\footnote{We can set $\b=1$ as temperature is
irrelevant for hard spheres. Note that the term $\int d\vec r \phi(r) g(r)=0$.}
\beq\begin{split}
\SS(m,\f) = \frac{\SS}{N} &= 
-\frac\r2 \int d\vec r \left\{ m g(r) [\ln g(r) -1] + m(m-1) \wt g(r) [\ln \wt g(r)-1]
+m^2 + m \phi(r) g(r) \right\}
- m (\ln \r-1) \\ & + \frac1{2\r} \int \frac{d\vec q}{(2\p)^d}
\Big\{ (m-1) \ln[1+\r (h(q)-\wt h(q))] + \ln[1+\r (h(q)+(m-1)\wt h(q))]  \\
& \hskip30pt - m\r h(q)
+\frac{m \r^2 h(q)^2}{2} + \frac{m(m-1) \r^2 \wt h(q)^2}2 \Big\} \ ,
\end{split}\eeq
where $h(q),\wt h(q)$ are the Fourier transforms of $h(r),\wt h(r)$. 
The HNC equations can be written as
\beq\label{HNCrepm}
\begin{split}
&\ln g(r) = -\phi(r) + W(r) \ , \\
&\ln \wt g(r) = \wt W(r) \ ,
\end{split}\eeq
where $W(r),\wt W(r)$ are defined by their Fourier transforms
\beq\begin{split}
&W(q) = \frac1m \frac{\r [h(q) + (m-1)\wt h(q)]^2}{1 + \r [h(q) + (m-1)\wt h(q)]}
+ \frac{m-1}{m} \frac{\r [h(q) -\wt h(q)]^2}{1 + \r [h(q) - \wt h(q)]}   \ , \\
&\wt W(q) = \frac1m \frac{\r [h(q) + (m-1)\wt h(q)]^2}{1 + \r [h(q) + (m-1)\wt h(q)]}
- \frac{1}{m} \frac{\r [h(q) -\wt h(q)]^2}{1 + \r [h(q) - \wt h(q)]} \ .
\end{split}\eeq
Using Eq.~(\ref{mcomplexity}), the free energy and complexity of the states are
\beq
\begin{split}
&
\begin{aligned}
 s^*(m) =  \frac{\partial \SS(m)}{\partial m}  &= 
-\frac\r2 \int d\vec r \left\{ g(r) [\ln g(r) -1] + (2m-1) \wt g(r) [\ln \wt g(r)-1]
+2 m + \phi(r) g(r) \right\}
- \ln \r+1  \\ & + \frac1{2\r} \int \frac{d\vec q}{(2\p)^d}
\Big\{ \ln[1+\r (h(q)-\wt h(q))] + \frac{\r \wt h(q)}{1+\r (h(q)+(m-1)\wt h(q))} \\
& \hskip30pt - \r h(q)
+\frac{ \r^2 h(q)^2}{2} + \frac{(2m-1) \r^2 \wt h(q)^2}2 \Big\} \ ,
\end{aligned} \\
&
\begin{aligned}
\Si(m) = -m^2 \frac{\partial (\SS(m)/m)}{\partial m}  &= 
\frac{\r m^2}2 \int d\vec r \left\{ \wt g(r) [\ln \wt g(r)-1] + 1 \right\}
 - \frac1{2\r} \int \frac{d\vec q}{(2\p)^d}
\Big\{ \ln[1+\r (h(q)-\wt h(q))] \\ & 
- \ln[1+\r (h(q)+(m-1)\wt h(q))]
+ \frac{m \r \wt h(q)}{1+\r (h(q)+(m-1)\wt h(q))}
+ \frac{m^2 \r^2 \wt h(q)^2}2 \Big\} \ .
\end{aligned}
\end{split}
\eeq
\end{widetext}

The advantage of this formulation is that equations~(\ref{HNCrepm}) are relatively
easy to solve numerically and give direct access to both the thermodynamic
property of the glass (entropy and pressure), the structure factor $g(r)$ and
the non-ergodic parameter $\wt g(r)$.

\subsection{Results}

These equations were used in \cite{MP96,CFP98,CFP98b} to compute the properties of amorphous states
of hard spheres. We reproduced these calculations for illustrative purposes\footnote{The HNC equations were
solved by an iterative Picard scheme, using an initial Gaussian guess for $\wt c(x)$.
We used a grid defined by imposing a cutoff $r < L = 8 D$ and discretizing space
with a step $a = D/128$. We checked the stability of the reported 
results by doubling the cutoff and inverse step.}.

\subsubsection{Phase diagram}

The $(m,\f)$ phase diagram for $d=3$ is reported in figure~\ref{fig:mstar}. 
The values of $\f_K=0.63$ and $\f_d=0.619$ are reasonable, 
even if $\f_d$ is a bit too large if compared to $\f_{MCT} \sim 0.58$~\cite{MU93}. Also the value
of $\f_{th} \sim 0.67$ that one can guess from the extrapolation of $m_d$
to $m=0$ is bigger than the accepted value $\f_{J}\sim 0.64$ from numerical 
simulations.

Moreover, one immediately notices that
the static line does not seem to extrapolate to zero at reasonable 
densities\footnote{It is worth to note at this point that on increasing the density
long range correlations seems to develop and short range singular behavior emerges in
$g(r)$, as expected from numerical simulations. Thus one is forced to increase the
cutoff and inverse step used in the discretization of the HNC equations. We checked
that in the range reported in figure~\ref{fig:mstar} the resulting $m_s$, $m_d$ are
not affected by discretization corrections.}. 
This was already observed in \cite{MP96} and is one of the major problems of the HNC
approximation. Before discussing this issue in detail, let us discuss the results for
the correlation functions.

\subsubsection{Correlation functions}

\begin{figure*} \centering
\includegraphics[width=8cm]{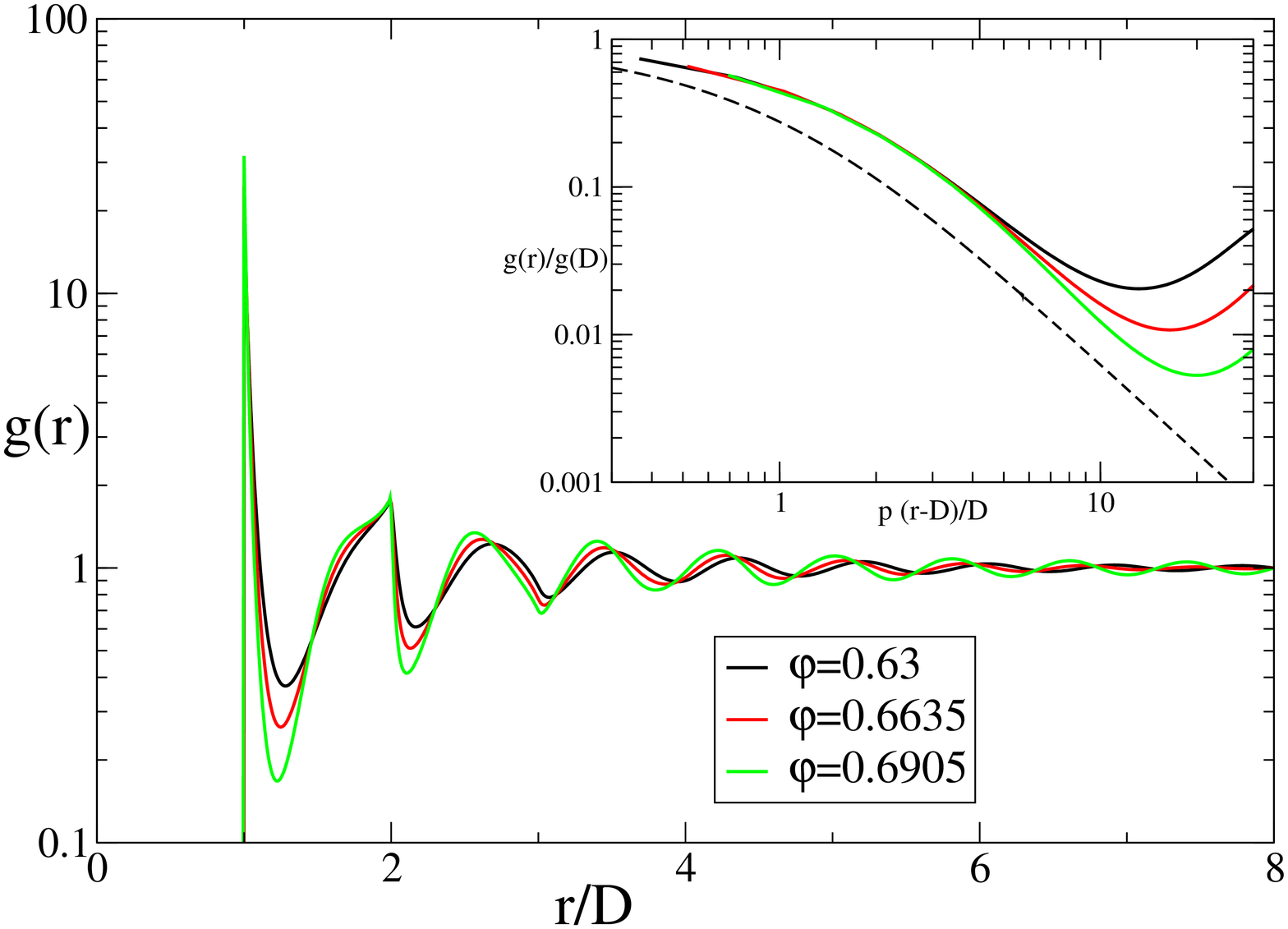}
\includegraphics[width=8cm]{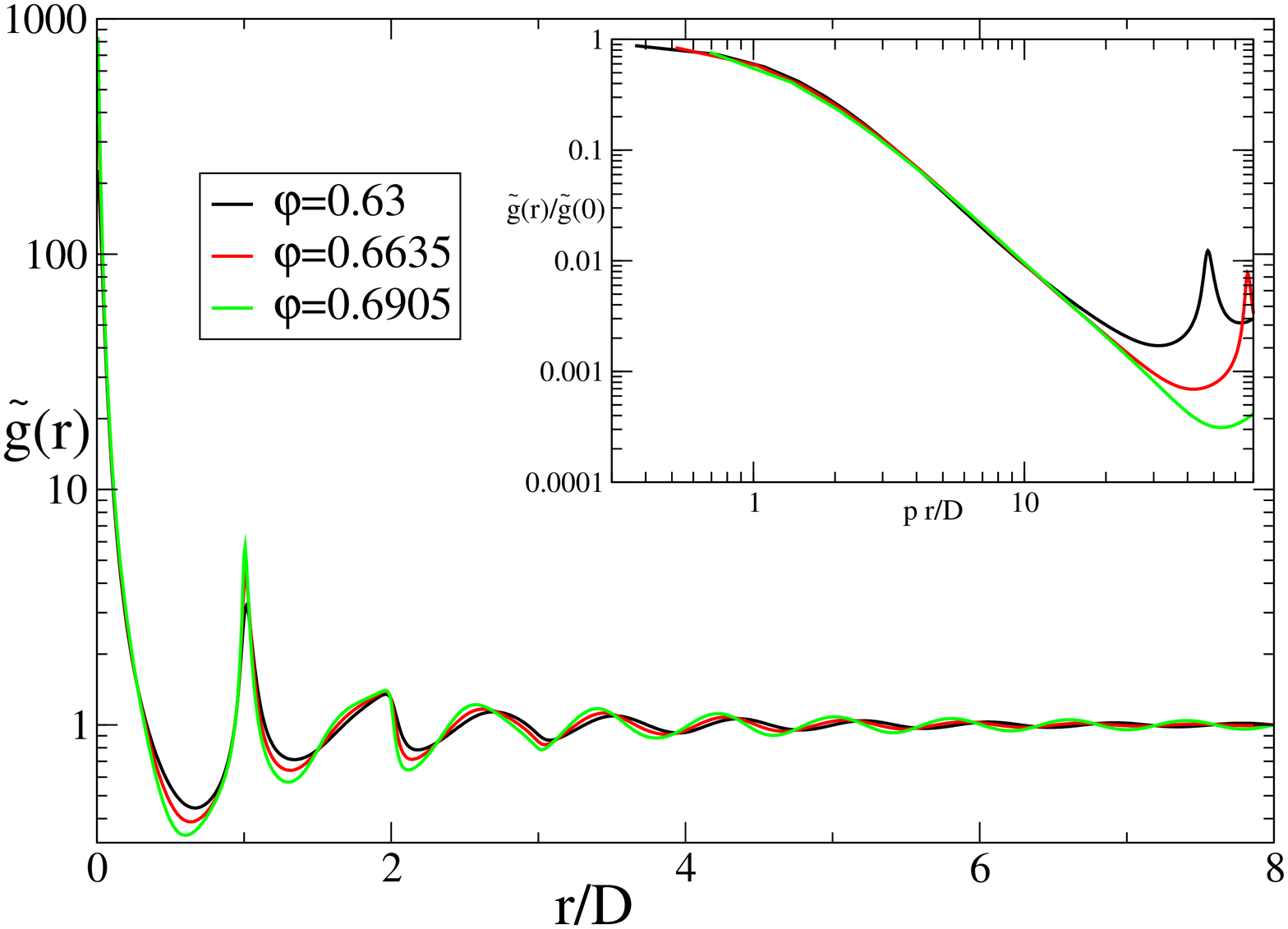}
\caption{$\gG(r)$ and $\wt g(r)$ along the ideal glass line $m_s(\f)$ at three different densities, computed
using the replicated HNC equations (section~\ref{sec:HNC}).
Inset: scaled plot of $\gG(r)/\gG(D)$ vs $p (r-D)/D$ 
and $\wt g(r)/\wt g(0)$ vs $p r/D$,
where $p =1 + 4 \f \gG(D)$.
}
\label{fig:gofrHNC}
\end{figure*}

\begin{figure} \centering
\includegraphics[width=8cm]{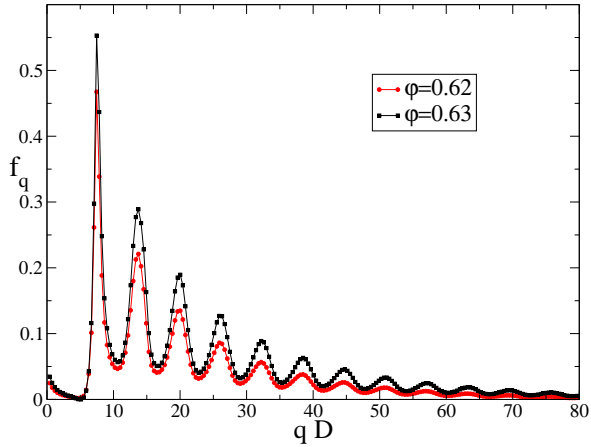}
\caption{
Nonergodicity factor, defined in Eq.~(\ref{eq_fqdef}), computed
using the replicated HNC equations (section~\ref{sec:HNC}) at
$\f = 0.62 \sim \f_d$ and $\f = 0.63 \sim \f_K$.
}
\label{fig:fofkHNC}
\end{figure}

Clearly in the liquid phase $m=1$ and the correlation function of the liquid coincides
with the one obtained within the non-replicated HNC approximation.
The result for $\gG(r)$ along the line $m_s(\f)$ (\ie for the ideal glass) 
is reported in figure~\ref{fig:gofrHNC}. Many interesting features expected from numerical
simulations (see section~\ref{sec:gofr}) 
are observed: first of all, $\gG(r)$ develops a delta peak close to $r=D$, whose
shape (inset of left panel) is quite similar to the one observed by Donev et al.~\cite{DTS05}.
Moreover, a dip for $r\sim 1.2$ is developed, due to particles of the first shell getting closer
to the reference one. Another interesting feature is the jump that develops in $r=2D$ and the
cusp in $r=3D$, which are in fact related to the delta peak in $r=D$.
Finally, we observe that on increasing the density $\gG(r)$ develops long ranged oscillations for large
$r$, even if a systematic study of these correlation would require the investigation of smaller
values of $m$ at which the approximation gives inconsistent results.

As for $\wt g(r)$, the most interesting feature is the peak close to $r=0$ that describes the
particle of replica $b$ which is in the same molecule of the reference particle of replica $a$.
Its shape has a clear scaling form on decreasing $m$: the cage radius, \ie 
the scale on which the delta peak is rounded, decreases and goes to $0$ for $m\to 0$, as anticipated
in previous sections.
The behavior for larger $r$ is very similar
to the one of $g(r)$, ``rounded'' by the fluctuations inside each molecule of particles of different
replicas. Finally, in figure~\ref{fig:fofkHNC} we report the nonergodicity factor $f_q$, Eq.~(\ref{eq_fqdef}),
which is a central object of Mode-Coupling theory. The shape of $f_q$ obtained with 
replicas is qualitatively very similar
to the one obtained within MCT~\cite{BGS84,Go99,MU93}
and experiments~\cite{MUP91}, although from the quantitative point of view marked
differences are present.

\subsection{Discussion}

The replicated HNC equation already gives interesting qualitative indications on the 
phase diagram and correlations functions, and is not too bad from the quantitative point
of view. However, the value $\f_d = 0.619$ is too large compared with numerical estimates,
and the nonergodicity factor is quite far from the measured values.
Another unsatisfactory feature of the HNC approximation is that it gives a complexity $\Si$ of
the order of $0.01$, which is two orders of magnitude smaller compared to what is observed in 
simulations and experiments, where $\Si \sim 1$.
As for the correlation of the glass, this approximation misses the peak at $r=\sqrt{3}D$ 
(which should be encoded in the diagrams that have been neglected)
and the square-root singularity close to $r=D$, but reproduces well the other characteristic
features of jammed disordered packings.

Unfortunately, the results of the HNC approximation become definitely bad at small values
of $m$, \ie deep inside the glass phase on approaching jamming.
This can be understood quite easily by inspecting the diagrams that have been neglected in
the HNC approximation, already at the lowest non-trivial order in density; these are shown
in figure~\ref{fig:diag_HNC}.
The argument goes as follows: as discussed above, when the pressure is high (or $m$ is small)
the cage becomes very small, 
and the peak in $\wt g(r)$ close to $r=0$ approaches a delta function. 
Consider the contribution to the diagrams in figure~\ref{fig:diag_HNC} when the replica
indeces are all different, $a \neq b \neq c \neq d$; then on each link there is a
factor $\wt g(x_i - x_j)$. We focus on the small $x_i - x_j$ behavior; then 
$\wt g(x_i - x_j) \sim \d(x_i - x_j)$ and the leftmost diagram has a contribution
\beq\begin{split}
\int dx_1 dx_2 & dx_3 dx_4 \d(x_1-x_2) \d(x_2-x_3) \times \\
& \d(x_3 - x_4) \d(x_4 - x_1) \sim \d(0)
\end{split}\eeq
since three delta functions are enough to constrain the four integration variables
to be close to each other. The notation $\d(0)$ here is a formal way to indicate that
the contribution above will diverge roughly as the maximum of the peak of $\wt g(r)$.
Now, the other two diagrams in figure~\ref{fig:diag_HNC} have more links; performing
the same computation as above, we find that they diverge as $\d(0)^2$ and $\d(0)^3$.
But these diagrams are not included in the HNC and this makes the approximation bad at high
pressure. Actually, this reasoning shows that the HNC is keeping the less divergent diagrams
at each order, while the maximally divergent diagrams are the completely connected ones,
at least as far as the contribution with different replica indeces are concerned.
Therefore to study the jamming limit we need to treat correlations inside a 
molecule in a more appropriate way. We will discuss in next section a 
different approximation scheme that fixes this problems by taking into account all
the correlations between different replicas (in some sense, resumming the contribution of
completely connected diagrams).

The argument above is supported by the fact that
we were not able to derive analytically the limit $m\to 0$ of the HNC equations,
and we have a strong feeling that they become pathological in this limit.
The same problem is present in the limit $d\to\io$, and should be connected to the fact
that the cage radius is very small in this limit, as we will see below. 
In summary, the HNC approximation does not work for small cage radius: 
multiple correlations functions in the replica sector become important.

\section{The replicated liquid: effective potentials}
\label{sec:effectivepotentials}

A way to compute
the replicated free energy for a system of Lennard-Jones particles in a more
accurate way, by taking into account multiple correlations inside a molecule, 
has been described in~\cite{MP99b}. 
The idea is to write the free energy as a function of the cage radius $A$ and expand
systematically in powers of $A$, which is assumed to be small. Thus we expect this method
to work well in the dense region and for $m \sim 0$, where the HNC approximation fails.
The method is successful but cannot be extended
straightforwardly to hard spheres, because at some stage in~\cite{MP99b} it was assumed that vibrations were
harmonic, an approximation that clearly breaks down for hard core potentials. In this section
we will discuss a general method that allows to map the replicated free energy
onto the non-replicated free energy of a liquid of particles interacting via
some effective potentials to be computed below.
The method is very similar in spirit to the self-consistent phonon 
theory used in~\cite{SW84,HW03,HW08} to study the stability of a glass state.
This will allow to derive in a simple way the small cage expansion for the
case of hard spheres. Moreover, the correlation function of the glass
turns out to be the correlation function of the effective liquid,
thus simplifying a lot its computation. Similar ideas have been recently used in~\cite{SWHW08}.

\subsection{Entropy as a functional of the single-molecule density}

We wish to compute the entropy as a functional of the order parameter $A$, the cage radius,
whose expected behavior has been schematically illustrated in figure~\ref{fig:sfere}.
Instead of expanding directly the partition function (\ref{Zstart}) for large $\ee$, and then Legendre
transform with respect to $\ee$ as in Eq.~(\ref{LegA}), 
we will first use standard liquid theory to Legendre transform Eq.~(\ref{Zstart})
with respect to the full function $z(\bar x)$. In this way we obtain the entropy as a functional of the
single-molecule density 
\beq
\r(\bar x) = \left\langle \sum_i \d(\bar x- \bar x_i) \right\rangle \ ,
\eeq
where $\d(\bar x- \bar x_i) = \prod_a \d(x_a - x_{ai})$,
and of the interaction function $\bar\chi(\bar x, \bar y)$:
\beq\label{Sdirho}
\begin{split}
\SS[\r(\bar x),\bar\chi(\bar x,\bar y)] &= \int d\bar x \r(\bar x) [1-\ln \r(\bar x)] \\
&+ \sum \left[ \text{a class of diagrams} \right] \ .
\end{split}\eeq
The class of diagrams contributing to (\ref{Sdirho}) 
is defined precisely in \cite{MH61,DM64,Hansen} but is not important 
here. What is important is that a diagram $\DD$ represents an integral of the form
\beq\label{diagramma}
\DD = \frac{1}{S} \int  \prod_i  \r(i) d \bar x_i \prod_\ell (\bar\chi(\ell)-1) \ ,
\eeq
where $i$ are the vertices of the diagram, $\ell = (i<j)$ are the links, and
to lighten the notation we defined $\r(i) = \r(\bar x_i)$ and
$\bar\c(\ell) = \bar\c(\bar x_{i},\bar x_{j})$. $S$ is the symmetry factor of the diagram,
\ie the number of equivalent relabelings of the vertices of the diagram.

The reason why we start from (\ref{Sdirho}) instead of (\ref{Zstart}) is that the single-molecule
density $\r(\bar x)$ is directly related to the order parameter $A$ we want to study. 
In particular, we can make a simple Gaussian {\it ansatz}\footnote{
This simple ansatz assumes that all particles have the same cage radius $\sqrt{A}$.
However, at very high pressure, it is well known that most of the particles are immobile,
while a small fraction (typically $\sim 5\%$) can vibrate in a cage that remains finite
even at infinite pressure. These particles are called ``rattlers'' in the literature on
jamming. Our ansatz completely neglects rattlers. One could think to improve it by
introducing two cage radii, one associated to jammed particles and the other to rattlers:
we did not explore this possibility.
}
\beq\label{rhomolgauss}
\begin{split}
\rho(\bar x) &= \frac{\r m^{-d/2}}{ (2\pi A)^{(m-1)d/2}}  
e^{-\frac1{2mA} \sum_{a<b} (x^a_i-x^b_i)^2} \\ &= 
\frac{\r}{ (2\pi A)^{m d/2}}
\int dX e^{-\frac1{2A} \sum_a (x_a-X)^2} \\
&= 
\r \int dX \prod_a \frac{e^{-\frac{1}{2A} (x_a-X)^2}}{ (2\pi A)^{d/2} }
\\ &\equiv \r \int dX \prod_a \g_A(x_a-X)
\end{split}
\eeq
where $\g_A(x)$ is a normalized Gaussian with variance $A$; 
we have then
$\int d\bar x \r(\bar x) = \r V = N$, therefore
$\r$ has the interpretation of the number density of molecules (note that having Legendre
transformed we are now working at fixed $N$).

The Gaussian {\it ansatz} for $\r(\bar x)$ is {\it not} equivalent to the
Gaussian {\it ansatz} (\ref{zmolgauss}) for the single molecule activity; in fact
$\r(\bar x)$ is a sum of diagrams containing $z(\bar x)$ and the interaction function,
so it is Gaussian at lowest order in $\ee$ but has corrections coming from higher
orders. The corrections
are singular because of the singularity of the interaction. For this reason it is convenient
first to Legendre transform with respect to $z(\bar x)$ and then to make the Gaussian
{\it ansatz} for $\r(\bar x)$: the resulting small $A$ expansion has a better behavior.

The Gaussian form (\ref{rhomolgauss}) allows to
compute exactly the ideal gas term of the entropy:
\beq\label{gasperfetto}
\begin{split}
 & \frac{1}{N} \int  d\bar x \rho(\bar x) [1 -\ln \rho(\bar x) ] = \\
&= 1 - \ln \rho  - \frac{d}{2}(1-m)\ln(2\pi A) - \frac{d}{2} (1-m-\ln m) \\ 
&\equiv
1 - \ln\rho + S_{harm}(m,A) \ .
\end{split}\eeq
where we defined
\beq
S_{harm}(m,A) = \frac{d}2 (m-1) \ln(2 \pi A) + \frac{d}{2} (m-1+\ln m) \ .
\eeq
We want to find a simple way to rewrite the replicated entropy exploiting the
fact that $A$ is small.

\subsection{The effective non-replicated liquid}

Before proceeding to the formal computation, it is better to understand intuitively what will
be the result. To this aim, note that we assumed 
that the vibrations of the $m$ copies of the particles,
$x=(x_1,\cdots,x_m)$,
are described by the Gaussian distribution (\ref{rhomolgauss}), corresponding to harmonic
vibrations.

The width $A$ is a variational parameter and we will maximize the entropy
with respect to it at the end; for the moment we assume that $A$ is small.
The idea is to choose a replica (say replica 1) as a reference and consider the
vibrations of the other $m-1$ particles around the reference one.
At the zero-th order in $A$, the $m-1$ copies essentially coincide 
with the reference one, and $\SS(m,\f;A) \equiv N^{-1} \SS[\r(\bar x),\bar\chi(\bar x,\bar y)]$ 
is given by
the entropy $S(\f)$ of the non-replicated liquid (corresponding to replica 1), 
plus the free energy of
$m-1$ harmonic oscillators of spring constant 
$A$:
\beq\label{SS0}
\SS^{(0)}(m,\f;A) = S(\f) + S_{harm}(m,A) \ .
\eeq
We see indeed that the ideal gas term (\ref{gasperfetto}) corresponds
to Eq.~(\ref{SS0}) where $S(\f)$ has been approximated by the ideal gas contribution.

A first order approximation is obtained by considering the effective
two-body interaction induced on the particles of replica 1 by the coupling
to the $m-1$ copies. We consider two particle $x_1$ and $y_1$ of replica $1$
belonging to two molecules $\bar x$ and $\bar y$.
Each particle of a given replica $a$ interacts with the other
particles of the same replica via the hard core potential $\phi(r)$. The effective
interaction between particles in replica 1 is obtained by averaging this
interaction over the probability distribution $\r(\bar x)$ of the two
molecules $\bar x$ and $\bar y$:
\beq\label{feff_def}\begin{split}
e^{-\phi_{eff}(x_1-y_1)} &= \int dx_{2,m} dy_{2,m} \r^{-2}\r(\bar x) \r(\bar y) 
\prod_{a=1}^m e^{-\phi(x_a - y_a)} \\ & \equiv e^{-\phi(x_1-y_1)} 
\left\langle \prod_{a=2}^m e^{-\phi(x_a - y_a)} \right\rangle_{x_1,y_1}
\ .
\end{split}\eeq
As an example, the potential $\phi_{eff}(r)$ for hard spheres in $d=3$ is reported in
Fig.~\ref{fig:feff} for some values of $A$ and $m$ (see Appendix~\ref{app:A} for its
calculation).
A first order approximation to $\SS(m,\f;A)$ is then obtained by substituting to the
entropy $S(\f)$ of the simple hard sphere liquid the free energy
of a liquid of particles
interacting via the potential $\phi_{eff}(r)$:
\beq\label{SS1}
\SS^{(1)}(m,\f;A) = -\b F[\f;\phi_{eff}(r)] + S_{harm}(m,A) \ .
\eeq
It is now evident that we can obtain better approximations of the true function
$\SS(m,\f;A)$ by considering also the three body interactions induced on particles of
the replica 1, and so on. 

This procedure can be justified more formally by
introducing a diagrammatic expansion in powers of $A$ and resuming a class of 
diagrams: this is discussed in detail in Appendix~\ref{app:linkexp}.
The result is the following: the replicated entropy (\ref{Sdirho}) can be
{\it exactly} rewritten as
\beq\label{Sresum}\begin{split}
\SS&(m,\f;A)  \equiv N^{-1} \SS[\r(\bar x),\bar\chi(\bar x,\bar y)] \\
&= S_{harm}(m,A) - \b F_{eff}[\f;\phi_{eff},\phi^{(2)}_{eff},\phi^{(3)}_{eff},\cdots] \ ,
\end{split}\eeq
where the effective potentials $\phi_{eff}^{(n)}$ depend on both $A$ and $m$.
The correlation function of the glassy states as function of $A$ and $m$
is just the correlation function of the liquid described by $F_{eff}$.

The effective potential $\phi_{eff}^{(n)}$ is constructed as follows: one
constructs all the
possible connected diagrams with $n$ links and such that one pair of vertices 
is not connected by more than one link. 
Then one numbers the vertices $x_1,x_2,\cdots$. The
effective potential $\phi_{eff}^{(n)}$ for one chosen diagram depends on the
variables $x_1,x_2,\cdots$ (whose number depend on the diagram).
For $n=1$, we have only one possibility,
$\phi_{eff}$ depends only on $\vec r=x_1-x_2$ by translation invariance and is given
by (\ref{feff_def}). For $n=2$ also we have only one possibility where the two
links share one vertex, the resulting potential depend on $\vec r=x_1-x_2$ and
$\vec s=x_1-x_3$ (denoting by $x_1$ the common vertex) and we have
\beq\begin{split}
&e^{-\phi_{eff}^{(2)}(x-y,x-z)}
= \\ &= \frac{\left\langle \prod_{a=2}^m e^{-\phi(x_a-y_a)}
 e^{-\phi(x_a-z_a)} \right\rangle_{x,y,z}
}{\left\langle \prod_{a=2}^m e^{-\phi(x_a-y_a)} \right\rangle_{x,y}
\left\langle \prod_{a=2}^m e^{-\phi(x_a-z_a)} \right\rangle_{x,z}}
\end{split}\eeq
where the brackets denote averages similar to the one in (\ref{feff_def}),
see Appendix~\ref{app:linkexp} for details. There 
it is also shown that the correlation function $g_G(r)$ of the glass is given
by the correlation function $g_{eff}(r)$ of the effective liquid.

\subsection{Properties of the effective potentials}

The effective potentials have the following general properties, that are discussed
in Appendix~\ref{app:linkexp}:
\begin{enumerate}
\item If for at least one of the links the distance $r = |x_i-x_j|$ is 
such that $|r-D| \gg \sqrt{A}$, the potential vanishes exponentially as $e^{-(r-D)^2/A}$.
\item For the reason above, we can argue that they contribute $O(A^{n/2})$ to the free energy, where $n$ is
the number of links in the potential. This is because the potential is non-vanishing only if, for all the 
links, $|r-D| \sim \sqrt{A}$, and this region has volume $O(A^{n/2})$.
\item One can show that the leading order of the 
potentials vanishes if all the unit vectors of the links are orthogonal.
Then, one can argue that in the limit $d\to \io$
where the links are almost always orthogonal, the potentials for $n \geq 2$ give vanishing contributions.
\item All the potentials are $O(m-1)$ for $m \to 1$. Thus in this limit they can be treated as perturbations
of the hard spheres liquid.
\item In the opposite limit $A \to 0$ with $A = \a m$, (jamming limit, see below) 
the potentials tend to become delta functions around $r = D$. 
\end{enumerate}
In the following we will use these properties to derive the glass equation of state and its
structure factor as function of the dimensionality. 

Before concluding this section it is worth to remark that in the case of a smooth potential
it might be preferable to use as reference positions the center of masses instead of
the coordinates of replica 1. The expansion of Appendix~\ref{app:linkexp} can be carried
out similarly in this case and leads to analogous expressions for the effective potentials.
The difference is that in the case of a smooth potential, the small cage expansion starts
with a term of order $A$ and one can show that if the center of masses are chosen as reference
positions, the $n$-body potential is of order $A^{n-1}$. Conversely, if one uses replica 1
as a reference, all the potentials are of order $A$.
The advantage of using replica 1 as a reference is that the correlation function of the
effective liquid is directly the correlation function of the glass. In the case of hard
spheres the expansion is well behaved because of the properties above.
One should keep in mind that depending on the problem at hand, 
one expansion might be better than the other.


\section{The limit of large space dimension}
\label{sec:larged}

We will begin by the study of the limit $d\to\io$,
because in this limit all the expressions
simplify a lot and we are able to obtain a consistent solution in all the regions
of the phase diagram.
Moreover, this limit is interesting because metastability effects should be less
important in high dimension: the limit $d\to\io$ is a kind of ``mean field'' limit
where the surface and the volume are of the same order of magnitude and nucleation 
becomes almost impossible.
As we briefly discussed in the introduction, the limit $d\to\io$ has also interesting 
applications in the digitalization of signals, see \cite{ConwaySloane}.

\begin{figure} \centering
\includegraphics[width=9cm]{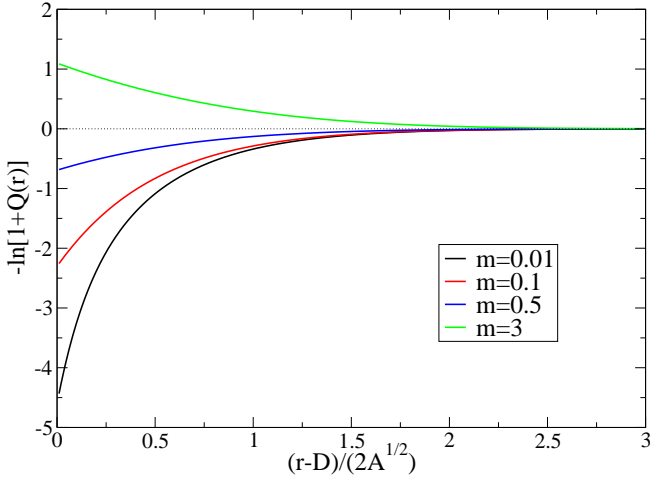}
\caption{The non-hard core part of the 
effective potential $\phi_{eff}(r) -\phi(r)= -\ln[1+Q(r)]$, Eq~(\ref{feff_def}), 
as a function of $(r-D)/\sqrt{4A}$ for $r\geq D$ for different values of $m=0.01,0.1,0.5,3$
(from bottom to top) in $d=3$. Note that the range of the potential is $O(\sqrt{A})$ and that it is
attractive for $m<1$ and repulsive for $m>1$. The strength of the potential diverges for 
$m\to 0$.
}
\label{fig:feff}
\end{figure}

\subsection{The liquid in $d\to\io$}

The problem of computing the entropy of the hard sphere liquid for
$d\to \io$ was addressed in \cite{FP99,PS00}, where the same result was obtained
in two independent ways.
In \cite{FP99} it was shown that the ring diagrams dominate the virial series order
by order in $\rho$ for large $d$. The resummation of these
diagrams gives
\beq\label{Sanelli}\begin{split}
S(\f) &= 1 - \ln \r + \frac\r2 \int d\vec r f(r) \\
&- \frac{1}{2\r} \int \frac{d\vec q}{(2\p)^d} \left[ \ln[ 1-\r \wh f(q) ] + 
\r \wh f(q) + \frac{(\r \wh f(q))^2}2
\right] \ ,
\end{split}\eeq
where $\wh f(q)$ is the Fourier transform of the Mayer function 
$f(r) = \chi(r)-1$,
\beq\label{fqHS}
\wh f(q) = -V_d \, \G\left(\frac{d}2+1\right) \left(\frac2q\right)^{\frac{d}2} J_{\frac{d}2}(q) \ .
\eeq
The key observation is that, up to a density
$\f^{HS}_{max} = \exp[d (1-\ln 2)/2] / 2^d$, the logarithm does not have
poles so that the above expression is well defined. If $2^d \f$ is not exponentially
large, the non-ring
diagrams are conjectured to give only exponentially small corrections~\cite{FP99}, and
one can show that the last term is exponentially small~\cite{FP99}. 
Therefore, {\it if $2^d \f$ does not grow exponentially with $d$}, 
$S(\f)$ is given by the ideal gas term plus
the first virial correction (\ie by the Van der Waals equation)
up to exponentially small corrections:
\beq\label{Fdgrande}
\begin{split}
&S(\f) = 1-\ln \r - 2^{d-1}\f + O(e^{-d}) \ , \\
&g(r) = \chi(r) [1+O(e^{-d})] \ .
\end{split}
\eeq
In \cite{PS00} simple equations for the pair correlation function $g(r)$ were introduced,
and solved in the limit $d\to\io$; it was shown that one obtains the same result, 
up to $\f^{HS}_{max}$, meaning that Eq.~(\ref{Fdgrande}) might be a reasonable analytic continuation
of the liquid equation of state up to $\f^{HS}_{max}$ which is exponentially larger than $2^d$.
At this value of the density a pole develops at finite $q$ that seems to correspond
to a liquid instability (the {\it Kirkwood instability} \cite{FP99}).

The physical meaning of this instability is not clear.
Although very interesting from the mathematical point of view (maybe also in
relation to the problem of finding the most dense lattices \cite{Pa07}), this
question is not physically relevant in this context. We will show later that
the glass transition indeed preempts this instability that is therefore in a non-physical region of
the density: this system becomes unstable toward replica symmetry breaking at a density $\f \sim 2^{-d} d$,
before reaching the Kirkwood instability.

\subsection{The effective liquid: the Baxter model in $d\to\io$}

To compute the free energy of the replicated liquid,
we will neglect all the potentials but the two-body one. This
is because as discussed above we have indications that the corrections 
coming from the many-body potentials vanish
in this limit. 
However this point deserves a more careful investigation.

\subsubsection{Two-body potential in large dimension}

The calculation of the two-body potential was already done
at first order in \cite{PZ05}; in Appendix~\ref{app:A} it is carried out
at any order. The resulting effective potential, Eq.~(\ref{feff_def}), is reported
in figure \ref{fig:feff} for $d=3$. Note that at fixed $m$ the effective potential is a
function of $(r-D)/\sqrt{4A}$ only.

In the limit $d\to\io$ the correlations of the liquid vanish and
its $g(r)$ approaches a step function $\th(r-D)$; on the contrary, as we will show
below, the cage radius in the relevant region is very small (of the order of $1/d$). Then,
as the two-body potential vanishes on a scale $\sqrt{A}$, it is very reasonable to approximate it
with a delta function in $r=D$: this leads to the Baxter model 
\cite{Ba68} where
\beq\begin{split}\label{Qdelta}
e^{-\phi_{eff}(r)} &= \c(r) [1+Q(r)] \\ & \to \chi(r)\left[ 1+ D G_m(A) V_d(1) \d(r-D)\right] \ ,
\end{split}\eeq
where $G_m(A)$ is related to the integral of $Q(r)$ by Eq.~(\ref{GmAdef}) of Appendix~\ref{app:A}:
\beq\label{1sutau}
G_m(A) = \frac{1}{V_d(D)} \int d\vec r \, \chi(r) Q(r)=
\frac{d}{D^d} \int_D^\io dr \, r^{d-1} Q(r) \ .
\eeq
We will show below that the natural scale for $A$ in this limit is $\wh A = d^2 A/D^2$ which is of
order $1$ at the glass transition. With this scaling it is possible to compute the leading 
expression in the limit $d\to\io$, see Eq.~(\ref{taudinf}) in Appendix~\ref{app:A}:
\beq\label{GmA}\begin{split}
\lim_{d\to\io} & G(m,D^2 \wh A d^{-2}) \equiv \GG_m(\wh A) \\
&= \int_{-\io}^\io dy \, e^{y}
\left[ \Th\left(\frac{y + \wh A}{\sqrt{4 \wh A}}\right)^m - \th(y)\right] \ ,
\end{split}\eeq
where $\Th(t) = \frac12 [1 + \erf(t)]$.

\begin{figure} \centering
\includegraphics[width=8cm]{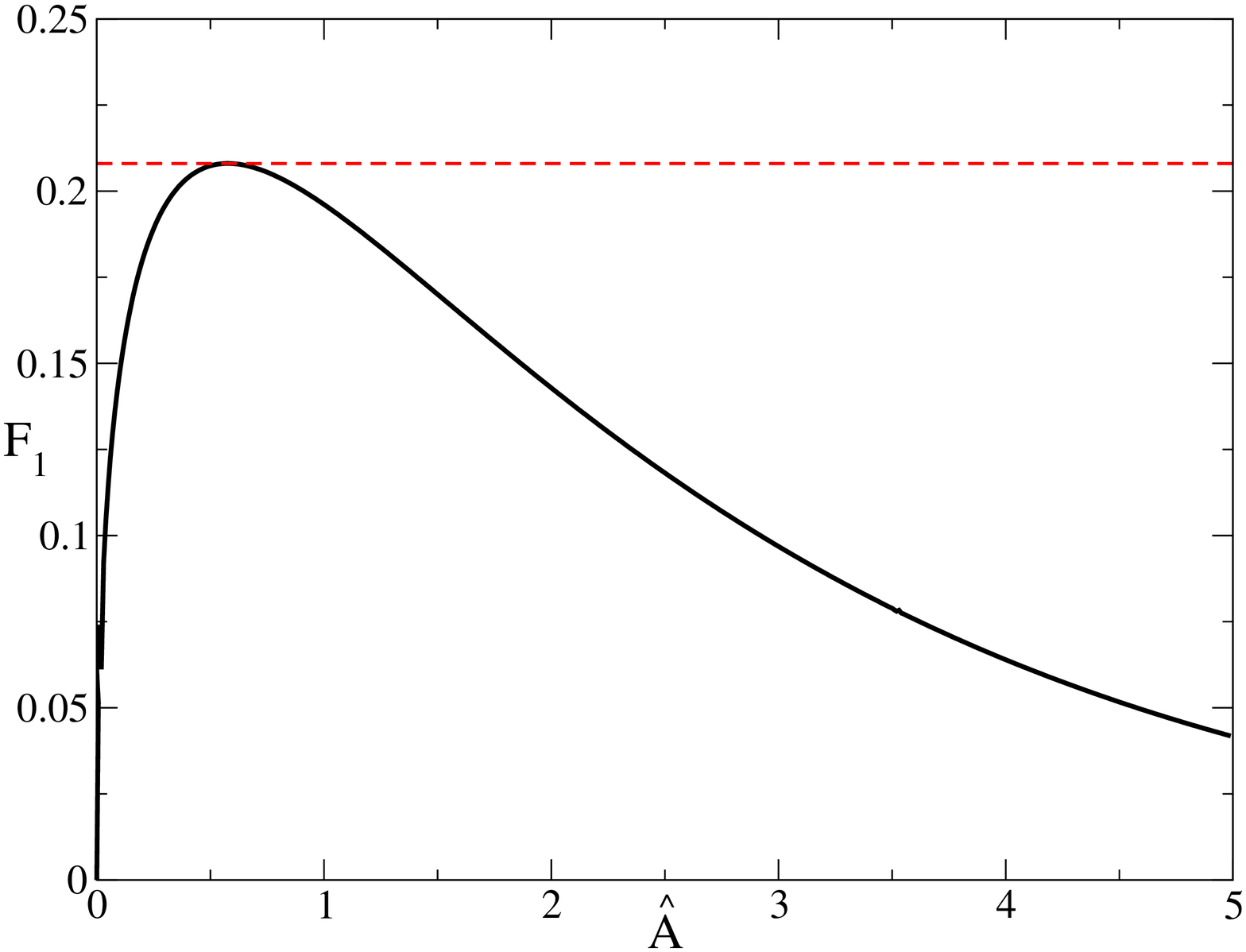}
\caption{
The function $\FF_1(\wh A)$.
}
\label{fig:FmA}
\end{figure}

\subsubsection{The Baxter liquid in large dimension}

We have then to solve the Baxter model in the limit $d\to\io$. To do this it is enough
to observe \cite{PS00} that the Fourier transform of $\d(r-D)$ and of $\th(D-r)$ coincide
at the leading order in the limit $d\to\io$. Using the results of \cite{PS00} it is easy to show that
\beq\label{fqeff}
\wh f_{eff}(q) = \left( 1- \GG_m(\wh A) \right) \wh f(q) \ ,
\eeq
where $\wh f_{eff}$ is the Mayer function corresponding to the Baxter potential (\ref{Qdelta}).
Note that this relation is exact for $q=0$.
It is possible to show (\eg by direct numerical computation) that, for all $m$ and $\wh A$,
$\GG_m(\wh A) < 1$; therefore the coefficient $1-\GG$ is positive and
the only difference between the Baxter liquid and the hard sphere liquid is in a $O(1)$ 
renormalization of the density. Therefore, up to a density 
$\f_{max}^{Baxter} = \f_{max}^{HS}/(1-\GG_m(\wh A))$, the free energy of the Baxter liquid is given by
\beq\label{Fbaxterdinf}
-\b F_{Baxter}(m,\f;\wh A) = 1-\ln \r - 2^{d-1}\f\left( 1-\GG_m(\wh A) \right) \ .
\eeq
We will see that the replica instability happens for $\f < \f_{max}^{Baxter}$, so this expression
will be enough for our purposes.
The replicated entropy is finally given by substituting (\ref{Fbaxterdinf}) and (\ref{GmA})
into (\ref{SS1}):
\beq\label{Srepdinf}
\begin{split}
\SS(m&,\f;A) =\\
&= S_{harm}(m,A)+ 1-\ln \r - 2^{d-1}\f\left( 1-\GG_m(\wh A) \right) \\
&=S(\f) + S_{harm}(m,A) + 2^{d-1} \f \GG_m(\wh A) \\
&= 1-\ln \r - 2^{d-1}\f 
 - \frac{d}{2}(1-m)\ln(2\pi \wh A/d^2) \\ 
&- \frac{d}{2} (1-m-\ln m) \\
&+2^{d-1} \f \int_{-\io}^\io dy \, e^{y}
\left[ \Th\left(\frac{y + \wh A}{\sqrt{4 \wh A}}\right)^m - \th(y)\right]
 \ .
\end{split}
\eeq
From this expression (that might be exact in the large dimension limit) it is possible to derive the
phase diagram of the system.

\subsection{The equation for $A$ and the clustering transition}

From Eq.~(\ref{Srepdinf}) we obtain the equation for $A$
from the condition
$\frac{\partial \SS}{\partial A}=0$:
\beq\label{EqAdinf}
\frac{d}{2^d \f} = \frac{\wh A}{1-m} \frac{\partial \GG_m(\wh A)}{\partial \wh A}
\equiv \FF_m(\wh A) \ .
\eeq
The solution to this equation must be substituted in Eq.~(\ref{Srepdinf}) to obtain
the replicated entropy $\SS(m,\f)$. For generic $m$, $\FF_m(\wh A)$ has the shape
reported in Fig.~\ref{fig:FmA} for $m=1$.
Therefore a solution exists only if 
\beq\label{fdmdinf}
\f \geq \frac{d}{2^d} \frac1{\max_{\wh A} \FF_m(\wh A)} \equiv \f_d(m) \ ,
\eeq
which defines the clustering transition density $\f_d(m)$. As $\max_{\wh A} \FF_m(\wh A)$
is a quantity of order $1$ for all $m$, the scale of the clustering transition is
$\f_d \sim d/2^d$. The same result was obtained in \cite{KW87} by means of density
functional theory.
Moreover, this result compares well with the results of Torquato et al.~\cite{TUS06}
who found that simple algorithms are able to construct packings up to this scale of
density.

For $\f > \f_d(m)$, the equation for $\wh A$ admits two solutions, 
the physical one being
the smallest, that correspond to a maximum of the free energy for $m<1$, as usual
in the replica computations. This solution has the right physical behavior as the cage
radius becomes smaller on increasing the density.
While the full curve $\f_d(m)$ can easily be computed numerically from (\ref{fdmdinf}),
we will focus in the following on the special cases $m=1$ and $m=0$ that define the equilibrium
clustering transition $\f_{d}$ and $\f_{th}$, respectively.

\subsubsection{The clustering (Mode-Coupling) transition}

For $m=1$, the clustering transition can be identified with the 
usual dynamical (or Mode-Coupling) transition. This is because when
the liquid splits into an exponential number of glassy states,
we expect the dynamics to become very slow as the system must cross barriers to change state.
In $d\to\io$ these barriers become very high and we expect a real dynamical transition
characterized by a divergence of the relaxation time at $\f_d$.
It would be very interesting to check this conjecture by investigating the Mode-Coupling
equations in the limit $d\to\io$.

We have from
Eqs.~(\ref{EqAdinf}), (\ref{GmA}), 
taking\footnote{Before taking the limit it is convenient to 
substitute in Eq.~(\ref{GmA}) 
$\th(y) \to \Th\left(\frac{y + \wh A}{\sqrt{4 \wh A}}\right)$; it can be shown that
the integral is unchanged. This is due to the property $\int d\vec r [q_A(r)-\chi(r)]=0$,
see Appendix~\ref{app:A}.}
the limit $m\to 1$
\beq
\FF_1(\wh A) = -\wh A \frac{\partial}{\partial \wh A} 
\int_{-\io}^\io dy \, e^{y}
 \Th\left(\frac{y + \wh A}{\sqrt{4 \wh A}}\right)
\ln  \Th\left(\frac{y + \wh A}{\sqrt{4 \wh A}}\right) \ .
\eeq
This function can be easily computed numerically: the result is shown in Fig.~\ref{fig:FmA}.
It has a maximum for $\wh A = 0.576$ where $\FF_1 = 0.208$; the corresponding value for the
clustering transition is
\beq\label{cluster_m1}
\f_d = \f_d(1) = 4.8 \frac{d}{2^d} \ , \hskip1cm d \to\io
\ .
\eeq
Note that this value is somehow larger (but not too much) than the saturation density 
$\f_{RSA} = d/2^d$ \cite{TUS06} of the Random Sequential Addition process in which spheres
are added sequentially at random.

\subsubsection{The $J$-point in $d\to\io$}

In the limit $m\to 0$ the cage radius $A$ that optimizes the free energy 
goes to $0$.
Therefore the jammed states
are obtained in the double limit $m\to 0$, $\wh A\to 0$, $\wh A=\wh\a m$,
as will be discussed in section~\ref{idgldinf}.
Using the results of Appendix~\ref{app:A}, in particular Eq.~(\ref{dtaujamming}),
one can show that the equation (\ref{EqAdinf}) for $\wh A$ becomes in this limit
\beq\label{alphadinf}
\frac{d}{2^d \f} = \FF_0(\wh \a) =
\frac{1}{4 \wh \a} \int_0^\io dy \, y^2 \, e^{-y-\frac{y^2}{4\a}} \ .
\eeq
The function $\FF_0(\wh \a)$ has the same qualitative shape of $\FF_1(\wh A)$, see
Fig.~\ref{fig:FmA}, and assumes its maximum for $\wh \a = 0.302$ with $\FF_0 = 0.160$.
This gives the leading order in $d\to\io$ of the clustering density for $m=0$:
\beq\label{cluster_m0}
\f_{th} = \f_d(0) = 6.26 \frac{d}{2^d}  \ , \hskip1cm d \to\io \ .
\eeq
As previously discussed this density correspond to the first appearance of jammed
states at infinite pressure and
we conjecture that it coincides with the $J$-point density, at least in this large dimension
limit.

\subsection{The ideal glass state}
\label{idgldinf}

Once Eq.~(\ref{EqAdinf}) has been solved, one can obtain the entropy and complexity of
glassy states from Eqs.~(\ref{mcomplexity}) and (\ref{Srepdinf}):
\beq \label{sigmadinf}
\begin{split}
s^*(m,\f) &= -d \ln d + \frac{d}2 \left[ \ln(2 \p \wh A) + 1 + \frac1m \right] \\
&+ 2^{d-1} \f \, \partial_m \GG_m(\wh A) \ , \\
\Si(m,\f) &= \frac{d}2 \ln d - \frac{d}2[1 + \ln(\wh A/m)] \\
&- 2^{d-1} \f [1 + m^2 \partial_m m^{-1} \GG_m(\wh A)] \\
&- \ln(2^d \f) - \frac12 \ln(\p d) + 1 \ ,
\end{split}
\eeq
where we used the Stirling formula to expand the Gamma function appearing in $V_d$.
In the region of the clustering transition we already showed that $2^d \f \sim d$, and $\wh A = O(1)$;
thus the first term in $\Si$ always dominates and gives a positive complexity.

To find the solution to $\Si(m,\f)=0$ we have then to choose $\f = \frac{d}{2^d} \wh \f$ and
look to $\wh \f \to \io$. In this case the solution for $\wh A$ vanishes. For small $\wh A$, the
function $\GG_m(\wh A) \sim \sqrt{4 \wh A} Q_0(m)$, see Appendix~\ref{app:A}. Substituting in
Eq.~(\ref{EqAdinf}) we get $\wh A \propto \wh \f^{-2}$, and substituting this in (\ref{sigmadinf}):
\beq
\Si(m,\wh \f) = \frac{d}2 \ln d -\frac{d}2 [\wh \f + O(\ln \wh \f)]
\eeq
Then we see that the leading behavior is $\wh \f = \ln d$ {\it independently of $m$}. The dependence
on $m$ comes from the subleading terms, see \cite{PZ06b} for an analysis of these terms.
Therefore, both the Kauzmann and glass close packing density scale as
\beq\label{Kauz_dinf}
\f_K, \, \fICP = \frac{d \ln d}{2^d}  \ , \hskip1cm d \to\io \ ,
\eeq
for $d\to \io$.
Note that in the limit $m\to 0$, $\wh A =\wh \a m$, the vibrational entropy goes to $-\io$, 
due to the term $\ln \wh A$ in (\ref{sigmadinf}). 
In summary, for $m\to 0$ the cage radius goes to $0$, 
the entropy goes to $-\io$ and the pressure diverges: the particles
get in contact and the system is jammed.

\subsection{The correlation function}

The correlation function of the glass is given by the correlation function of the effective
Baxter liquid. In the ring resummation for Hard Spheres, Eq.~(\ref{Sanelli}), we get
\beq\label{grdinf}
\begin{split}
&g(r) = \frac{2}\r \frac{\d S}{\d \ln\chi(r)} = \chi(r) y(r) \ , \\
&y(r) = 1 + \int \frac{dq}{(2\p)^d} e^{-iqr} \frac{\r \wh f(q)}{1 - \r \wh f(q)} \ ,
\end{split}\eeq
where $\wh f(q)$ is given by (\ref{fqHS}).
For the Baxter effective liquid one has to substitute $\wh f(q) \to \wh f_{eff}(q)$ and
$\chi(r) \to e^{-\phi_{eff}(r)}$. In both cases, for $d\to \io$, the analysis of \cite{PS00}
or a direct investigation of (\ref{grdinf}) shows that $y(r) = 1 + e^{-\frac{d}2 \wh y(r)}$,
where the function $\wh y(r)$ can be determined, see \cite{PS00}. Ignoring
this exponentially small correction\footnote{That however is necessary to ensure that
$S(q) \geq 0$ for all $q$ and that $S(0)=0$ \cite{PS00}.} we are left with
$\gG(r) = e^{-\phi_{eff}(r)}$. Therefore the correlation function of the glass is essentially
given by a step function plus a peak in $r=D$, which becomes a delta function in the
jamming limit $m\to 0$, $A = \a m$. We can compute the weight of the delta peak $\chi(r) Q(r)$, which
is related to the average number of contacts per sphere $z$ by:
\beq
z = \r \int dr \chi(r) Q(r) = \frac{\r\Omega_d D^d}{12 \t}
= 2^d \f G_m(A) \ .
\eeq
In the jamming limit $\GG_m(\wh A) \to \GG_0(\wh \a)$, see Eq.~(\ref{G0adef}),
and $\wh \a = \wh\a(\f)$ is the solution of Eq.~(\ref{alphadinf}), so that
\beq\label{zdinf}
z = d \left. \frac{\GG_0(\wh\a)}{\FF_0(\wh\a)} \right|_{\wh\a = \wh\a(\f)} \ ,
\eeq
where $\FF_0(\wh\a) = \wh\a \frac{d \GG_0(\wh\a)}{d \wh\a}$ is defined in (\ref{alphadinf}).
In the limit $\wh\a \to 0$, that corresponds to random close packing, we have 
$\GG_0(\wh\a) \propto \sqrt{\wh\a}$ which implies $z \to 2 d$, \ie the packings are
{\it isostatic}. However, for $\wh\a > 0$ Eq.~(\ref{zdinf}) gives $z>2d$, with $z \sim 3.6 d$
at the threshold density. 

This strange result might be due to different reasons: {\it i)} subleading corrections
that we neglected might affect the result in some subtle way; {\it ii)} the states 
corresponding to finite $\wh\a$ might be unstable; it is possible that
only states with $\wh\a \sim (\ln d)^{-2}$ are stable; {\it iii)} 
it is possible that other contributions, \eg those of the square root singularity
in lower dimension, are merged with the true contacts on the scale of the delta peak.
This point deserves further investigation.

\subsection{Discussion}

We showed that for $d\to\io$ the model can be almost
completely solved, at least within our initial assumptions, \ie that there
is a clustering transition where the configuration space splits into many disconnected
clusters without further structure\footnote{In replica jargon this corresponds
to a 1RSB solution. Unfortunately we cannot study the instability of this solution
towards more complicated replica symmetry breaking solutions.},
and the phase diagram computed in full detail.

To further check the consistency of the small cage expansion, it is interesting to estimate the
Lindemann ratio in the glass phase, when $2^d \f \sim d \ln d$.
The Lindemann ratio $L$ for a given solid phase is the ratio between the typical amplitude 
of vibrations around the equilibrium positions and the mean interparticle distance. In our
framework it can be defined as
\beq
L \equiv \rho^{1/d} \sqrt{A} \ .
\eeq
Using $\sqrt{\wh A} = d \sqrt{A}/D \propto 1 / \wh\f \propto (\ln d)^{-1}$ as derived in 
section~\ref{idgldinf},
and $2^d \f = \rho V_d(D) \sim d \ln d$, one has
\beq\label{Lindedinf}
L \sim \frac{1}{\sqrt{d} \ln d} \ll 1 \ ,
\eeq
which is consistent with the assumption that vibrations are very small.

We can compare our prediction for the glass close packing
$\fICP \sim 2^{-d} d \ln d$ with the best available bounds on the density of crystalline
packings. 
A classical lower bound for the close packing density is the Minkowsky bound, 
$\f \sim 2^{-d}$. It has been improved in~\cite{Ba92} where, for the case of
lattice packings, it is proved that $\f \geq 2 d \, 2^{-d}$; see also~\cite{KLV04} 
where a procedure to construct packings achieving this bound has been discussed.
The best currently known upper bound is reported in~\cite{KL78}, and has the
asymptotic 
scaling $\f \sim 2^{-0.5990\ldots \, d}$
Our result for $\fICP$ lies between these bounds (it is only a factor $\ln d$ better
than the lower bound) so
we cannot give an answer to the question whether the densest packings of hard spheres
in large $d$ are amorphous or crystalline. Hopefully better bounds on the density
of crystalline packings will address this question in the future.
The values of densities of crystalline laminated lattices \cite{ConwaySloane}
up to $d$=50 seem to suggest that there are lattices where $2^d \f$ grows exponentially in $d$.
It is however quite possible that this is a preasymptotic effect;
see \cite{Pa07} for a detailed discussion.
It would be very interesting to
find the density of laminated lattices in larger dimensions. 
Finally, it is worth to note that in~\cite{TS06b} it has been proposed that
it is possible to achieve packings having density exponentially higher
than the Minkowsky lower bound, and actually very close to the upper bound cited above.
Trying to prove (or disprove) this conjecture is a challenge for future research.


\section{Finite dimension: first-order in the small cage expansion}
\label{sec:smallcage}

In this section we will discuss the simplest approximation that works well in finite dimension, 
namely a
first-order expansion in the cage radius $\sqrt{A}$. Despite its simplicity,
we will see that it is able to reproduce many of the available numerical data
with good accuracy. Its main drawback is that it does not allow to access
the clustering transition, nor the full shape of $g(r)$ in the glass phase.
In section \ref{sec:beyond}, we will discuss different
ways to try to improve over this approximation.

\subsection{First order replicated free energy}

We will focus on the first correction in the expansion of Eq.~(\ref{Sresum})
in powers of $\sqrt{A}$. Then, we can neglect all the $n\geq 3$-body
potentials that give contributions $O(A)$, see Appendix~\ref{app:linkexp}.
Moreover, at fixed $m$ the two-body potential is $O(\sqrt{A})$ and we can
treat it as a perturbation.
Note that in the limit $m\to 0$ this is not true, as in fact the expansion
parameter is $A/m$ and not $A$.
However we are interested in the limit $m\to 0$ with $A= \a m$. 
Taking the limit $m\to 0$ of the first order term in $\sqrt{A}$ will then give
the first order term in $\sqrt{\a}$.

\subsubsection{The small cage expansion}

Taking into account only the two-body potential, Eq.~(\ref{Sresum}) reduces to
(\ref{SS1}), where
\beq\label{Feff2}
-\b F_{eff}[\f;\phi_{eff}(r)] = N^{-1} \SS[\f;\c(r) (1+Q(r))] \ ,
\eeq
where $\SS[\f;b(r)]$ is the entropy functional (\ref{Sdirho}) for a non-replicated
liquid with interaction $b(r) = e^{-\phi_{eff}(r)}$. It is easy to show, using
standard liquid theory \cite{Hansen}, that for a translationally invariant system
\beq
\frac{1}{N} \frac{\d\SS[\f,b(r)]}{\d \ln b(r)} = \frac{\r}{2} g(r) \ ,
\eeq 
where $g(r)$ is the pair distribution function of the liquid.
Using this relation to expand (\ref{Feff2}) for small $Q(r)$, and calling 
$S(\f) = N^{-1} \SS[\f;\chi(r)]$ the entropy of the non-replicated hard
sphere liquid, the first order expression
for the entropy is obtained from (\ref{Sresum}):
\beq\label{Sfirstorder}\begin{split}
&\SS(m,\f;A) = N^{-1} \SS[\r(\bar x),\bar\chi(\bar x,\bar y)] \\ &=
S_{harm}(m,A) + S(\f) + \frac{\r}2 \int dr g(r) Q(r) \ .
\end{split}\eeq

To compute the correlation function of the glass, 
we start from Eq.~(\ref{corrglassv}); we use 
Eq.~(\ref{Sfirstorder}) substituting $\chi \to \chi e^{-v}$ and we get
\beq\label{gQg}
\frac{\r^2}2 \gG(x,y) = \frac{\r^2}2 g(x,y) + \frac{\r^2}2 \int du dv \frac{\d g(u,v)}{\d \ln \c(x,y)} Q(u,v)
\ ,
\eeq
and using Eq.~(\ref{dgdchi}) of Appendix~\ref{app:deriva}:
\beq
\frac{\d g(u,v)}{\d \ln \c(x,y)} = g(x,y) \d(x-u)\d(y-v) + \text{other terms} \ ,
\eeq
(where the other terms contain basically the connected 4-point function),
we get
\beq\label{gfirst}
\gG(r) = g(r) [1 + Q(r)] + \text{other terms}
\eeq
which is the result we already obtained in \cite{PZ05} in a completely different way.
One can argue that the contribution of the other terms is negligible for $r\sim D$;
therefore Eq.~(\ref{gfirst}) allows to access the shape of the contact peak in
$\gG(r)$.

If $\sqrt{A} \ll D$ we can assume that the function $g(r)$
is essentially constant\footnote{In fact this assumption is false:
it will turn out at the end of the computation that $\sqrt{A}/D \sim 0.01$
but on this scale $g(r)$ still has structure close to contact.
However this assumption gives consistently the first order in the small
cage expansion.}
on the scale $\sqrt{A}$, $g(r) \sim g(D) \chi(r)$ 
for $r -D \sim O(\sqrt{A})$. 
Then we have
\beq
\int dr g(r) Q(r) \sim g(D) \int dr \chi(r) Q(r) \equiv g(D) V_d(D) G_m(A) \ ,
\eeq
and
\beq\label{S_fo_nondev}
\SS(m,\f;A) =
S_{harm}(m,A) + S(\f) + g(D) 2^{d-1}\f G_m(A) \ ,
\eeq
which resembles the expression for the Baxter model in $d\to\io$ (\ref{Srepdinf}), 
except for the factor $g(D)$ in front of the last term.
The expansion in powers of $\sqrt{A}$ is easily obtained,
from Eq.~(\ref{taufirstorder}) in Appendix~\ref{app:A}:
\beq\begin{split}
\label{Fcorrect}
\SS(m,\f;A) &=
S_{harm}(m,A) + S(\f) + 2^d \f g(D) \frac{d \sqrt{A}}D Q_0(m)  \ .
\end{split}\eeq
This result has been obtained in \cite{PZ05,PZ06a}.

\subsubsection{Optimization with respect to the cage radius}

We finally have to optimize with respect to $A$ in
Eq.~(\ref{Fcorrect}) to get the free energy at the minimum.
Note that the derivative of (\ref{Fcorrect}) is a linear
function of $\sqrt{A}$, therefore it always allows a solution independently
of the values of $\f$ and $m$. Therefore investigating the dynamic
transition, \ie the point at which the solution for $A$ disappears, is not
possible within this approximation. This is because the function $F_m(A)$ has
a shape similar to the one in figure \ref{fig:FmA}:
if we expand it for small $A$ and keep only the leading $\sqrt{A}$ term, 
we lose the maximum in figure \ref{fig:FmA} and are left with a continuously
increasing function. Therefore the equation $F_m(A)$=const. will have solution
for all values of the constant, \ie of the density.

Keeping this in mind, we obtain for the replicated free energy:
\beq\label{repF}
\begin{split}
&\SS( m , \f)= S(\f) - \frac{d}2(1-m) \ln[2\p A^*] - \frac{d}2(m-1-\ln m) \ , \\
&\sqrt{A^*(m)} = \frac{1-m}{Q_0(m)} \frac{D}{\rho V_d(D) g(D)} = \frac{1-m}{Q_0(m)} \frac{D}{2^d \f Y(\f)}
\end{split}
\eeq
where\footnote{The reason for this notation is that $y(r) = e^{\phi(r)} g(r)$ is continuous
in $r=D$ \cite{Hansen} and $Y(\f) = y(D) = g(D)$.}
 $Y(\f) \equiv g(D)$.
Applying Eq.~(\ref{mcomplexity}) we get
\beq\label{siSi}
\begin{split}
s^*(m,\f) &= \frac{\partial \SS(m,\f)}{\partial m} 
= \frac{d}2 \ln[2\pi A^*(m)] \\
&+ d (1-m) \frac{Q'_0(m)}{Q_0(m)} + \frac{d}{2}\frac{m+1}m \\
\Si(m,\f) &= \SS(m,\f)-m s^*(m,\f) \\ &= 
S(\f) - \frac{d}2 \ln[2\pi A^*(m)] \\ &- d m (1-m) \frac{Q'_0(m)}{Q_0(m)}
+\frac{d}2 \ln m -d  m 
\end{split}
\eeq
The important remark is that the only input in the expressions above is the entropy of the
liquid, $S(\f)$; the only other quantity is $Y(\f)$ appearing in $A^*(m)$, see Eq.~(\ref{repF}),
but the latter is related to $S(\f)$ by the general relation
\beq\label{pressione}
p \equiv \frac{\b P}{\r} = 1 + 2^{d-1} \f g(D) = -\f \frac{dS(\f)}{d\f}
\eeq
where $p$ is the {\it reduced pressure}.

Given $S(\f)$, the density $\fICP$ is the solution of 
$\Si_j(\f) =\lim_{m\to 0} \Si(m,\f)=0$; using
Eq.~(\ref{siSi}) and the asymptotic behavior of $Q_0(m) \sim \sqrt{\pi/(4m)}$ 
(see Appendix~\ref{app:A}), we obtain the condition:
\beq\label{phiRCPdef}
\Si_j(\f) = \lim_{m\to 0} \Si(m,\f) = S(\f) -d \ln \left[\frac{\sqrt{8}}{2^d \f Y(\f)}\right] + 
\frac{d}2 = 0 \ ,
\eeq
while the Kauzmann density is the solution of $\Si_{eq}(\f) = \Si(1,\f) = 0$, 
which gives the condition
\beq\label{phiKdef}
\Si_{eq}(\f) = S(\f) - d \ln \left[ \frac{\sqrt{2\pi}}{2^d Q_0 \f Y(\f)} \right] =0 \ ,
\eeq
with $Q_0 \equiv -Q_0'(m=1) = 0.638\ldots$.

As $Q_0(m) \sim m^{-1/2}$ for $m\to 0$, from the second Eq.~(\ref{repF}) one can show
that in the limit $m\to 0$, $A \propto m$, and the first Eq.~(\ref{siSi}) shows that 
the vibrational entropy and the pressure both diverge.
This is consistent with the identification
of the limit $m\to 0$ and the jamming limit,
as we already discussed in section~\ref{idgldinf}.

It has been shown in \cite{PZ05,PZ06a} that these first order expressions already give 
quantitative predictions that are in very good agreement with
numerical simulations. In the following we will review these results.

\subsection{Equation of state of the glass and complexity of the liquid}

\begin{table}[t]
\centering
\begin{tabular}{c|c|c|c|c|c|c|}
\hline
$d$ & $\f_K$  & $\fICP$ & $\f_J$ & $\f_{MRJ}$ & $\f_K$ & $\fICP$ \\
& (theory) & (theory) &  (num.) & (num.) & (extr.) & (extr.) \\
\hline
2 &  0.8165 & 0.8745 & --- & --- & --- & --- \\
3 &  0.6175 & 0.6836 & 0.640 & 0.64 & ---& --- \\
4 &  0.4319 & 0.4869 & 0.452 & 0.46 & 0.409 & 0.473 \\
5 &  0.2894 & 0.3307 & --- & 0.31 & ---& --- \\
6 &  0.1883 & 0.2182 & --- & 0.20 & ---& --- \\
7 &  0.1194 & 0.1402 & --- & --- & ---& --- \\
8 &  0.0739 & 0.0877 & --- & --- & ---& --- \\
\hline
$\io$ & $2^{-d} d \ln d$ & $2^{-d} d \ln d$ & --- & $2^{-d} d$ ? & --- & --- \\
\hline
\end{tabular}
\caption{
Values of $\f_K$ and $\fICP$ from the first-order small cage
expansion of section~\ref{sec:smallcage} (only values
for $d\leq 8$ are reported for brevity, 
values for $d\geq 8$ are in Fig.~\ref{fig:scaling}) 
compared with the available numerically measured
values of $\f_J$ \cite{OSLN03,XBO05,SO08}
and of $\f_{MRJ}$ \cite{SDST06}. The last two columns give the values of
$\f_K$ and $\fICP$ extrapolated from the fits of the
data of \cite{SDST06} reported in figure~\ref{fig:phi_gamma}.
The large-$d$ scaling of $\f_{MRJ}$ has been conjectured in \cite{SDST06}.
}
\label{tab:I}
\end{table}

Once an equation of state for the liquid is given, Eq.~(\ref{siSi}) give access
to the phase diagram, except for the dynamic transition. In this section we discuss
the results in different space dimensions.

\subsubsection{$d=1$}

In dimension $d=1$ we do not expect any glass transition. Indeed, the particles
are always in a ``state'': apart from the trivial symmetries (permutations and
global translations), they vibrate around the ``configuration'' 
$x_1 < x_2 < \cdots < x_N$. The packing fraction is $\f = \r D$.
The diameter of the cages is given by
$2 R= D \frac{1-\f}{\f}$, and it is easy to show that the distribution
of the cage radii is exponential. Therefore in this case we expect our Gaussian
approximation to be very inaccurate.

The entropy can be computed exactly and is given by 
$S(\f)= 1 + \ln(2R) = 1 - \ln(\f/D) + \ln (1-\f)$.
The value of the correlation function at contact is $Y(\f) = \frac1{1-\f}$
and diverges at close packing $\f=1$.

\begin{figure}[t]\centering
\includegraphics[width=8.5cm]{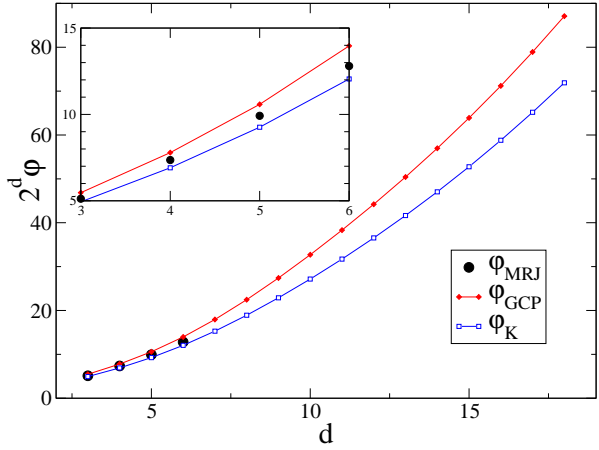}
\caption{Plot of $\f_K$ (open squares, obtained solving Eq.~(\ref{phiKdef})), $\fICP$ (full diamonds,
Eq.~(\ref{phiRCPdef})), and $\f_{MRJ}$ (full circles, numerical
  estimate of \cite{SDST06}) as a function of the
  dimension. Both $\f_K$ and $\fICP$ scale as $2^d \f \sim d \ln d$
  for large $d$, while their distance scales as $2^d [\fICP-\f_K] \sim d$.
In the inset the same plot for $3\leq d \leq 6$ (compare with Fig.6 in
\cite{SDST06}).}
\vskip-10pt
\label{fig:scaling}
\end{figure}

In our approach, as there is only one state, we expect to 
have $Z_m =\exp [ mNS(\f)]$, then $\SS(m,\f) = -N^{-1} \ln Z_m = -m S(\f)$,
and $s^*(\f) = - \partial_m \SS(m,\f) = S(\f)$, \ie the vibrational
entropy is equal to the total entropy and the complexity is $\Si=0$. 
Indeed, substituting the expressions above for
$S(\f)$ and $Y(\f)$ in (\ref{siSi}) we easily find:
\beq
\begin{split}
&2 \sqrt{A^*(1)} = D \frac{1}{0.638} \frac{1-\f}{\f} \ , \\
&s^*(\f) = \frac{1}{2} \ln [2\pi A^*(1)] \\ &= 
\ln \frac{\sqrt{2 \pi}}{2 \cdot 0.638} - \ln (\f/D) + \ln (1-\f) \ .
\end{split}
\eeq
Thus the theory reproduces the exact result apart from a (small) constant shift
of the entropy (corresponding to a multiplicative factor in front of the
cage radius). This is probably due to the Gaussian approximation, that is
clearly wrong as the distribution of the cage is exponential, and to the small
cage approximation. Indeed, for $\f \sim 1$, where the cages are small,
the leading term of the entropy is correctly reproduced by the expressions
above.

\begin{figure*}[t] \centering
\includegraphics[width=8cm]{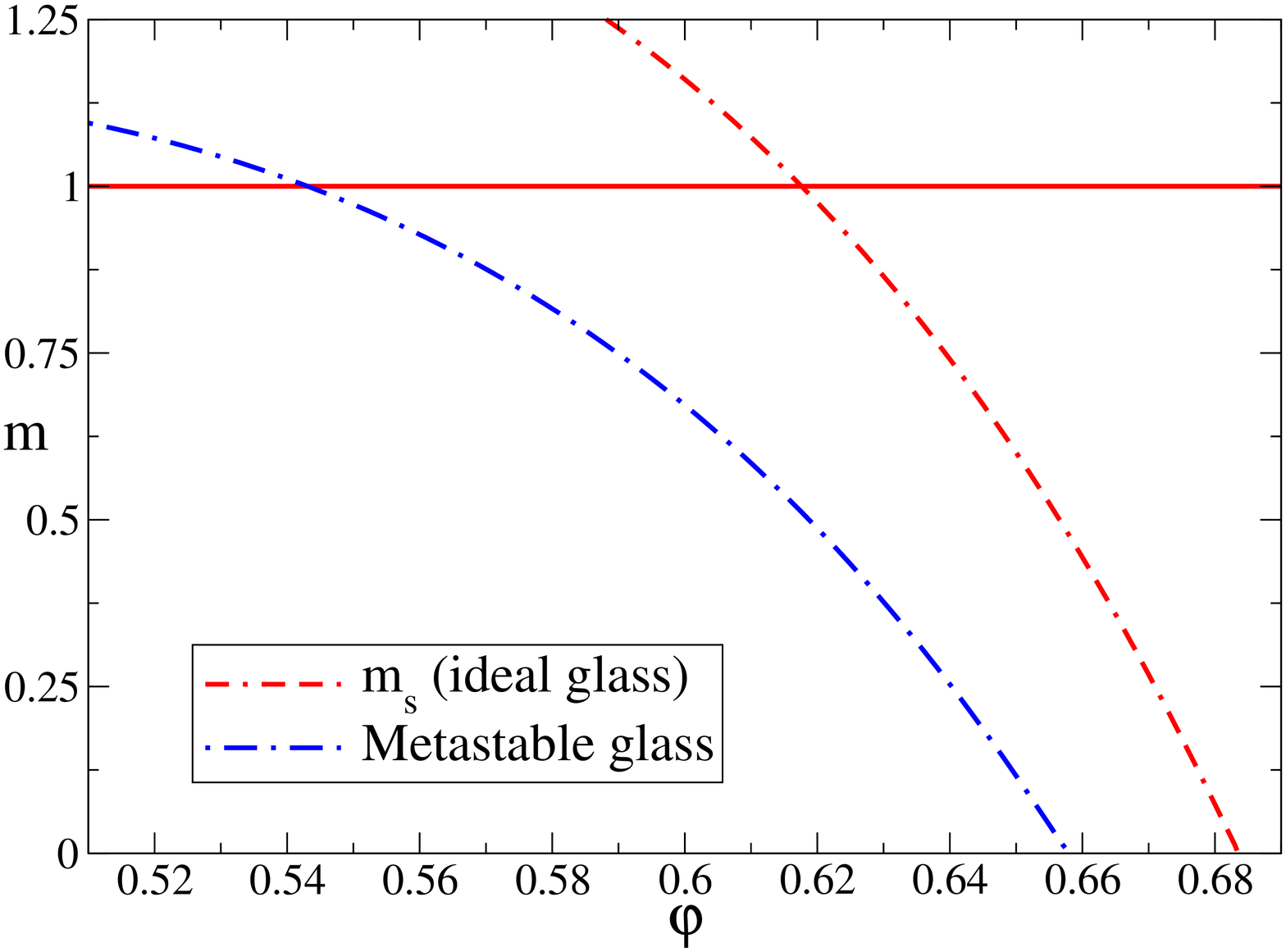}
\includegraphics[width=8cm]{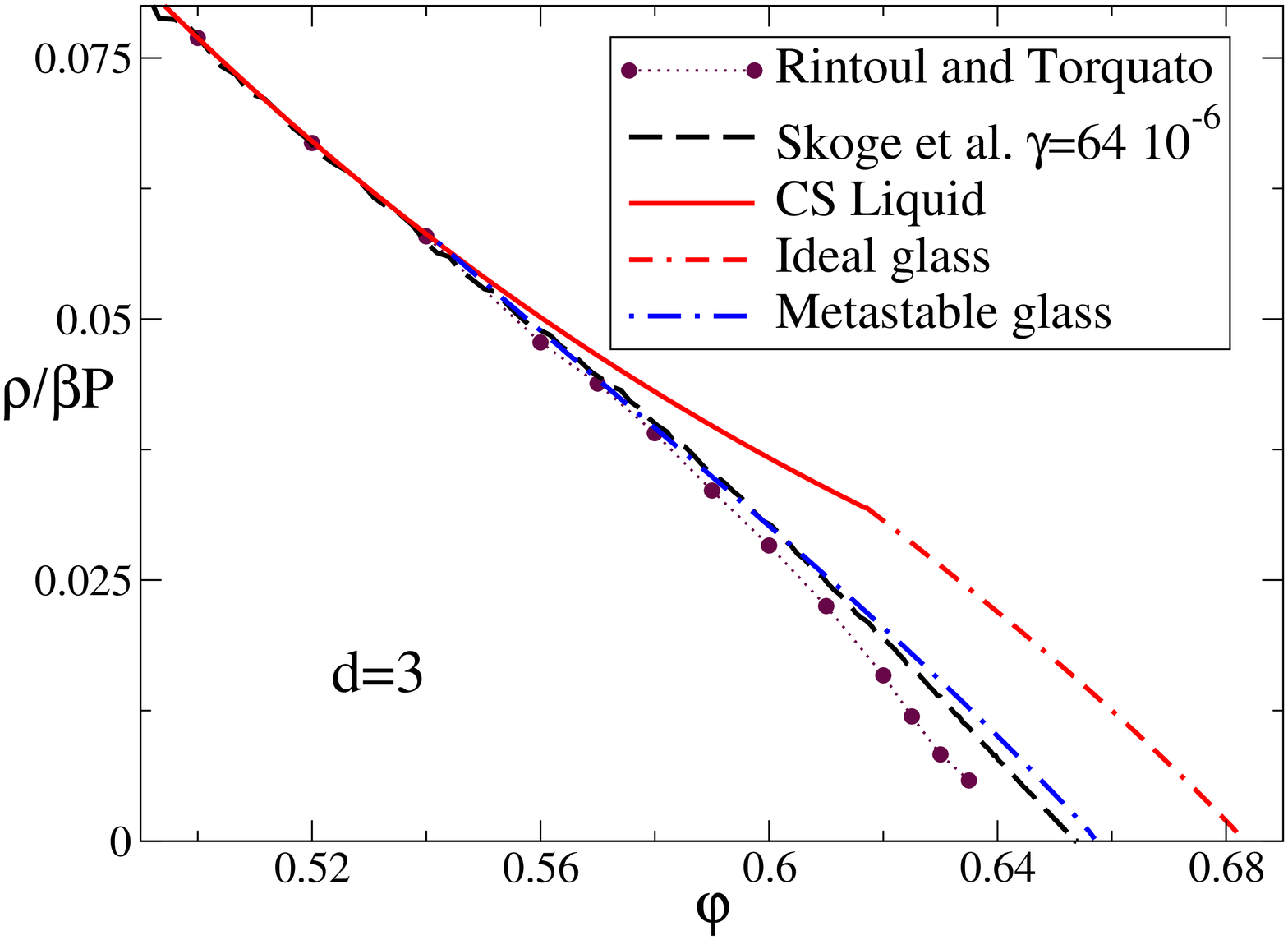}
\caption{Equation of state in $d=3$ using the Carnahan-Starling equation of
  state to describe the liquid, and in the small cage approximation (section~\ref{sec:smallcage}).
(Left) $(m,\f)$ phase diagram (compare with figure~\ref{fig:mstar}).
The line $m_s(\f)$ (dot-dashed-dashed red line) separates the liquid
and glass regions. The glass transition density is defined by $m_s(\f_K)=1$ and
the glass close packing density by $m_s(\fICP)=0$:
we find $\f_K = 0.62$ and $\fICP = 0.68$.
The blue dot-dashed line refers here to
a metastable glassy state defined by $\Si(m,\f)=\Si_j$ (here we chose $\Si_j=1.2$) 
with jamming density $\f_j = 0.659$.
(Right) Inverse reduced pressure as a function of density. 
Red full line is the CS equation of state; red dot-dashed-dashed line is the ideal
glass; blue dot-dashed line the same metastable glass of left panel.
Data from~\cite{RT96} (full circles) and more recent ones 
from~\cite{SDST06} (dashed black line,
same data as in figure~\ref{fig:Skoge}, left, for $\g = 64 \cdot 10^{-6}$)
are reported for comparison.
}
\label{fig:diad3}
\end{figure*}

\subsubsection{$d=2$}

In dimension $d=2$ we can use either the Henderson expression for $Y(\f)$ \cite{He75}:
\beq
Y_H(\f) = \frac{1-7 \f/16}{(1-\f)^2}
\eeq
or the improved expression of Luding \cite{Lu01}:
\beq
Y_L(\f) = Y_H(\f)-\frac{\f^3}{2^7 (1-\f)^4}
\eeq
with very small quantitative differences.
In the first case we find\footnote{Note that in Ref.~\cite{Za07} it was
  erroneously stated that in $d=2$ the present method predicts the absence of a
  glass transition. We wish to thank F.~Caltagirone for pointing out the error
  and providing the correct values of $\f_K$ and $\fICP$.}
$\f_K=0.816$ and
$\fICP=0.874$, while in the second we find 
$\f_K=0.811$ and
$\fICP=0.873$.

Note that in $d=2$ the existence of an ideal glass transition has been recently the object of an
intense debate \cite{SK00,SK01,BR04,DTS06,Ta07}. 
In fact, amorphous packings of monodisperse two-dimensional spheres are
particularly unstable~\cite{BW06}. For this reason, in the following we will
focus on $d \geq 3$.

\subsubsection{$d \geq 3$}

\begin{figure}[t] \centering
\includegraphics[width=8cm]{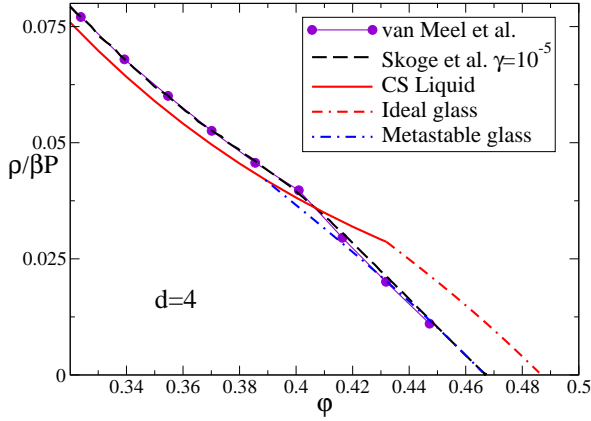}
\caption{Inverse reduced pressure as a function of density
in $d=4$ using the Carnahan-Starling equation of
  state (red full line) to describe the liquid, 
  and in the small cage approximation (section~\ref{sec:smallcage}).
Data from~\cite{SDST06} (black dashed line, 
same data as in figure~\ref{fig:Skoge}, right, 
for $\g = 10^{-6}$) and from~\cite{MFC08}
are reported for comparison. 
The reported metastable glass (dot-dashed blue line) corresponds to $\Si_j=1.44$ 
and has a jamming
density $\f_j = 0.467$. In this case the Carnahan-Starling equation is less
accurate; nevertheless, the overall quantitative agreement is still good.
The ideal glass branch is reported as a dot-dashed-dashed red line.
}
\label{fig:diad4}
\end{figure}

In $d=3$ we used the Carnahan-Starling expression \cite{Hansen} for the entropy $S(\f)$, 
which reproduces very well the numerical data for the equation of state of the
hard sphere liquid. 
In $d>3$ the Carnahan-Starling equation of state can be generalized \cite{SMS89}:
\beq\begin{split}
&Y(\f) = \frac{1-\a\f}{(1-\f)^d} \ , \\
&\a = d - 2^{d-1} (B_3/b^2) \ ,
\end{split}
\eeq
where $Y(\f)=g(D)$ is the value of the radial distribution function at contact,
and $b$ and $B_3$ are the second and third virial coefficients, whose exact
expression is known \cite{SMS89}. The entropy of the liquid $S(\f)$ is obtained by
integrating the exact expression (\ref{pressione}).
In $d=4$ this equation is not very accurate, see figure~\ref{fig:diad4}, and an
equation of state based on Pad\'e approximants \cite{BW05} seems more accurate,
see figure~\ref{fig:Skoge}. Still the error is not so large and 
for simplicity in the analytic computations
we will use the Carnahan-Starling approximation for all $d\geq 3$.

Using this expression for $S(\f)$, Eqs.~(\ref{phiRCPdef}) and (\ref{phiKdef}) 
can be easy solved
numerically to get the values of
$\fICP$ and $\f_K$ {\it for any given value of} $d$.
 The results are reported in
table~\ref{tab:I} for $d \leq 8$, and compared with $\f_{MRJ}$ as reported in
\cite{SDST06}. The latter quantity is the density of the 
{\it maximally random} (according to some measure of ``order'') 
{\it collectively jammed packings} 
of the system, see \cite{TTD00} for the precise
definition; it is estimated in \cite{SDST06} by the jamming
density $\f_j(\g)$ for finite but small $\g$ (see sec.IV in \cite{SDST06}
for a detailed discussion). As $\f_j(\g)$ is expected to increase on
decreasing $\g$ and $\fICP = \lim_{\g\to 0} \f_j(\g)$
(see figure \ref{fig:phi_gamma}),
it follows that $\f_{MRJ}$,
as estimated in \cite{SDST06},
should be strictly lower than $\fICP$, but close to it,
consistently with the data in table \ref{tab:I}.
A plot of $\f_K$ and $\fICP$ for $d$ up to $19$ is reported 
in Fig.~\ref{fig:scaling}. 
Note that it has been suggested in \cite{SDST06} 
that $2^d \f_{MRJ} \sim c_1 + c_2 d$; and indeed,
given that the compression rates used in numerical simulation are not very
small, $\f_{MRJ}$ should be quite close to $\f_{th}$, that scales
as $d/2^d$, see Eq.~(\ref{cluster_m0}).
Recalling that $2^d \f_K, 2^d \fICP \sim d \ln d$,
we expect that at some point
$\f_{MRJ}$ will become smaller than $\f_K$, even if this is not observed for $d \leq 6$.

The very nice data for $d=4$ reported in \cite{SDST06}, and
reproduced in figure~\ref{fig:Skoge}, allow for a more
precise comparison of the numerical and theoretical results: the value of
$\f_j(\g)$ has been measured for 5 different values of
$\g=10^{-3},10^{-4},10^{-5},10^{-6},10^{-7}$, see figure~\ref{fig:phi_gamma}. 
A standard procedure to extrapolate to $\g \to 0$ is to fit the data to a
Vogel-Fulcher law:
\beq\nonumber
\g(\f_j) = \g_0 10^{-\frac{D}{\fICP-\f_j}} \Leftrightarrow
\f_j(\g) = \fICP + \frac{D}{\log_{10}[\g/\g_0]} \ .
\eeq
Such extrapolations are often ambiguous; however the fit is good 
(see figure~\ref{fig:phi_gamma}) and gives
$\fICP = 0.473$. A similar extrapolation of
$\f_g(\g)$ (defined roughly as the point where the curves in figure~\ref{fig:Skoge}
leave the liquid
equation of state) to $\g=0$ gives $\f_K = 0.409$.
The final results
differ by $\sim 10\%$ from the theoretical values, see table~\ref{tab:I}.
This is a very good result given the ambiguities that are present both in the
numerical data (numerical error and extrapolation) and in the theory
(the choice of a particular expression for the equation of state of the liquid
that is not exact, and the small cage expansion).
Note that a similar extrapolation is not possible in $d<4$ due to
crystallization, and for $d>4$ due to lack of numerical data. 
Hopefully new data for $d>4$ will allow for a similar
comparison also in this case. 

In figures~\ref{fig:diad3} and \ref{fig:diad4} we report the inverse
reduced pressure as function of the packing fraction for $d=3$ and $d=4$.
We plot the numerical data of \cite{SDST06} already reproduced in figure~\ref{fig:Skoge},
and we choose two among the smallest compression rates available.
Still we see that the numerical data are far from the ideal glass pressure, and are better
described by metastable glassy states corresponding to a complexity $\Si \sim 1$. Indeed, it is well
known from numerical simulations of structural glass formers that the system falls out of
equilibrium when $\Si \sim 1$. The overall agreement of the theoretical predictions with the
data of \cite{SDST06} is very good.

Finally, in figure~\ref{fig:Sicomp} we report the equilibrium complexity
$\Si_{eq}(\f) = \Si(1,\f)$ as function of $\f$. Numerical data from
\cite{AF07} and \cite{Sp98} are available, and the agreement is again very good.

\subsection{Scaling close to jamming}

In this section we will derive asymptotic relations for the behavior of the glassy states
at $\f \to \f_j$. This is interesting because close to $\f_j$ the
correlation function shows interesting features that have been studied in detail numerically,
see the discussion in section~\ref{sec:gofr}.

\subsubsection{Scaling of the pressure}

First we compute the scaling of the pressure close to $\f_j$. It is convenient to recall the
procedure already discussed in section~\ref{sec:metastable}.
To each group of glassy states of jamming density $\f_j$ we can associate a complexity
$\Si_j$ given by Eq.~(\ref{phiRCPdef}). Then one has to solve 
Eq.~(\ref{Sijsolve}) to obtain $m(\f,\Si_j)$ or equivalently $m(\f,\f_j)$.
The function $m(\f,\f_j)$ represents a group of glassy states in the $(m,\f)$ plane
of figure~\ref{fig:mstar}, left panel. Once substituted in equation~(\ref{siSi}), 
it gives the entropy of the states\footnote{We recall 
that the case $\Si_j = 0$ correspond to the ideal glass.} labeled by $\f_j$ as a function of $\f$.

We are interested in the equation of state
of these states close to $m = 0$, or $\f=\f_j$.
Then we need to compute $m(\f,\f_j)$ at first order in $\f_j-\f$.
To this aim we have to linearize $\Si(m,\f)-\Si_j(\f_j)$ at first order in $m$ and
$\f_j-\f$; after a tedious computation (see Appendix~\ref{app:scaling}) we get
\beq\begin{split}
\Si&(m,\f) -\Si_j(\f_j) \sim \\ 
&\sim -\left[S'(\f_j) + d\frac{Y'(\f_j)}{Y(\f_j)} + \frac{d}{\f_j} \right]
(\f_j-\f) - d m \ ,
\end{split}\eeq
that gives, expressing $S'(\f)$ in terms of $Y(\f)$ by means of Eq.~(\ref{pressione})
\beq\begin{split}\label{mJ}
m(\f,\f_j) &= \frac{1}{d} \left[2^{d-1} Y(\f_j) - d\frac{Y'(\f_j)}{Y(\f_j)} + \frac{1-d}{\f_j} \right]
(\f_j-\f) \\ &\equiv \mu(\f_j) (\f_j-\f) \ .
\end{split}\eeq
The latter result must be substituted in $s^*(m,\f)$ to get the behavior close to $\f_j$. It is easy
to see from Eqs.~(\ref{siSi}) that for $m\to 0$, $s^*(m,\f) \sim d (1-m) \ln m$; therefore
\beq
s(\f,\f_j) \sim d \ln(\f_j-\f) - d \m(\f_j) (\f_j -\f) \ln (\f_j-\f) + \cdots
\eeq
close to $\f_j$, and the (reduced) pressure diverges as
\beq\label{pjamming}\begin{split}
p(\f,\f_j) &= -\f \frac{\partial s(\f,\f_j)}{\partial \f} \\
&\sim \frac{d \f_j}{\f_j-\f} - d \f_j \mu(\f_j) \ln (\f_j-\f) + \cdots \ .
\end{split}
\eeq
It is important to remark that the corrections to the leading order are quite large
(logarithmically divergent), as observed in
numerical simulations~\cite{SDST06}. 

\begin{figure}[t] \centering
\includegraphics[width=8cm]{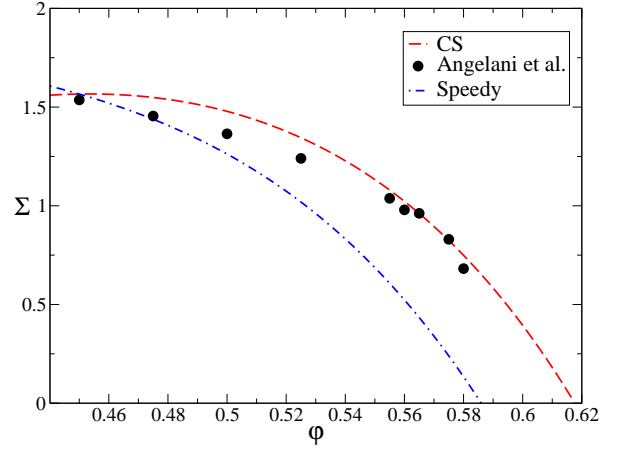}
\caption{Equilibrium complexity $\Si_{eq}(\f)$ as function of $\f$
in $d=3$ using the Carnahan-Starling equation of state in the small cage expansion (section~\ref{sec:smallcage}).
The prediction is compared with data from Angelani and Foffi \cite{AF07} and Speedy
\cite{Sp98}.
}
\label{fig:Sicomp}
\end{figure}

\subsubsection{Correlation function}

The most interesting results on the scaling close to $\f_j$ are obtained by investigating
the pair correlation function of the spheres in the glass state. 
At first order we obtained Eq.~(\ref{gfirst}); the first term describes the delta
peak, while the other terms do not contribute close to $r=D$.
It would be very interesting to evaluate them, but this seems a very complicated
task (see Appendix~\ref{app:deriva} for a complete list of these terms). Here we will
focus on the first term.
We have then to study the asymptotic behavior of $Q(r)$ in the jamming limit. This is done
in Appendix~\ref{app:Qasy}, and the result is
\beq\label{gglass}
\gG(r) = g(r) [1 + Q(r) ] =
g(r) \left[1 + \frac1m \D_m\left(\frac{r-D}{\sqrt{4mA}}\right)\right] \ ,
\eeq
where $g(r)$ is the (extrapolated) correlation function of the liquid at the
same density, and $m=m(\f,\f_{j})$, $A=A^*(m(\f,\f_{j}))$; the function $\D_m(t)$ is
defined in Eq.~(\ref{Dmt}) and plotted in figure~\ref{fig2}. 

\begin{figure}[t] \centering
\includegraphics[width=8cm]{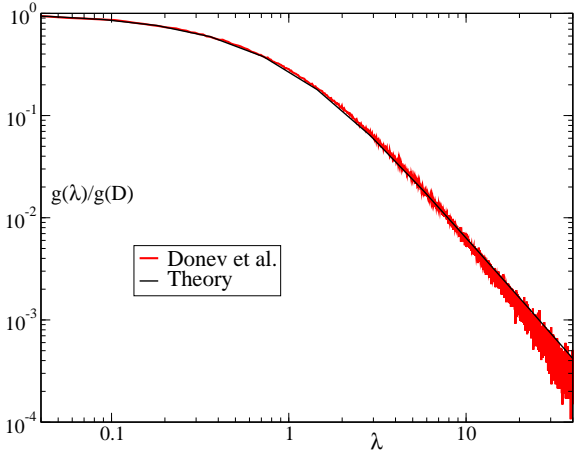}
\caption{Scaling limit of the delta peak. Numerical data, reproduced from~\cite{DTS05},
refer to a packing $\f_j \sim 0.64$ and $\f_j-\f \sim 10^{-12}$. $\gG(r) / \gG(D)$ is
plotted against $\l = \frac{r-D}D p$. The full line is
the theoretical prediction (\ref{deltapeak}). Compare with figure~\ref{fig:gofrHNC} for
a similar result obtained in the HNC approximation.
}
\label{fig:deltapeak}
\end{figure}

The function $\D_m(t)$ gives the leading contribution to $\gG(r)$ close to contact,
that comes from neighboring particles that are in contact with the reference
particle exactly at $\f_j$.
This contribution is nonvanishing only for $r-D \lesssim \sqrt{A} \sim \sqrt{\f_j-\f}$,
and becomes a delta peak at $\f_j$.
Other corrections are encoded in the remaining terms in (\ref{gfirst}) or are missed by
the first order small cage expansion.
A first consequence of this fact is that the integral of $\gG(r)-1$ does
not vanish at $\f=\f_j$ as it should, being proportional to the compressibility. A second
consequence of our approximations
is that Eq.~(\ref{pressione}) is not satisfied: this is because from 
Eq.~(\ref{gglass}), for $\f\to \f_j$
\beq
\gG(D) = Y(\f) [1+Q(D)] \sim \frac{Y(\f_j)}{m} = \frac{Y(\f_j)}{\m(\f_j) (\f_j-\f)} \ ,
\eeq
and therefore
\beq
1 + 2^{d-1} \f \gG(D) \sim \frac{2^{d-1} \f_j Y(\f_j)}{\m(\f_j) (\f_j-\f)} 
\neq \frac{d \f_j}{\f_j-\f} = p(\f,\f_j)
\ .
\eeq
This is not a surprise since the relation (\ref{pressione}) is violated
in most of the approximate theories for the liquid phase~\cite{Hansen}.
Note however that the difference
is not so big, because the first term in $\m(\f_j)$, given by Eq.~(\ref{mJ}), 
dominates as $Y(\f)$ is quite large ($\sim 20$) 
in the density range of interest.

Using again $m(\f,\f_j) = \m(\f_j)(\f_j-\f)$ and the expression of $A^*(m)$ we get
from Eq.~(\ref{gglass}) close to $\f_j$:
\beq\label{deltapeak}\begin{split}
\frac{\gG(r)}{\gG(D)} 
&\sim \D_0\left( \frac{(r-D) \sqrt{\pi} 2^{d} \f_j Y(\f_j)}{ D \m(\f_j) (\f_j-\f)} \right) \\
&= \D_0 \left( \frac{ \sqrt{\pi}}2 \frac{r-D}{D} (1+2^{d-1}\f \gG(D)) \right) \\
&= \D_0 \left(  \frac{ \sqrt{\pi}}2 \l \right) \ , \\
\end{split}\eeq
having defined the variable $\l = \frac{r-D}D (1+2^{d-1}\f \gG(D)) \sim \frac{r-D}D p$
(the last equality holds only if (\ref{pressione}) holds). 
Note that the same scaling of the delta peak was already predicted in the HNC approximation,
see figure~\ref{fig:gofrHNC}; thus it is a very robust feature that emerges from the replica
method independently of the approximations.

\begin{figure*}[t] \centering
\includegraphics[width=8cm]{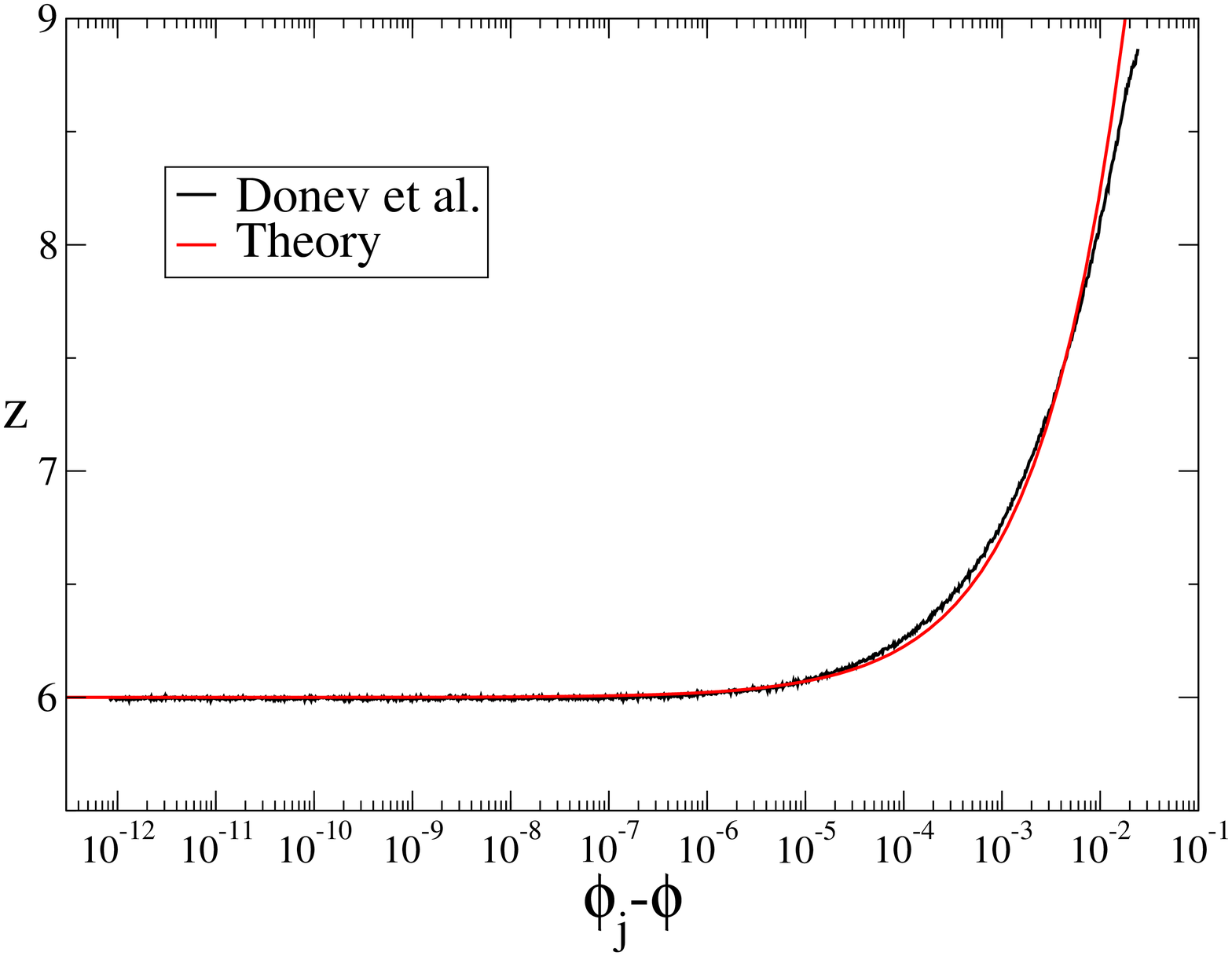}
\includegraphics[width=8cm]{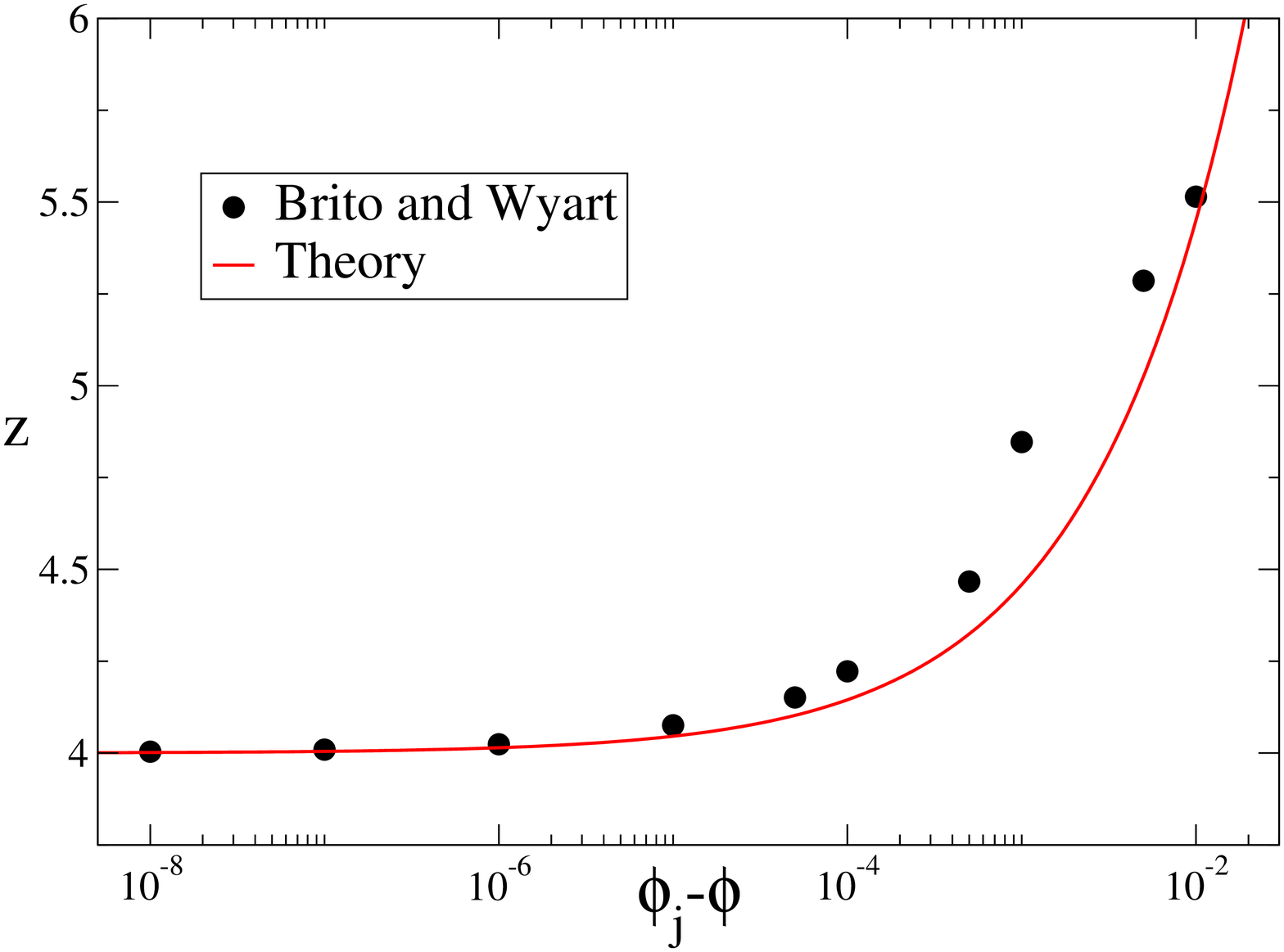}
\caption{
Number of contacts from Eq.~(\ref{zclosej}). (Left) Data from Ref.~\cite{DTS05} in $d=3$ and $\f_j \sim
0.64$. We have $z \sim 6 + 20 \sqrt{\f_j-\f}$ for both curves.
(Right) Data from Ref.~\cite{BW06} in $d=2$ and $\f_j \sim 0.83$.
The best fit gives $z=4 + 31 \sqrt{\f_j-\f}$ while the theory gives
$z=4 + 14.5 \sqrt{\f_j-\f}$.
}
\label{fig3}
\end{figure*}

This scaling form of the delta peak of $\gG(r)$ close to jamming has been observed 
numerically in \cite{DTS05}.
In Fig.~\ref{fig:deltapeak} we report numerical data from \cite{DTS05} on the delta-peak contribution
for a packing in $d=3$ with $\f_j \sim 0.64$ and $\f_j-\f \sim 10^{-12}$.
We see that our result, given by Eq.~(\ref{deltapeak}) for $\f_j=0.64$, is in excellent agreement
with the numerical data.

In summary, from Eq.~(\ref{gglass}) and Appendix~\ref{app:Qasy},
we get the following scaling for the delta-peak of $\gG(r)$:
\begin{widetext}
\begin{equation}\label{scalag}
\gG(r) \sim \begin{cases}
\frac1{\f_j-\f} \hskip120pt r-D \sim \f_j-\f \ , \hskip10pt \l \sim 1 \ , \\
\frac1{(\f_j-\f) \l^2} \sim \frac{\f_j-\f}{(r-D)^2} \hskip62pt
\f_j-\f < r-D \sim \sqrt{\f_j-\f} \ ,  \hskip10pt \l \sim 1/\sqrt{\f_j-\f} \ , \\
\exp[-(r-D)^2/(\f_j-\f)] \hskip35pt
r-D \gg \sqrt{\f_j-\f} \ , \hskip10pt \l \gg 1/\sqrt{\f_j-\f} \ ,
\end{cases}
\end{equation}
\end{widetext}
where the last line comes from the cutoff on the function $\D_m(t)$ at finite $m$,
see Appendix~\ref{app:Qasy}.

In numerical simulations a $(r-D)^{-\a}$ divergence of $\gG(r)$ close
to $D$ is observed, with $\a \sim 0.5$. 
This leads to a difference with respect to (\ref{scalag}), with the $(r-D)^{-2}$
regime extending only up to $r-D \sim (\f_j-\f)^{2/3}$. This feature is not
present in our calculation; this discrepancy deserves further investigation.

\subsubsection{Number of contacts}

The interpretation of Eq.~(\ref{scalag}) is that for $\f \to \f_j$ there is a
shell of width $\f_j -\f$ around a given particle where the probability of finding
other particles is very high. 
These particles become {\it neighbors} of the particle in the origin
at $\f=\f_j$, and their number is finite as 
the integral of $\gG(r)$ on this shell is finite.

There is however a second shell 
$\f_j -\f \leq r-D \leq \sqrt{\f_j -\f}$ where the probability of finding
particles is very high. The integral of $\gG(r)$ over this shell is of
order $1$ so also in this shell there is a finite number of
particles that will become neighbors of the particle in the origin.

The integral of $\gG(r)$ on a shell $D \leq r \leq D + O(\sqrt{\f_j-\f})$ gives
then the number of contacts for $\f \to \f_j$:
\beq\label{zdef}
z = \Omega_d \rho D^{d-1} g(D) \int_D^{D+O(\sqrt{A})} dr \, [1+ Q(r)] \ .
\eeq
Recall that in the expression above, as everywhere in the paper, 
$g(D) = Y(\f)$ is the contact value
of the liquid correlation, that is finite; this should not be confused
with $\gG(D)$ which diverges at jamming.
For $\f_j - \f$ finite but small this number can be interpreted as the number
of particles that collide with the particle in the origin during a finite but
long time $\t$ \cite{BW06}.

Remarkably, the integral of the second term in Eq.~(\ref{zdef}) can be
computed exactly using Eq.~(\ref{Dintegrale});
since the function $\D_m(t)$ contains an exponential cutoff 
at $r-D \sim \sqrt{A}$ (see figure~\ref{fig2}), the upper integration limit
can be extended up to $\io$ and we obtain
\beq
\Omega_d \rho D^{d-1} g(D) \int_D^{\io} dr \, Q(r) = 2d(1-m)
\eeq
\ie $z=2d$ for $\f=\f_j$.
Close to $\f_j$ there is a correction $\d z \propto m \sim \f_j -\f$ coming
from the equation above, plus a second correction $\d z \propto
\sqrt{\f_j-\f}$ coming from the integral of 1 in Eq.~(\ref{zdef}). The
second correction dominates, then we have
\beq
z = 2d + O(\sqrt{\f_j-\f}) \ ,
\eeq
as found in~\cite{BW06}. We can try to give a quantitative
estimate of the coefficient by observing that the function $\D_m(t)$ starts 
to drop exponentially at $t=\sqrt{m}$ and drops to 0 for $t=\t \sqrt{m}$ with
$\t$ very close to $2$.
Clearly the upper limit of integration $\t$ plays the role of a fitting
parameter, however the value $\t=2$ is very reasonable from Fig.~\ref{fig2},
given also the uncertainty involved in the numerical estimate.
Thus we assume that the integral is done up to $r = D+2\sqrt{A}$ and we have
\beq\label{zclosej}
\begin{split}
z &= 2d + \Omega_d \rho D^{d-1} g(D) 2\sqrt{A} \\ &=
2 d + 2 \Omega_d \rho D^{d-1} g(D)  \frac{1-m}{Q_m} \frac{D}{\rho V_d(D) g(D)} \\ 
&= 2 d + 2d \frac{1-m}{Q_m} \sim 2 d + 2d \sqrt{\frac{4 m}{\pi}} \\ &=
2 d + 2 d \sqrt{\frac{4 \mu(\f_j)}{\pi}} \sqrt{\f_j-\f} \ .
\end{split}
\eeq
The result is compared with numerical data in Fig.~\ref{fig3}. The value of
$\mu(\f_j)$ is given in Eq.~(\ref{mJ}). Despite the good agreement, note that
the square root growth of $\d z$ was attributed in \cite{BW06} to the square
root singularity of $\gG(r)$, which is a different mechanism from the one
we discussed above.

\subsubsection{Force distribution}

In \cite{DTS05} the force distribution of dense amorphous packings generated
by using the Lubachevsky-Stillinger algorithm in $d=3$ was investigated.
The interparticle forces are defined in this way: one takes a packing of density
$\f_j$, slightly reduces the density and measures the average momentum
exchanged by two neighboring particles over a large time $t$. 
Note that on very large times and if the
volume is large enough the packing will be unstable and relax toward a more compact structure
(either at $\fICP$ or at $\f_{FCC}$), see Fig.~5 in \cite{DTS05}. However before this happens
there is enough time to measure the forces with sufficient accuracy.

\begin{figure}[t]
\centering
\includegraphics[width=8.5cm]{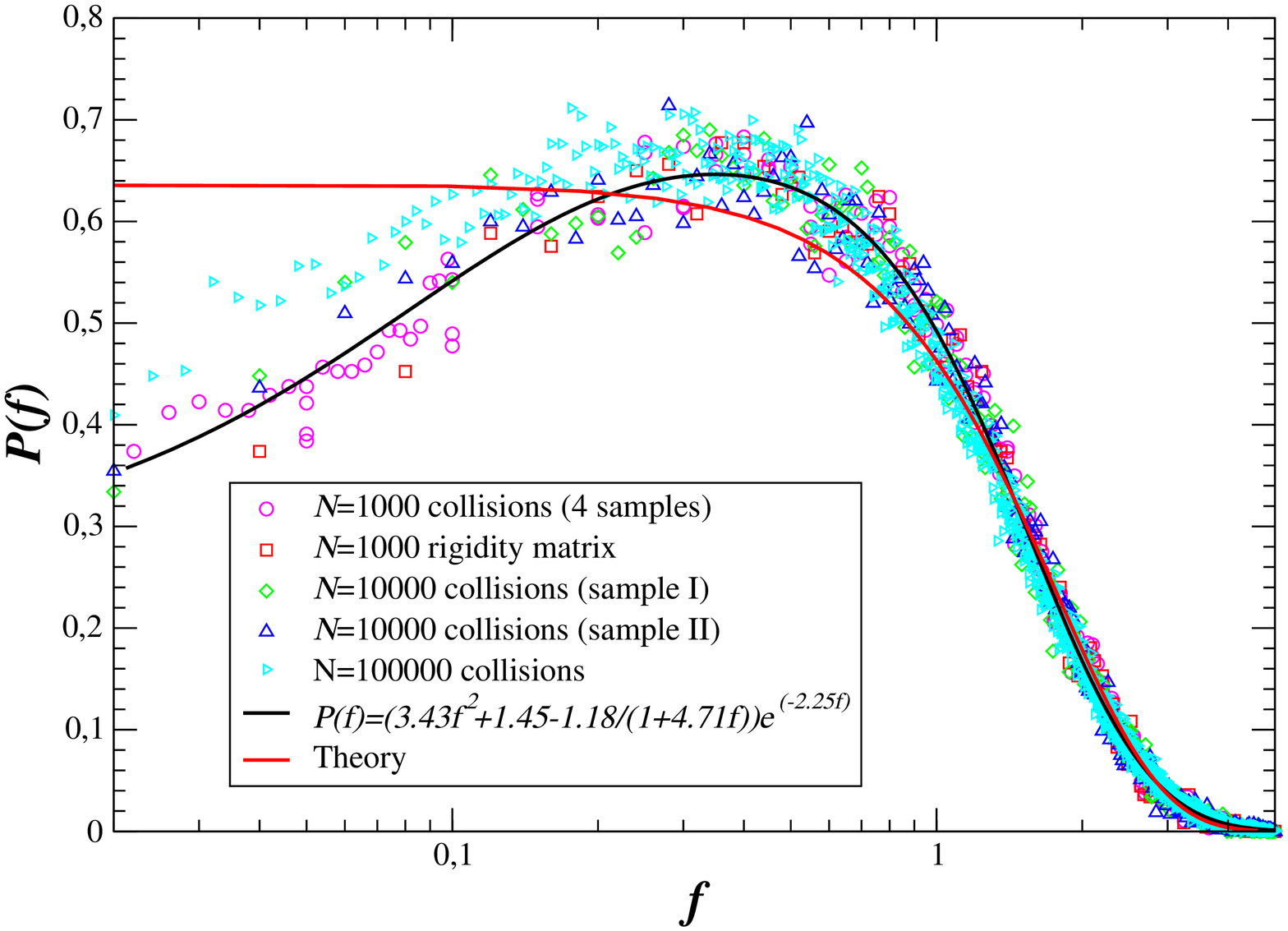}
\caption{Probability distribution of the forces, from~\cite{DTS05}.
The full red line correspond to the Gaussian form~(\ref{Pfgauss}).
The black line is a fit performed by the authors of~\cite{DTS05}.
}
\label{fig:forze}
\end{figure}

The interparticle forces are then normalized such that $\langle f \rangle = 1$ and the distribution
$P(f)$ is investigated. The theory of \cite{DTS05} relates this quantity to the shape of the
delta peak close to contact, Fig.~\ref{fig:deltapeak}, by the relation
\beq
\frac{\gG(\l)}{\gG(D)} = \int_0^\io df \, f \, P(f) \, e^{-\l f} \ ,
\eeq
where $\l = (r/D-1)p$ as in the previous sections. The previous equation expresses the fact
that for small interparticle separation the gap $h \propto 1/f$, as discussed also in \cite{BW06}. 
Eqs.~(\ref{deltapeak}) and (\ref{Dlimite}) give:
\beq\label{gscaling}\begin{split}
\frac{\gG(\l)}{\gG(D)} &=  \D_0\left(\frac{\sqrt{\pi}}2 \l \right) \\ &=  
2 \int_0^\io dy \, y \, e^{-y^2 -
  \sqrt{\p} \l y} \\&=
 \frac{2}{\p} \int_0^\io df \, f \, e^{-\frac{f^2}{\p} - \l f} \ ,
\end{split}\eeq
and finally
\beq\label{Pfgauss}
P(f) = \frac2\pi e^{-\frac{f^2}{\p}} \ .
\eeq
In Fig.~\ref{fig:forze} we can see that this form reproduces well the
numerical data for large forces.
The discrepancy at small forces can be explained by {\it i)} the fact that the
small $f$ behavior of $P(f)$ is related to the large $\l$ behavior of $\gG(\l)$,
and as discussed above in the large $\l$ region there are corrections to
$\gG(\l)$ that we are missing, and {\it ii)} a possible finite size effect
(indeed it seems that the data for larger samples are in better agreement with
the theory).

We wish to stress again that here we focused only on the results of~\cite{DTS05}
that refer to packings produced using slow compressions; these should be related
to the infinite pressure glassy states that are the object of this paper. 
Interestingly, similar results for $P(f)$
have been obtained in \cite{SVHS04,SVEHL04} by using an ensemble approach for the force network
based on Gaussian random matrices.
In~\cite{OLLN02} a Gaussian tail in $P(f)$ was found for packings close to the J-point;
moreover, it has been shown that finite size fluctuations of the jamming density can introduce
self-averaging problems when averaging over many configurations that in turn produce exponential
tails in $P(f)$. 
Therefore some care should be taken when comparing
our prediction of a Gaussian shape of $P(f)$ with numerical data.
Moreover, the force distribution has been studied in great detail for packings of 
frictional
spheres produced using different protocols and seems to depend strongly
on the protocol used. As discussed in the introduction, covering the relevant literature
here is impossible. In particular friction strongly affects the force network
and is expected to have a dramatic effect on $P(f)$.

\section{Beyond first order in the small cage expansion}
\label{sec:beyond}

In order to improve over the first order expansion, one could try to perform
a second order (and third, etc.) expansion systematically. However, the convergence
of this procedure is not clear, and moreover, the calculation of higher order corrections
becomes increasingly difficult and is already very difficult at second order.
Following the tradiction of theoretical physics, we then try to use resummation techniques
for our expansion. Unfortunately, the results presented here are not satisfactory, but
still we believe that they are interesting since they illustrate a possible direction
one can follow to improve the theory. We believe that a more accurate treatment of the
resummation we will discuss could indeed lead to much better results.
The reader interested only in ``stable'' results is then encouraged to skip this section.

The simple resummation we will discuss in this section has the
advantage of being fully analytically solvable,
but unfortunately
gives inconsistent results for the thermodynamics of the ideal glass. 
The main idea is to fully exploit the resummation discussed in 
section~\ref{sec:effectivepotentials}, that has been particularly useful to solve
the model in the limit $d\to\io$.
Note that the small cage expansion of section~\ref{sec:smallcage} does not give access to
the pair correlation function of glass, which is one of the most interesting
quantities to compute, see the discussion in section~\ref{sec:gofr}.
The advantage of the resummation we discuss here is that it allows, 
by finding a good approximation to compute the properties of the effective liquid,
a full computation of the correlation function of the glass.

\subsection{Percus-Yevick approximation for the Baxter model}

To resum a class of terms in the small cage expansion we start from
Eq.~(\ref{SS1}), \ie we neglect all the $n\geq 3$-body potentials but try to treat
the two-body one as exactly as possible. Then we have to compute the entropy of a liquid 
with an hard core repulsion and a small tail
given by $\phi_{eff}(r) = -\ln[1+Q(r)]$. The tail is {\it attractive} for $m<1$ and repulsive
for $m>1$, see Fig.~\ref{fig:feff}. 

The advantage of this formulation is that the correlation function of the glass is simply
\beq
 \gG(r) = - \frac{2}{\r} \left. \frac{\d S[\r,\chi(r)e^{-v(r)} (1+Q(r))]}{d v(r)} \right|_{v=0} =
g_{eff}(r) \ ,
\eeq
where $g_{eff}(r)$ is the pair correlation of the liquid with the effective potential $\phi_{eff}$.
Therefore $\gG(r)$, the correlation of the glass, turns
out to be equal to the correlation function of the effective liquid and then automatically has
many of the reasonable properties of a correlation function for hard particles
(\eg it is positive, it vanishes inside the hard core, the structure factor is positive, etc.).
We can use different approximation schemes (PY, HNC, etc.) to compute the free energy and
correlation function of the effective liquid.

To further simplify the problem, we observe that
in the limit $m\to 0$ the cage radius vanishes
and the strength of the attractive part of the potential diverges. Then we can approximate
$Q(r)$ with a delta function as in (\ref{Qdelta}) and the effective liquid becomes a Baxter
model. 
Note that the approximation (\ref{Qdelta}) strictly holds only for $m \to 0$, where $A \to 0$. 
However it might be reasonable for any $m < 1$, as in this case the interaction is attractive
and its range is anyway much smaller than the hard core diameter.
We run into problems, however, for $m > 1$, as in this case the interaction is repulsive and
cannot be described by (\ref{Qdelta}). Therefore in the following discussion
we always assume that $m < 1$.\footnote{Anyway, 
for $m>1$ the strength of the interaction is always finite (and much smaller than
for $m<1$).
In this case it seems very reasonable to use the first 
order approximation derived in the previous sections.}

In the notation of Baxter \cite{Ba68} we have:
\beq\label{Qdelta3}
e^{-\phi_{eff}(r)} = \c(r) [1+Q(r)] \to \chi(r)\left[ 1+ \frac{4 \pi D}{12 \t} \d(r-D)\right] \ .
\eeq
Comparing this with (\ref{GmAdef}), we see that
in $d=3$, the effective liquid is finally reduced to a Baxter model at the same density
$\f$ and of interaction strength $\t = 1/(4 G_m(A))$.

\begin{figure*}[t]
\centering
\includegraphics[width=8cm]{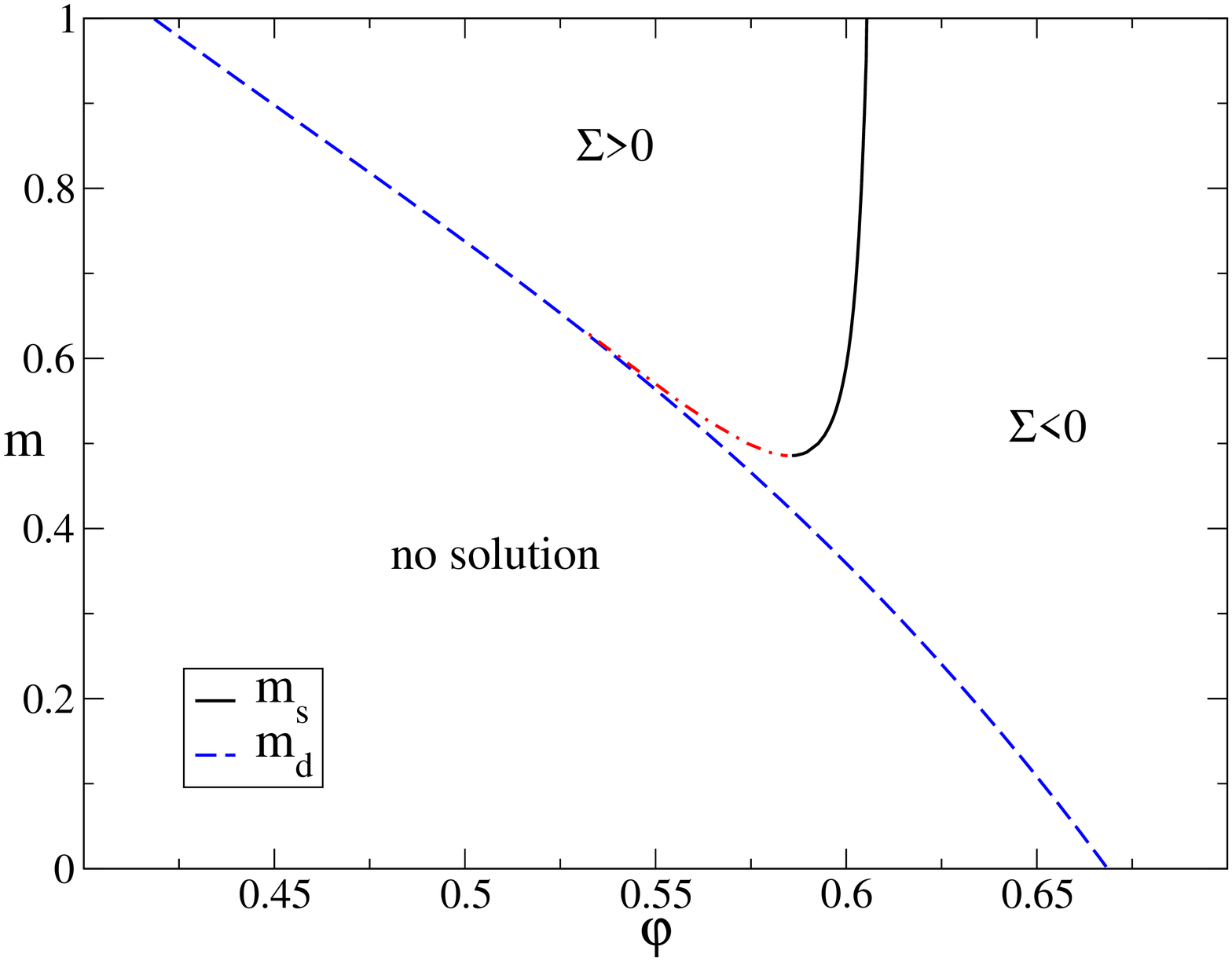}
\includegraphics[width=8cm]{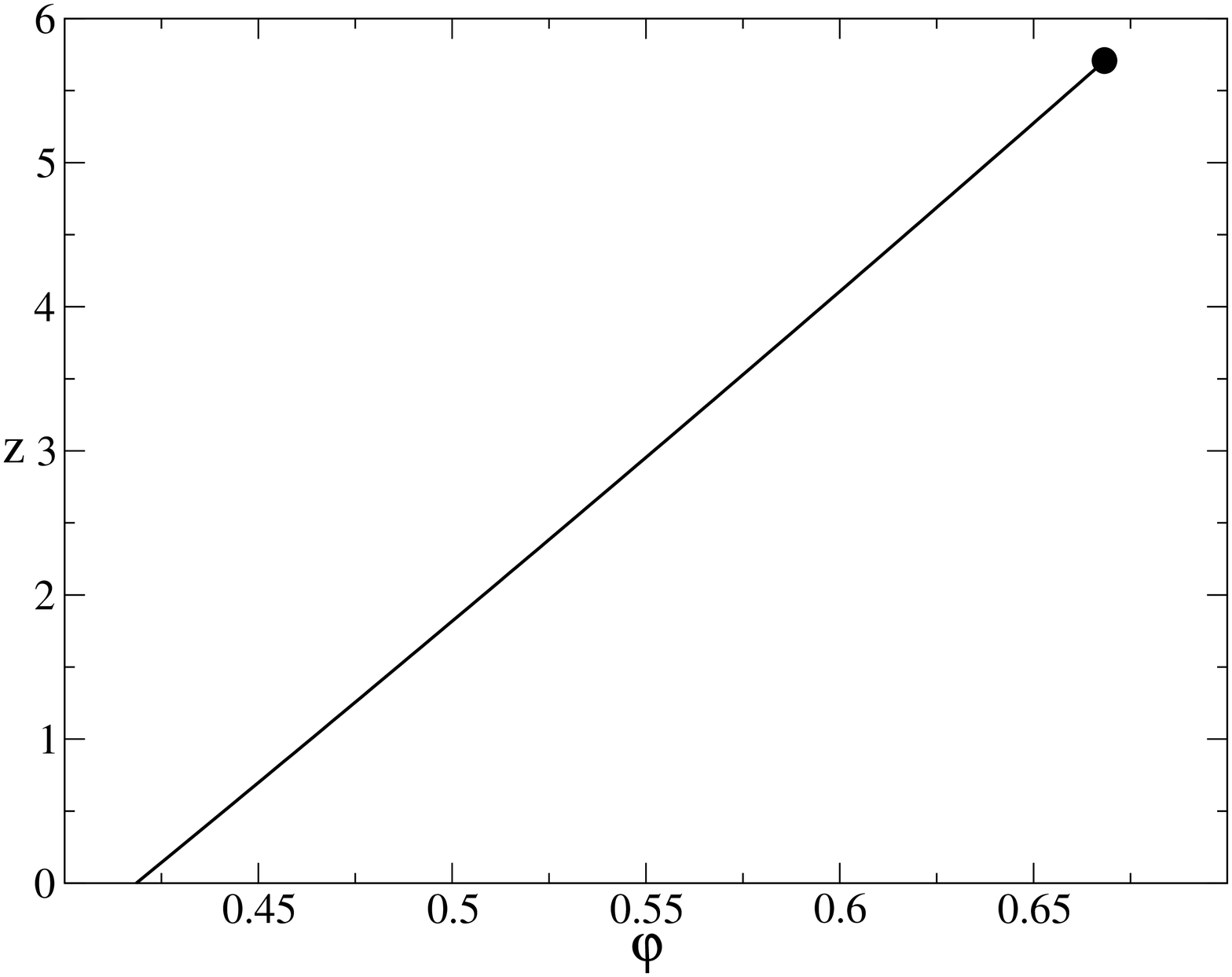}
\caption{(Left) Phase diagram of the Baxter resummation (section~\ref{sec:beyond}). The dashed (blue)
  line is the boundary of the region where a solution for $A$ is found. The
  full (black) and dot-dashed (red) lines are the boundary of the region where
  the complexity is positive. (Right) Number of contacts $z$ as a function of
  density along the line $m_d(\f)$. The black dot corresponds to $m_d(\f)=0$.
  $z$ increases almost linearly from $0$
  (liquid state) to $z \sim 6$, the value expected for an amorphous jammed
  phase. Note however that the upper part of the line correspond to negative complexity.
}
\label{fig:Baxter}
\end{figure*}

\subsubsection{Free energy and the equation for $A$}

Remarkably, the Percus-Yevick approximation for the Baxter model has been solved 
exactly in $d=3$~\cite{Ba68}: an analytical expression for the free energy has been given in
Eqs.~(2.5) and (2.7) of Ref.~\cite{TB93}, 
while the correlation function can be computed using the method of~\cite{We63}.

The function $\t(m,A)$ can be computed in $d=3$ by a numerical integration of 
Eq.~(\ref{GmAdef}), (\ref{F03d}).
Then we get an analytical (although complicated) expression for $\SS(m,\f;A)$:
\beq\label{SSBaxter}
\SS(m,\f;A) = S_{harm}(m,A) + 1 - \ln \frac{6 \f}{\pi} -
\psi\left[\f,\frac{1}{4 G_m(A)} \right] \ ,
\eeq
where $\psi[\f,\t]$ is given in Eq.~(2.7) of \cite{TB93}.
The equation for $A$ reads, using the definition of Appendix~\ref{app:def}:
\beq
1 = -\frac{8}{3} \frac{d\psi}{d \t^{-1}}\left[\f,\frac{1}{4 G_m(A)} \right] F_m(A) \ ,
\eeq
and it has to be solved numerically for generic $m$, $\f$.

However some simplifications are possible. First of all, for $m \to 1$
we have $G_m(A) \to 0$ and $\t \to \io$. It is possible to show that
\beq\begin{split}
&S_{PY}(\f) = 1 - \ln \frac{6 \f}{\pi} - \lim_{\t\to\io}\psi[\f,\t] \ , \\ 
&\lim_{\t\to\io} \frac{\partial\psi[\f,\t]}{\partial \t^{-1}} = - \f Y_{PY}(\f) \ .
\end{split}
\eeq
Therefore the equation for $A$ becomes
\beq
1 = \frac{8}{3} \f Y_{PY}(\f) F_1(A) \ .
\eeq

In the jamming limit $m\to 0$, $A= \a m$, $G_m(A) \to G_0(\a)$ that can be computed easily, see 
Eq.~(\ref{taujamming}). Using Eq.~(\ref{dtaujamming}) the equation for $\a$ is
\beq
1 = - \frac83 \frac{d\psi}{d \t^{-1}}\left[\f,\frac{1}{4 G_0(\a)}\right] \, F_0(\a) \ ,
\eeq

The expression for the complexity can be simplified similarly.

\subsubsection{Results}

Unfortunately, the results of the Baxter resummation are very poor in $d=3$.
The main problem is that the static value of $m$, $m_s(\f)$ is not a
decreasing function of $\f$. This is inconsistent since it would imply that the
ideal glass state at has smaller density than the liquid state at
$\f=\f_K$. On the contrary, the behavior of the line $m_d(\f)$ is
reasonable even if the value of the transition at $m=1$, $\f_d =0.418$, 
is exceedingly smaller than the conjectured one, $\f_d \sim 0.58$.
For these reasons the phase diagram in this approximation seems
unreliable, see Fig.~\ref{fig:Baxter}.

More interesting is the result for the pair correlation function. Indeed, the
pair correlation function of the Baxter liquid shows many characteristic features
which are observed in jammed packings of hard spheres, as it has been
discussed in \cite{MF04}. The most interesting among them are the peak at
$r/D = \sqrt{3}$ and the jump in $r/D = 2$ which have been observed in 
\cite{DTS05} for random jammed packings of monodisperse spheres.
Unfortunately, the values of $\t$ which comes from the analysis above seems to
be too high for these features to be present in the $\gG(r)$ with the correct
order of magnitude so that a quantitative comparison with numerical
simulations seems to be impossible at this stage.

Note also that the $g(r)$ of the Baxter liquid is characterized by a delta
peak at contact (due to the adhesive potential). From the amplitude of the
latter peak, which is analytically known in the Percus-Yevick solution
\cite{Ba68}, we can compute a number of neighbors as a function of density.
The result is similar to what expected: the number of neighbors
increases from $0$ to $z \sim 6 = 2 d$, \ie an isostatic packing, see Fig.~\ref{fig:Baxter}.
However the upper part of the curve corresponds to negative complexity so that
the corresponding packings do not exist.

\subsection{Discussion}

This particular very simple resummation scheme seems not able to describe 
correctly the glassy states of hard spheres. 
However, the idea of resumming all the terms
corresponding to the two-body interaction seems promising, as the resulting
$\gG(r)$ shows some features that are observed in real packings and the behavior
of the delta peak (number of contacts) is correct.

It is possible that three (and more) body interactions play a major role,
however it would be worth to try to keep only the simplest two-body
interactions and use more refined equations for the effective liquid.
For instance, the delta-function approximation for $Q(r)$ might be too rough
as $\sqrt{A} \sim 0.01$ and on this scale the $g(r)$ of hard spheres has
strong variations at least close to contact. One could then try to solve
numerically the Percus-Yevick approximation for the exact potential
$\phi_{eff}$.
Alternatively, a more precise equation of state (such as the one proposed in
\cite{MF04b}) could be used for the Baxter model.

One might be worried because of the fact that in the limit $m\to 0$ we do
not recover the small cage expansion results (\eg the complexity
was positive for $\f < 0.68$ at first order in $\sqrt{A}$ while here we always
found a negative $\Si$). The reason is that for
$m\to 0$, the small cage parameter is $\a=A/m$ and not $A$. Therefore, even
if $A\to 0$ in this limit, the small cage expansion is not recovered because
$\a$ remains relatively big.

\section{Binary mixtures}
\label{sec:binary}

In this section we report the recent extension of the theory to binary mixtures
that was obtained in~\cite{BCPZ09}. The extension to binary system is important
since in these systems crystallization can be easily avoided; this allowed to
study their glass transition in great 
detail, see \eg~\cite{GV03,FGSTV03,FGSTV04,DTS06,SK00}
and in particular the recent works~\cite{BW08,BW09,HD09}.
Jamming of binary mixtures has also been extensively investigated,
see \eg~\cite{Do75,Do80,OT81,OLLN02}.
Their investigation allows to test the prediction of the theory concerning the variation
with composition of packing density, structure, etc. 
The details of the computation of the function $\SS(m,\f)$ for a general
multicomponent mixture, following~\cite{CMPV99}, can be found in~\cite{BCPZ09} 
in the framework of the first-order small cage approximation.
Here we only report the results for a binary mixture of two types of three-dimensional
spheres $\mu=A,B$ in a volume $V$, 
with different diameter $D_\mu$ and density $\r_\mu = N_\mu/V$. 
We define $r = D_A / D_B > 1$ the diameter ratio and $x= N_A/N_B$ the concentration ratio;
$\f = \r_A V_3(D_A)+\r_B V_3(D_B)$ the packing fraction;
$\eta = \r_B V_3(D_B) / \f = 1/(1+ x r^3)$ the volume fraction of the small ($B$) component.

As in the monodisperse case, once an equation of state for the liquid has been chosen,
one can obtain for a given $\Si_j$ the equation of state of the corresponding metastable
glass, as well as its jamming density $\f_j$.
The equation of state used in \cite{BCPZ09} is a generalization of
the Carnahan-Starling equation. The theoretical results were compared with numerical
results obtained from the Lubachevsky-Stillinger algorithm discussed in 
section~\ref{sec:algorithms}.
\begin{figure}
\includegraphics[width=8cm]{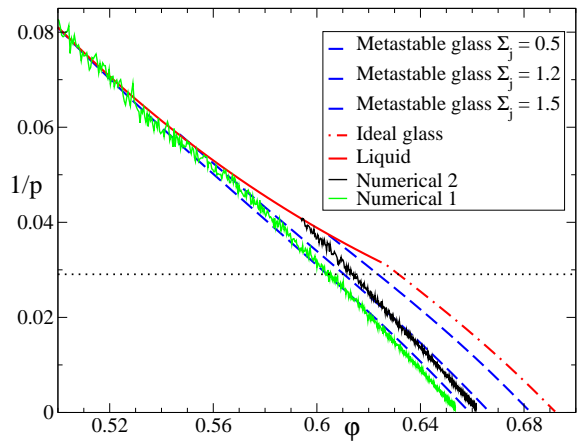}
\caption{
Inverse reduced pressure, $\rho/(\b P)$ as a function of the packing fraction $\f$ for
a mixture with $r=1.4$ and $x=1$.
Numerical data are obtained using two different protocols. In the first, compression
is started at low density. In the second, compression is started from an equilibrated configuration 
at $\f=0.58$.
The equation of state of different metastable glasses, corresponding to different $\Si_j$, are
reported as dashed lines. The dot-dashed line is the pressure of the ideal glass, corresponding
to $\Si_j = 0$. A numerical estimate~\cite{BW09} of the Kauzmann pressure, $p_K = 34.4$, is reported
as a dotted horizontal line.
}
\label{fig:pressure}
\end{figure}
In figure~\ref{fig:pressure} the evolution of the inverse
reduced pressure
during compression
is reported for a mixture
with $x=1$ and $r=1.4$, that has been recently
studied in great detail~\cite{BW08,BW09,OLLN02}.
Overall,
one observes the same behavior already discussed in the monodisperse case.
The curve ({\it Numerical 1}) corresponds to a relatively
fast compression ($\g = 10^{-2}$).
The curve ({\it Numerical 2}) 
has been obtained starting the compression (at the same compression rate) 
from a carefully equilibrated
liquid configuration of the same mixture at $\f = 0.58$
(see \cite{BW09} for details on how this configuration was produced and equilibration was checked). 
In the latter case, since the relaxation time of the liquid at that density 
is already very long compared to the compression rate,
the system falls immediately out of equilibrium and
the pressure increases fast until jamming occurs at a higher density compared to the previous case.
The numerical equation of state
is compared with that of a glassy states corresponding to $\Si_j = 0.5, 1.2, 1.5$.
Starting the compression from a high-density liquid equilibrium configuration produces
a glassy state with lower $\Si_j$. This is a nice confirmation
of a prediction of the theory, that different glassy states jam at different density.
Finally, we report, for the same system, 
the numerically extrapolated value of the reduced pressure $p=\b P/\r$ at the 
ideal glass transition,
$p_K = 34.4$, obtained in \cite{BW09}.
Again, this corresponds well (within 10$\%$)
to the computed value $p_K = 31.8$ from the
theory, see figure~\ref{fig:pressure}. 

\begin{figure}
\includegraphics[width=8cm]{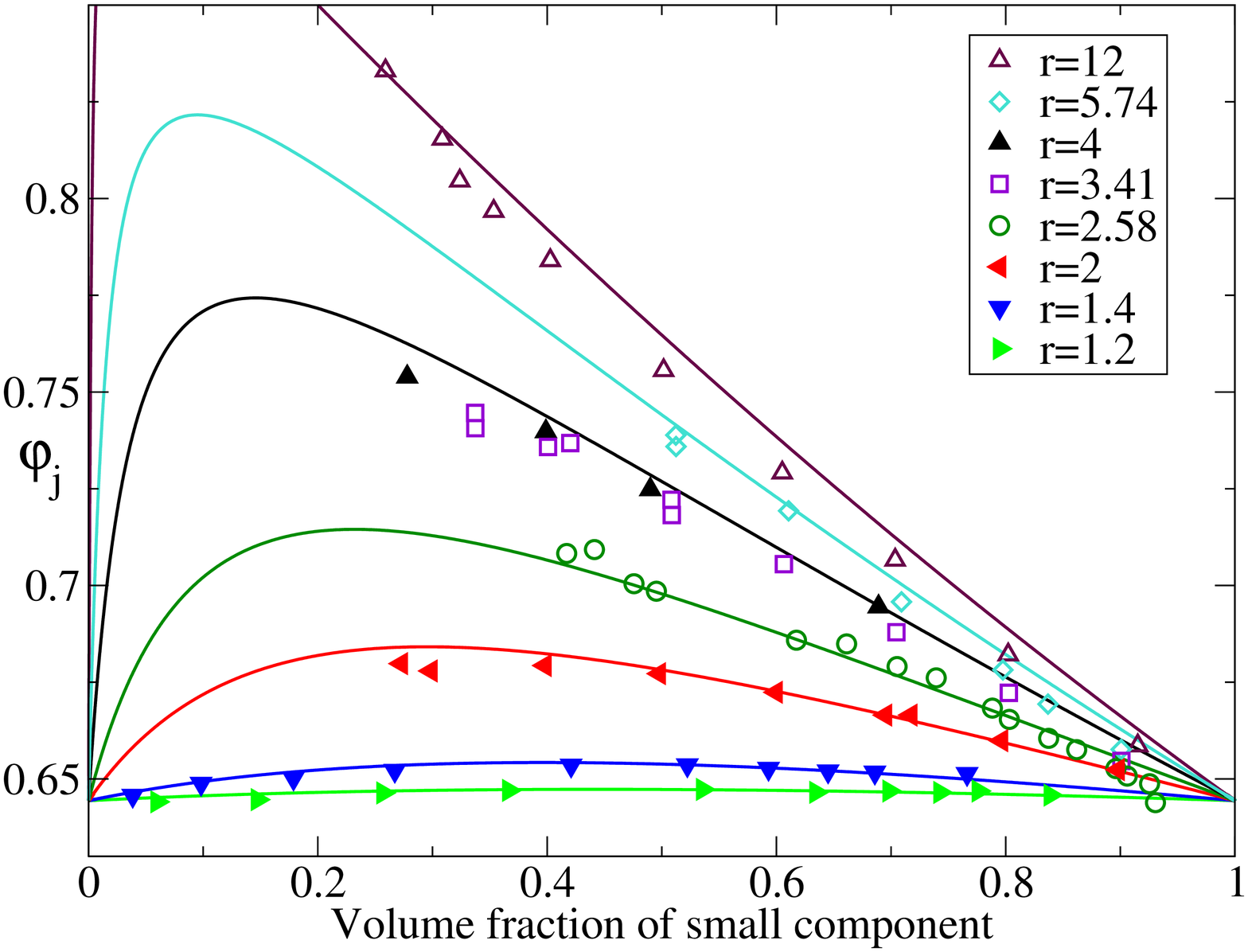}
\caption{
Packing fraction $\f_j$ as a function of 
$\eta = 1/(1+x r^3)$ at fixed $r$.
Full symbols are numerical data from~\cite{BCPZ09}. Open symbols are experimental results from~\cite{YCW65}. 
Lines are predictions from theory,
obtained fixing $\Si_j=1.7$. Note that the large $r$-small $\h$ region cannot be explored, since for
such very asymmetric mixtures the large spheres form a rigid structure while
small spheres are able to move through the pores and are not jammed~\cite{Do80,OT81}.
}
\label{fig:phij}
\end{figure}

\begin{figure*}
\includegraphics[width=7cm]{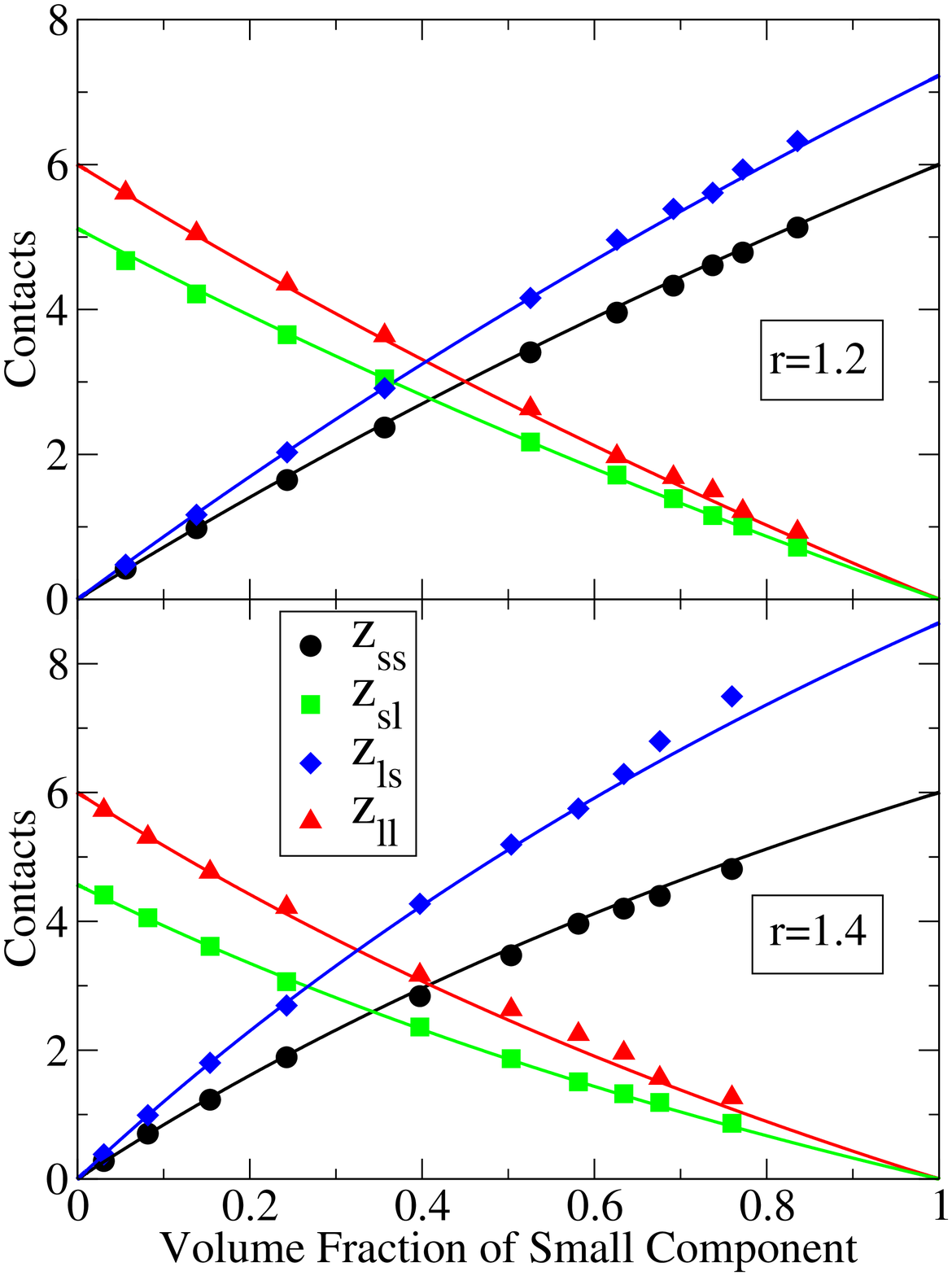}
\includegraphics[width=7cm]{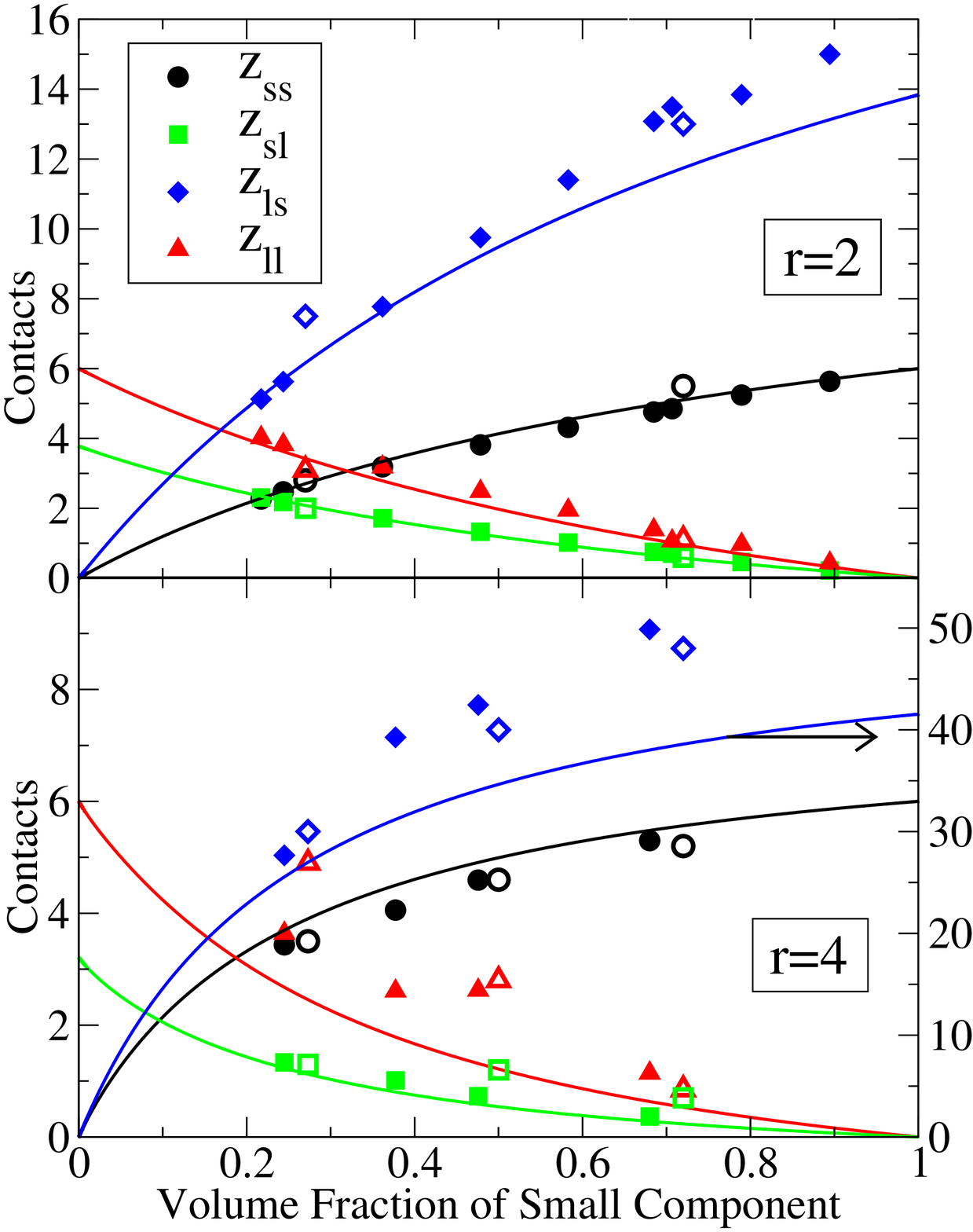}
\caption{
Partial average coordination numbers 
(small-small, small-large, large-small, large-large)
as a function of
volume fraction of the small particles $\h= 1/(1+x r^3)$ for different values of $r$.
Full symbols are numerical data from~\cite{BCPZ09}. Open symbols are experimental
data from \cite{PZYZM98}.
Note that in the lower right panel a different scale is used for $z_{ls}$.
}
\label{fig:contacts}
\end{figure*}

In figure~\ref{fig:phij}, the jamming density $\f_j$ is reported for different mixtures, putting
together numerical results~\cite{BCPZ09}, experimental data~\cite{YCW65}, 
and the theoretical results. The latter have been obtained by fixing $\Si_j = 1.7$
for all mixtures, which gives the best agreement.
Note that a single ``fitting'' parameter $\Si_j$, that is strongly constrained, allows
to describe different sets of independent numerical and experimental data.
The prediction of the theory are qualitatively similar to previous ones~\cite{Do80,OT81},
but the quantitative agreement is much better. Interestingly, a similar qualitative behavior
for $\f_{MCT}$ has been predicted by Mode-Coupling theory~\cite{GV03,FGSTV03}; 
although there is no {\it a priori}
reason why the variation with mixture composition of $\f_{MCT}$ and $\f_j$
should be related, it is reasonable to
expect that they show similar trends~\cite{FGSTV03}.

The average coordination numbers at $\f_j$ are denoted $z_{\m\n}(\f_j)$, 
and it was checked in~\cite{BCPZ09} that their variations with $\f_j$ are negligible.
They are reported in figure~\ref{fig:contacts} for different mixtures. 
The numerical values have been obtained in~\cite{BCPZ09}
by removing the rattlers from the packing.
Experimental data from~\cite{PZYZM98} are also reported
in the right panel of figure~\ref{fig:contacts}.
As discussed above, the total coordination
is close to the isostatic value $z=6$, which is the value predicted by the theory
also for binary mixtures~\cite{BCPZ09}.
As it can be seen from figure~\ref{fig:contacts},
the computed values agree very well with the outcome of the numerical simulation, at
least for $r$ not too large, while some discrepancies are observed in the contacts of the large
particles for large $r$.

\section{Conclusions}

This paper is based on the main assumption that {\it amorphous jammed packings of hard spheres can
be identified with the infinite pressure limit of glassy states.}
In addition we assumed that the mean field scenario for the glass transition holds also in finite 
dimension, at least on the time and length scales that are currently investigated in numerical
simulations (and sometimes also in experiment on colloids and granular systems when the number of particles
is not so large)~\cite{BB04}.

The mean field picture leads to a non-trivial structure of the phase diagram, 
whose main consequence is {\it the existence of amorphous jammed packings
in a range of densities $[\f_{th},\fICP]$}. The random close packing density can be any density
within this interval and its precise value depends on the details of the protocol used to construct
the packings~\cite{KK07}. 

Based on these assumptions we used the replica method to compute the properties of these
glassy states. Before concluding we will briefly summarize our results and discuss some possible
future developments, including how the picture is modified in finite dimensional systems.

\subsection{Summary of our results}

We used different approximation schemes for the replicated liquid. The HNC approximation
seems to work well in the moderately dense phase close to $\f_d$ and $\f_K$ at $m=1$.
On the contrary, the small cage approximation works better in the regime where cages are
small, namely for $m \sim 0$ close to the ideal glass line and in large dimension.
In dimension $d=3$ the mapping of the replicated liquid onto the Baxter adhesive hard sphere
model seems a promising way to obtain a satisfactory description in all the phase diagram
but for the moment gives poor quantitative results.

We list here the main results we discussed in this paper; many of them have been compared
with numerical results in the figures.
\begin{itemize}
\item We presented a consistent description of the glass transition for hard
spheres in $d\to\io$: in particular we give predictions for the clustering 
-Eqs.~(\ref{cluster_m1}) and (\ref{cluster_m0})- and
Kauzmann -Eq.~(\ref{Kauz_dinf})- densities.
We are able to compute the correlation function, Eq.~(\ref{grdinf}), and
the number of contacts, Eq.~(\ref{zdinf}), which we find equal to $2d$ at least
close to $\fICP$.
\item We computed the Kauzmann and glass close packing densities in any finite dimension
in the small cage expansion, see table~\ref{tab:I} and figure~\ref{fig:scaling}.
\item In $d=3$ we obtain an expression, Eq.~(\ref{gscaling}), for the scaling
of the contact peak of $g(r)$ close to jamming. This expression describe very well
the numerical results, see figure~\ref{fig:deltapeak}.
From this expression 
it follows that the distribution of contact forces is Gaussian, see Eq.~(\ref{Pfgauss})
and figure~\ref{fig:forze}, and that the number of neighbors is $z=2d$ at jamming (\ie 
packings are isostatic), see
Eq.~(\ref{zclosej}) and figure~\ref{fig3}. 
The form of the scaling function seems particularly robust as it is found
both in the small cage expansion and in the HNC approximation.
\item We are able to reproduce the equation of state of the glass for slow
compression rates, see figures~\ref{fig:diad3} and \ref{fig:diad4}, and the
equilibrium complexity of the liquid, see figure~\ref{fig:Sicomp}.
\item In the HNC approximation we have indications for the development in $g(r)$ of a jump
in $r=2 D$ and of long range correlations; these features cannot be studied within
the small cage approximation but should be present also in the Baxter resummation.
\item For binary mixtures, we showed that the theory correctly predicts the variation
with mixture composition of the jamming density and the partial coordinations.
\end{itemize}
Finally, very recently strong evidence has been reported for a divergence of the equilibrium
relaxation time of an hard sphere liquid at a density $\f_0$ at which the pressure of the liquid
stays finite~\cite{BEPPSBC08,BW08,BW09}. 
This seems consistent with the existence of a Kauzmann transition at $\f_0 = \f_K$ of the kind
discussed in this paper, even if a computation of the complexity for this system is still missing.
Still these results seem to exclude the possibility that the equilibrium relaxation time diverges
only at random close packing as proposed by some authors.

\subsection{Discussion and perspectives}

There are still many important points that deserve investigation.
A tentative list is the following:
\begin{itemize}
\item Our theory is based on an ``equilibrium'' computation (where we also include
metastable long lived states). Hence, a key role is played by entropy. We wish to
stress that in the context of metastable glassy states, there are two very important
concepts related to entropy: the first one is the {\it complexity}, that counts the
number of such states; the second is the {\it surface tension} between different amorphous
states, which must also be of entropic nature since we are dealing with hard spheres.
The existence of a surface tension (a free energy cost) to match different glassy states
is necessary for those states to be well defined; 
see Appendix~\ref{sec:metastability} and \cite{XW01a,BB04,Ca09} for a more complete discussion. 
While the complexity
can be computed, at least approximately, within the theory presented here, 
a first principle method to compute the surface tension is still 
missing, despite preliminary attempts in simplified models~\cite{Fr05}.
However, the study of the latter has very recently become 
the subject of intense numerical effort~\cite{BBCGV08,CCGGV09}, 
which will hopefully lead to a more complete
understanding of the properties of glassy states.
\item Clearly, a major open problem is whether a thermodynamic glass transition really exists in finite
dimensional systems. Our mean-field theory has not much to say on this problem. It might
seem that in our theory the existence of a Kauzmann transition is a major assumption. This is
indeed not true: going carefully over the discussion, it will become clear that what really matters for
us is the existence of long-lived metastable glassy states. The glass transition might be avoided in finite
dimension because of some still unknown mechanism. 
Still the existence of metastable glassy states seems well established and mean field theory
provides a precise description of their properties.
\item Indeed, as discussed in Appendix~\ref{sec:metastability}, in finite dimension, 
in the limit of infinite volume
and if one waits an infinite time, only the ideal glass state (if any) remains stable at finite (but arbitrarily large)
pressure; therefore in a finite dimensional system stable amorphous packings exist strictly only 
at $\fICP$. Here we say that a packing is stable if the system remains close to it for an infinite time when
the pressure is made finite by slightly decompressing the particles. It might be useful to elucidate the 
relation of this notion of stability with the jamming categories of \cite{TTD00,DCST07}.
On the other hand, the time scale needed to observe the instability should diverge exponentially in the
distance from $\fICP$, with an associated diverging length scale \cite{BB04}; therefore for all practical
purposes we expect the mean field picture to hold in finite systems. A systematic investigation of the
stability of packings as a function of system size would be very important in this respect.
\item The existence of a further phase 
transition has been recently recognized in the context of mean field models defined on random graphs,
including for instance the hard sphere model of \cite{BM01}. 
It has been called {\it freezing} or {\it rigidity} transition and is characterized by the appearance
of a finite fraction of frozen particles in the system~\cite{ZK07,Se08}. 
In other words, at densities below the freezing
transition, even if the structure is frozen, particles can still diffuse out of their cages, while above the freezing transition
a finite fraction of them is really stuck inside his cage and can only vibrate; the diffusion coefficient is strictly zero
for these particles. This transition is very peculiar to mean field models: indeed, it has been shown~\cite{Os98}
that for finite dimensional hard spheres the diffusion constant of a tagged particle is always strictly positive at finite pressure.
Even if this transition is avoided in finite dimensional system, it might play some role for finite volumes and finite times. Its
role in the context of jamming should be clarified.
\item 
An important result that attracted a lot of attention in the last years is the presence of an anomalous
bunch of soft modes in jammed packings at $\f_J$~\cite{OSLN03,SLN05,WNW05,MSLB07}. These modes are related to
isostaticity, they are associated
to a diverging length scale and have
been related to the Boson peak observed in glasses at low temperature.
It would be interesting to find out the origin of these soft modes within the approach
presented here.
\item
Another subject of discussion in the community is the role of rattlers (particles that are not
blocked by their neighbors) in jammed packings. In our theory the presence of rattlers is ignored,
due to the simple form we chose for the single molecule density, Eq.~(\ref{rhomolgauss}).
A more refined {\it ansatz} for this quantity could allow to compute the fraction of rattlers, and
it would be interesting to compare this with numerical simulations. Also, in presence of rattlers
the internal entropy of the glass should remain finite at jamming (even if its derivative, the pressure,
should diverge). It would be important to check this explicitely.
\item
An important role might be played by 3 and more body interactions in the
  effective replicated liquid, at least
in finite dimension. In this paper we always ignored these interaction. The inclusion
of three body interaction will allow to compute the second order in the small cage
expansion and might cure the unsatisfactory behavior of the Baxter resummation.
We believe that this is the most important technical point to be studied in the future.
\item The result for the number of neighbors in $d\to\io$ is strange and should be
reconsidered with more care. In particular one should check whether the 1RSB solution
is stable and look more carefully to subleading corrections, also coming from many
body interactions.
\item Our result for the clustering transition for $d\to\io$, $\f_d \sim d/2^d$, 
suggests that a dynamical glass transition happens around this density.
It would be very interesting to study the Mode-Coupling equations in the limit $d\to\io$
to investigate this possibility.
\item The Baxter resummation can be improved in different ways as discussed in
the last section. In particular there are some features of the $g(r)$ like 
the square root singularity and the peak in $r=\sqrt{3}D$ that we are not able
to reproduce at present and might be captured by a more careful resummation.
\item An extension to non-spherical particles such as
ellipsoids should be possible and very interesting, since these systems
show a non-trivial behavior of packing density with aspect 
ratio~\cite{DCSVSCTC04}.
\item Potentials made by an hard-core part plus a short-range attractive potential
can be investigated: at the Mode-Coupling level these potentials 
show an interesting phase diagram characterized by a reentrant glass transition and
a glass-glass transition line (characterized by higher-order Mode-Coupling singularities).
These results have been partially reproduced
by a static HNC computation in \cite{VPR06}, still with no evidence of a glass-glass
transition. It would be interesting to see if
better approximation schemes (such as the small cage approximation) 
could describe also the glass-glass transition.
\item Soft repulsive potentials such as those used in \cite{OLLN02,OSLN03,BW08,BW09} could
be studied with this method, but this will require matching between the small cage expansion for
hard spheres discussed here with the harmonic expansion of \cite{MP99}. This might be technically 
difficult.
\end{itemize}
We hope that future work will address at least some of these issues.

\vskip1cm

{\large \bf Acknowledgments}

\vskip.5cm

We wish to thank in particular the authors of \cite{DTS05,SDST06,AF07,BW06,SO08,BW09,MFC08} 
for sending us their numerical
data, giving the permission to reproduce them here and answering with patience to all our
question about the data.

FZ benefited a lot from the stimulating environment of two workshops:
the ``Jamming'' workshop in Aspen Center
for Physics (August 2007) and the ``Dynamical Heterogeneities'' workshop
held in Lorentz Center, Leiden (September 2008). 
He wishes to thank both centers for hospitality and all the participants
for many useful discussions.

Finally, 
it is a great pleasure for us to thank L.~Angelani, L.~Berthier, I.~Biazzo, 
G.~Biroli, J.-P.~Bouchaud, 
F.~Caltagirone, A.~Cavagna, P.~Charbonneau, 
B.~Coluzzi, O.~Dauchot, A.~Donev, W.~Ellenbroek, G.~Foffi, S.~Franz,
A.~Giuliani, P.~Goldbart, C.~O'Hern,
W.~Krauth, F.~Krzakala, J.~Kurchan, J.~L.~Lebowitz, A.~Liu, R.~Mari, M.~M\'ezard, D.~Ruelle,
M.~Schr\"oter, B.~Scoppola, G.~Semerjian, M.~Tarzia, 
S.~Torquato, P.~Verrocchio, P.~G.~Wolynes, M.~Wyart,
L.~Zdeborov\'a, for many 
interesting discussions that were fundamental for the development of this work.

\appendix

\section{Metastable glassy states in finite dimension}
\label{sec:metastability}

In this Appendix we discuss 
how the concepts of {\it metastable glassy state} and of {\it complexity}
should be modified in finite dimensional systems.
The discussion will be brief and we refer to the original literature for more details.

\subsection{Metastable states in ferromagnetic systems}

As discussed in the body of the paper, the key feature of mean field theory is the existence of
an exponential number of metastable glassy states at high density. In mean field these states can be 
defined as minima of a suitable free-energy functional: the TAP functional \cite{MPV87,TAP77} for spin systems,
or a suitable density functional for particle systems~\cite{KW87,CKDKS05,DV99}.
These states are stable because in the infinite range case the barriers may be divergent.
However, in finite dimensional systems, states that have a free energy (per particle) greater than the ground state can always decay towards the ground state by a 
nucleation process, by  crossing a barrier that may diverge only when the energy difference with the ground state goes to zero.

Therefore we should find a suitable definition of these states. 
Let us first recall the way this is done, for instance, in a ferromagnetic system.
In positive magnetic field, the state (let us call it $-$) 
with negative magnetization is metastable with respect to the state $+$. 
However one can study the state $-$ by preparing
the system in negative field where this state is stable, and then increasing slowly the field towards 
its final positive value. 
After some time the state $+$ will nucleate. 
Still, if one is far enough from the spinodal of the
state $-$, 
the nucleation time is long enough and one is able to follow the state $-$ into the
positive field region and measure its properties (\eg its magnetization). In dimension $d$ 
the life-time of the metastable state diverges for small $h$ as $\tau(h)=\exp(Ah^{1-d})$. 
This example shows that once the correct order parameter
has been identified, one can study metastable states by adding a suitable external field coupled to the order
parameter in order to stabilize this state, and then change the external field following the evolution of the
state into the region where it is metastable.

Actually this can be turned into a practical computational scheme as follows. Suppose we fix the magnetic
field to its positive value, but then perform an expansion of the free energy in $1+m$, assuming that the
system is in the metastable state $-$ and therefore its magnetization $m$ is close to $-1$. 
One can check explicitly that below the critical temperature for ferromagnetism the free energy obtained
in this way displays a minimum at $m=-m^*$, that gives the magnetization of the $-$ state. One might wonder
why a minimum appears, as the free energy for positive magnetic field should be convex and have a single
minimum at positive $m$. The key observation is that
the decay of the metastable state is 
non-perturbative in $1+m$, so that it is missed at any order in perturbation theory\footnote{Technically speaking the free energy is non-analytic at the 
point $h=0$, 
being however a $C^{\infty}$ function of $h$. This fact is practically invisible from the expansion around $m=\pm 1$. If we analytically continue the free-energy in the region of 
negative magnetic field, it should acquire an imaginary part proportional to $\tau(h)^{-1}$, signaling that a sharp definition of the properties of the metastable phase is not 
possible. See Chapter 10.5 of \cite{ParisiBook} for a more detailed discussion.} around $m=-1$.
Thus, the expansion around $m=-1$
stabilizes the metastable minimum also for a finite dimensional system. 
This might seem an artifact of the approximation, 
but in some sense it reflects a physical
property of the system, the existence of a metastable state, that is relevant for numerical simulations
and experiments as far as they are able to probe this state. This is exactly
what we are interested in. 

\subsection{Free energy functional for metastable glassy states: order parameter and coupled replicas}

In the case of a 1RSB transition to an amorphous state, it is not possible to identify a simple order parameter
since the density profile of each state is amorphous and depends on the state. 
For each state, one should add to the Hamiltonian a specific external potential that favors the specific
density profile of that state, but this is impossible since the density profiles are not known {\it a priori}.

A precise definition of the complexity was given in~\cite{FP97,Me99,MP00} 
and it is based on the same idea that we discussed above in the case of the ferromagnet: 
one couples an external field to
the order parameter in order to prepare the system in the metastable state.
In this case one needs to consider a replicated system, as discussed in section~\ref{sec:method},
since the order parameter is the overlap between replicas (or equivalently the
cage radius).
The derivation is the same as in section~\ref{sec:ZrepLeg}. We introduce a parameter $\ee$
which is coupled to the average distance between replicas and allows to compute the entropy as a function of the cage radius
(see figure~\ref{fig:sfere}) via a Legendre transform. The physical values of $A$ for a given $\ee$ are the solutions of
$\partial_A \SS(m,\f;A) = d (m-1) \ee$ and we are ultimately interested in the case $\ee =0$.
The function $\SS(m,\f;A)$
is therefore the entropy of the system of $m$ copies with the constraint that each pair
of copies is at fixed (in the thermodynamic limit) distance $A$ given by Eq.~(\ref{cagedef}). 

The typical shape of $\SS(m,\f;A)$ is reported in Fig.~\ref{fig:sfere}. 
In a mean field model, at densities below $\f_K$, the correlated phase is metastable and the system is
liquid; at the glass transition it becomes stable as discussed above. In mean field
all this is well defined an one can perform
explicit computations~\cite{Mo95,Me99}.
In a finite dimensional system, however, $\SS(m,\f;A)$ must be a convex function of $A$: then the
minimum at small $A$ should disappear below $\f_K$. However, if we compute $\SS(m,\f;A)$ in a power 
series expansion around $A=0$, we will find
a stable minimum at small $A$. This is exactly the same effect that we discussed for the ferromagnetic 
case. In this way, the properties of the metastable state can be studied also in finite dimensional
systems.
Let us focus on the region where the equation for $A$
has two solutions that correspond to a local minimum of 
$\SS(m,\f;A) - d (m-1) \ee A$.
At fixed density, by varying $\ee$ one of 
the two solutions looses its stability and it disappears: these two curves are the equivalent of the spinodal lines in 
usual first order transition. 
The dynamical transition is the point where for the first time the small-$A$ solution exists at $\ee=0$:
only at higher density the two coupled replicas may remain at an high value of the overlap 
in absence of a force that keeps them together. On the contrary the static transition is 
characterized by the fact that the coexistence line of the two solutions touches the axis $\ee=0$.

\begin{figure}
\includegraphics[width=8cm]{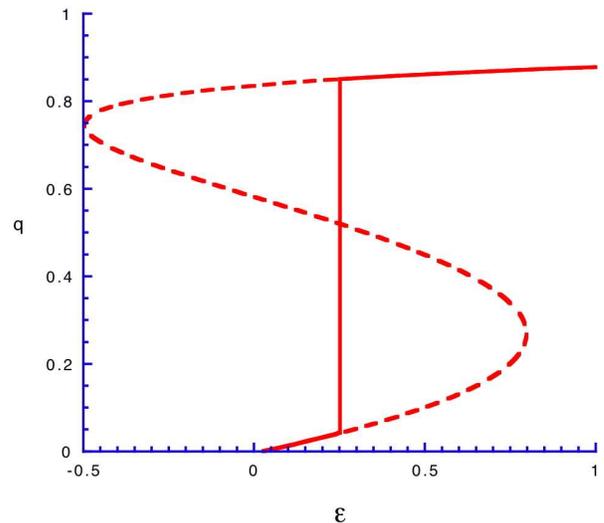}
\caption{
The shape of the function $q = A_0/A$ for $\f > \f_d$: the full line is the correct result and the dashed line
is the output of a mean field approximation.
}
\label{qeps}
\end{figure}

In finite dimension, general arguments tell us that  the free energy is a convex  
function of $A$, so that the correct shape of the function $\SS$ can be obtained by 
the Maxwell construction.
To discuss the consequences of this fact on the definition of the complexity let us 
consider the function $q(\ee) = A_0/A(\ee)$ (for some reference value of $A_0$)
for densities bigger than $\f_{d}$, that is shown in figure~\ref{qeps}. 
As can be seen from the figure, the point where we evaluate the complexity (i.e. $\ee=0$ and high $q$) is always in 
the metastable region for $\f<\f_K$ where the equilibrium complexity is non zero.  This causes an intrinsic ambiguity in 
the definition of complexity because the free energy in not defined with infinite precision in the metastable phase.  
However we can use the fact that the free energy is a $C^{\infty}$ function of $\ee$ near the discontinuity point to 
extrapolate the high $\ee$ free energy in the metastable region.  This ambiguity becomes smaller and smaller the more we 
approach the Kauzmann density 
and in general it is rather small unless we are very near to the clustering phase transition. 
This ambiguity is not important from practical 
purposes; however it implies that there is no sharp, infinitely precise definition of the 
equilibrium complexity.  If we forget this intrinsic ambiguity in the definition of the complexity we may arrive to 
contradictory results.

It is important to remark once again that this discussion can be turned into a practical computational scheme to obtain the complexity
analytically, via the small cage expansion, or in numerical simulations, where
it has been used for both Lennard-Jones \cite{CMPV99} and hard sphere \cite{AF07} systems,
see the original papers for details.

\subsection{The physical meaning of the small $A$ minimum of the free energy}

The previous discussion was based only on the analytic properties of the replicated
free energy $\SS(m,\f;A)$. To complete the description, it is desirable to have a 
understanding of
the mechanism that is beyond the metastability of the small $A$ minimum.
This is a key point because, as we will see, the
mechanism is completely different in mean field and in finite dimension.

\subsubsection{Mean field}

Let us first discuss what happens in mean field. In this case, glassy
states are really stable, in the sense that they are local minima of a 
well defined free energy functional, separated by barriers whose height
diverges in the thermodynamic limit. In a dynamical perspective, once the
system is prepared in a glassy state, it will remain there forever.
There is a real ``ergodicity breaking'' in the liquid state above $\f_d$,
corresponding to the ideal Mode-Coupling dynamical transition.
So why is the small $A$ minimum metastable with respect to the liquid one
between $\f_d$ and $\f_K$?

The reason is that the number of glassy states is exponentially large in
system size between $\f_d$ and $\f_K$. Consider for instance the case of
two coupled replicas, $m=2$. If the limit of zero coupling, if both replicas
are in the same state, the total entropy will be twice the internal entropy
of a state, plus the complexity $\Si$ that represents the contribution of
all possible states in which the two copies can be.
On the other hand, if the two replicas are uncorrelated, the total entropy
is twice the internal entropy, plus twice the complexity, since now each
replica can be in each state independently.
It is clear that by forcing the two replicas in the same state, one 
{\it loses entropy}. This is why the small $A$ minimum has lower entropy,
or higher free energy, and is metastable, as long as the complexity $\Si > 0$.

Therefore, in mean field, glassy state are stable but their huge number
is responsible for the fact that the replicated system finds more convenient
to have each replica in a different state. This situation is very strange: 
indeed it is very specific of the mean field structure and it changes completely
when one considers finite dimensional systems.

\subsubsection{Finite dimension}

What goes wrong in the above picture when applied to finite dimensional systems?

The main problem is that, in a finite
dimensional system, an exponential number of equilibrium states cannot exist.
This can be argued as follows:
\begin{enumerate}
\item Pure states are defined by 
taking the thermodynamic limit with a specified sequence of boundary conditions; and the number of
different boundary conditions scales as $e^{\kappa L^{d-1}}$,
\ie as the exponential of the surface of the system, 
so that the number of states cannot be exponential in the volume. The picture of the liquid
state split in an exponential number of stable pure states does not make sense in finite
dimension.
On the contrary, at and above $\f_K$ the number of states is not exponential
in the system size even in mean field. Therefore in this case the picture does not need to be 
modified in finite dimensions: a sub-exponential number of pure states are perfectly
allowed as the number of possible boundary conditions is still very large.
\item
A more sharp argument can be done for soft particles if we consider the minima of the energy. 
Of course there can be an exponentially large number of minima of the energy. A true 
metastable state would be a configuration, whose energy cannot be decreased by moving a finite number of particles.
It is easy to argue that such a configuration must have the same free energy of the ground state. 
On the other hand we can consider local minima of order $k$, i.e. configurations whose energy cannot be decreased by moving at most $k$ particles.
It is quite evident that the associated complexity densities $\Sigma_{k}(E)$ will be different from zero, 
although they will tend to zero when $k$ goes to infinity~\cite{BM00}. 
Note that analytic computations based on the small cage expansion discussed in this paper
will likely predict something similar to $\Sigma_{1}(E)$, and the effective value of $k$ will increase with the precision of the computation. However 
for not too large times, and in particular in the experiments it is quite possible that $\Sigma_{1}(E)$ or $\Sigma_{2}(E)$ are the relevant quantities.
\end{enumerate}
The two arguments above strongly suggest that the metastability of the small $A$ minimum in finite dimensions
is connected to the large scale properties of the states and in particular to large-scale rearrangements.
This is at the origin of the ambiguity in the definition of the complexity that we discussed in the previous section.

Then, in what sense an exponential number of metastable glassy states, giving a finite
complexity, can exist in this case?
A consistent ``real space'' formulation of the problem has been recently discussed in a series of
papers~\cite{XW01a,XW01b,BB04,Fr05}. One first assumes that for a finite (and small)
system it is possible to define metastable states and that they initially are 
exponentially many in system size. Then, one can ask whether these states remain stable
on increasing the system size. The key observation is the following~\cite{BB04}.
Consider a ball of size $R$ inside the system. The rest of the system will influence
the ball through its boundary, therefore favoring one particular state $\a$ of the ball
among the exponentially many, through a surface contribution $F_\a = -\k R^{d-1}$
to the free energy of the ball when the latter is in state $\a$. 
On the other hand, the ball might be in any other state $\b \neq \a$, and in this
case it will gain entropy because of the large multiplicity of states: 
$F_{\text{not } \a} = F(\text{all possible } \b\neq \a) = -T \Si R^d$. 
The ball will choose whether to
stay in state $\a$ or not according to which is the largest between $F_\a$ and
$F_{\text{not } \a}$. It is easy to see that for small enough $R$, $F_\a$ wins, while
for large enough $R$, the other term wins and the ball will choose not to stay in
the state $\a$ suggested by its boundary. This reasoning shows that, for large $R$,
the assumption of the existence of an exponential number of states is inconsistent,
since for entropic reasons
subsystems will always ignore the influence of their boundary and escape from the state.

See~\cite{XW01a,XW01b,Fr05} and in 
particular~\cite{BB04} for a much more detailed discussion. In this series of papers
the discussion above was precisely formulated in a nucleation theory where the driving
force is of {\it entropic} nature.
Remarkably, methods to compute the length of the
critical nuclei (droplets) and their relaxation time have been proposed;
this is a first step towards a quantitative description of activated
processes between $\f_d$ and $\f_K$, even if a complete theory is still lacking.

\subsubsection{Finite dimension, finite volume}

The last point to be discussed is the assumption made in the previous section, that
for finite and small enough systems, an exponential number of states still exist and
can be taken as the starting point for the nucleation theory discussed in \cite{XW01a,BB04}.

We will discuss this crucial point in the specific case of hard spheres, that is of
interest here. Consider a system of hard spheres enclosed in a finite box with
(for simplicity) hard walls.
It is very easy to see that for a small box and high enough density, there will be 
disconnected sets of configurations, in the sense that there are pairs of 
configurations that cannot be transformed one into the other simply 
by moving continuously the spheres. Therefore we can group each set of configurations
that can be continuously transformed one into each other in a ``state''.
We can say that configurations inside a state are ``blocked'' in the sense
that once the system is prepared in a state it will never escape\footnote{We use the
word ``blocked'' in order to avoid confusion with the concept of ``jamming''. In a jammed
configuration no particle can move. Instead, in a blocked configuration particles can move
a little but the whole system is unable to visit all the phase space; the existence of blocked configurations was shown
long time ago by Ruelle (Ruelle, private communication).}.
We can define a finite-volume complexity as $\Si_V = \frac{1}V \ln \NN_V$,
where $\NN_V$ is the number of such states in a volume $V$.

Now we can increase the volume of the box while at the same time adding particles in order
to keep the density fixed. On a general ground, we expect $\Si_V$ to be a decreasing function
of $V$ since by increasing the volume one will open new channels to ``unblock'' the configurations
and connect some states that will be merged in a bigger state. 
On the basis of the analysis in \cite{BB04}, we expect
three possible behaviors of $\Si_V$:
\begin{enumerate}
\item for $\f < \f_d$, in the un-clustered liquid phase, we expect $\Si_V$ to fall very rapidly
to 0, as for large enough volumes all the configurations will be connected and form the unique
liquid state.
\item for $\f_d < \f < \f_K$ we are in the clustered liquid phase; then we expect that $\Si_V$
first decreases to a finite {\it plateau} for small enough volume. The plateau correspond to
the mean field value $\Si(\f)$. On larger length scales, once the nucleation effect described
in the previous section becomes effective, the glassy states are ``unblocked'' and the complexity
will drop to zero; only the liquid state survives.
\item for $\f > \f_K$, the plateau value goes to zero since the mean field complexity vanishes.
Therefore in this case the complexity will go to zero fast as in case 1. Still, even in the
$V\to\io$ limit a large number of amorphous states will survive (but their number is not exponential).
\end{enumerate}
It is interesting to remark that in case 2, ``unblocking'' a configuration will require a very large
time since a large number of particles have to rearrange together. Therefore the dynamics in this
region will be very slow and the system will stay for a long time close to a configuration that is
made by patches of locally blocked configurations, the ``mosaic state'' of \cite{XW01a}.
Conversely, in case 3, the dynamics will be completely frozen even in the thermodynamic limit.

A consequence of this discussion is that an infinite system of hard spheres 
at finite pressure, even if very large,
will always relax to the ideal glass state, \ie at density $\fICP$ (if crystallization is avoided).
Therefore the ideal glass states at $\fICP$ are the only jammed states that remain stable if pressure
is made very large but finite, in the limit $V \to \io$. Of course for finite sizes or finite times
this will be true also for metastable glasses.

\section{The ``link'' expansion for Hard Spheres}
\label{app:linkexp}

We will describe here a formal way to justify the introduction of the effective potentials
to integrate over $m-1$ replicas and obtain an expression for the replicated entropy which
is formally equal to the one of a non-replicated liquid.
We use one replica, say replica 1, as
a reference, and integrate over the small displacements of the
other $m-1$ replicas. Note that we are {\it not} breaking the replica symmetry: we are only
looking for a way to expand the entropy for small $A$.
In the following we will use the notation $\xx=(x_2,\cdots,x_m)$.

The expansion of a given diagram proceeds as follows.
At the zeroth order ($A=0$),
the function $\rho(\bar x)$ is a product of delta functions, 
therefore all the replicas coincide. Then we have 
\beq
\DD=\DD_0 = \frac{1}{S} \int \prod_i \r d x_i \prod_\ell ( \chi(\ell) -1) \ ,
\eeq
\ie $\DD_0$ is the corresponding diagram of the non-replicated system.

The quantities $x_1-x_a$ are of order $\sqrt{A}$. Then
the crucial observation is that, for $a\geq 2$,
the function $\c(x_{ai}-x_{aj})$ in Eq.~(\ref{diagramma}) is essentially constant
if $|x_{1i}-x_{1j}|$ differs from $D$ by a quantity $\gg \sqrt{A}$, 
in fact it becomes simply equal to $\c(x_1-y_1)=\th(|x_1-y_1|-D)$.
This means that
the integration region in the space $x_{1i}$ such that $||x_{1i}-x_{1j}| - D| \gg \sqrt{A}$ 
for all links $\ell=(i<j)$ does not give any contribution
apart from the zeroth order one.
Let us call a link $\ell$ such that 
$||x_{1i}-x_{1j}| - D |\sim \sqrt{A}$ a ``critical link'':
the idea is to isolate the contribution of the regions such that $k$ links are critical, whose
volume is of order $A^{k/2}$. 

\subsection{Expansion of the replicated entropy}

We define $\DD x_1 = \prod_i \r dx_{1i}$ and 
$\DD \xx = \prod_i (\r(i)/\r) d\xx_i$ and note that using Eq.~(\ref{rhomolgauss}) we have
$\int \DD\xx = 1$.
Defining $\c_a(\ell) = \c(x_{ai}-x_{aj})$, we have for a diagram $\DD$ contributing to $\SS$:
\begin{widetext}
\beq\label{Zconti}
\begin{split}
\DD &= \frac1S \int \prod_i \r(i) dx_{1i} d\xx_i \prod_{\ell} (\bar\chi(\ell)-1) =
\frac1S \int  \DD\xx \DD x_1 \prod_{\ell} [\chi_1(\ell)-1 + \bar\chi(\ell) -\c_1(\ell)] \\
&= \frac1S \int \DD x_1  \prod_{\ell} (\chi_1(\ell)-1) + 
\frac1S \int  \DD x_1 \sum_{\ell} \prod_{\ell'\neq \ell} (\chi_1(\ell')-1) 
\int \DD \xx [\bar\chi(\ell)-\chi_1(\ell)] \\
&+ \frac1S \int \DD x_1 \sum_{\ell<\ell'} \prod_{\ell''\neq \ell,\ell'} (\chi_1(\ell'')-1) 
\int \DD \xx [\bar\chi(\ell)-\chi_1(\ell)][\bar\chi(\ell')-\chi_1(\ell')] + \cdots \\
&= \DD_0 + 
\int \DD x_1 \sum_\ell \wt Q(\ell) \prod_{\ell'\neq \ell} (\chi_1(\ell')-1) 
+ \int \DD x_1 \sum_{\ell<\ell'} \wt Q_2(\ell,\ell') \prod_{\ell''\neq \ell,\ell'}
( \chi_1(\ell'')-1)  + 
\cdots \\
&= \DD_0 + \sum_{n \geq 1} \int \DD x_1 \sum_{\ell_1 < \cdots <\ell_n} \wt Q_n(\ell_1,\cdots,\ell_n)
\prod_{\ell \neq (\ell_1,\cdots,\ell_n)}( \chi_1(\ell)-1) \ .
\end{split}
\eeq
having defined the functions $\wt Q_n(\ell_1,\cdots,\ell_n)$ from
\beq\label{wtQdef}
\begin{split}
&\wt Q_1(\ell) = \wt Q(\ell) = 
\int \DD \xx [\bar\chi(\ell)-\chi_1(\ell)] = \int d\xx_i d\xx_j \frac{\r(i)\r(j)}{\r^2} 
[\bar\chi(i,j)-\chi_1(i,j)] \ , \\
&\wt Q_n(\ell_1,\cdots,\ell_n) = \int \DD \xx \prod_{i=1}^n [\bar\chi(\ell_i)-\chi_1(\ell_i)] =
\int \prod_{i\in \partial L_n} d\xx_i \frac{\r(i)}\r \prod_{\ell\in L_n} [\bar\chi(\ell)-\chi_1(\ell)]
 \ . 
\end{split}
\eeq
\end{widetext}
The integral over $\DD \xx$ contains all the vertices of the original diagram. However, the 
integrand depends only on the vertices that are adjacent to one of the links 
$L_n = (\ell_1,\cdots,\ell_n)$; calling this set $\partial L_n$, and using that the integrals over the
other vertices give 1, we obtained the last equality in (\ref{wtQdef}).

Note that, for example, the function $\wt Q_2(\ell,\ell')$ depends on three or four 
$x_i$ depending whether the two links $\ell,\ell'$ are adjacent or not. But if the two links
are non-adjacent, then $\wt Q_2(\ell,\ell') = \wt Q(\ell) \wt Q(\ell')$. This motivates the 
introduction of the connected functions, defined by $\wt Q^c(\ell) = \wt Q(\ell)$ and
\beq\label{wtQcon}
\begin{split}
\wt Q_2(\ell,\ell'&) = \wt Q(\ell) \wt Q(\ell') + \wt Q_2^c(\ell,\ell') \ , \\
\wt Q_3(\ell,\ell'&,\ell'') = \wt Q(\ell)\wt Q(\ell')\wt Q(\ell'') \\ &+ 
\{ \wt Q_2^c(\ell,\ell')\wt Q(\ell'') + \text{perm.}\} + \wt Q_3^c(\ell,\ell',\ell'') \ , \\
\end{split}
\eeq
and so on, as usual in statistical mechanics. The important property of these functions is that
they are non-vanishing {\it only if} the sub-diagram identified in $\DD$ by the links in $L_n$ 
is connected.

Now we insert the expression of the connected functions in (\ref{wtQcon}) in (\ref{Zconti}). It is
not difficult to check that we can resum the contributions containing $\wt Q(\ell)$ to obtain
\beq
\begin{split}
\DD &= \frac1S \int \DD x_1  \prod_{\ell} (\chi_1(\ell) + \wt Q(\ell) - 1) \\
&+ \frac1S \int \DD x_1 \sum_{\ell<\ell'} \wt Q^c_2(\ell,\ell') \prod_{\ell'' \neq \ell,\ell'}
( \chi_1(\ell'')+ \wt Q(\ell) -1) + \cdots
\end{split}
\eeq
The interpretation of the above equation is the following: the original diagram of the replicated
liquid generates a diagram in which the vertices carry a factor $\r$ and the links carry a function
$\chi_1+\wt Q -1$, plus a sum of contributions in which a subdiagram made of links $L_n$ 
is marked and {\it on each of its connected parts} an interaction $\wt Q^c$ is placed. The
unmarked links carry again the function $\chi_1+\wt Q -1$, and the vertices a factor $\r$.
In fact, it is easy to check on specific examples that the explicit sum over the links in $L_n$ can
be rearranged by grouping together, as usual, diagrams with the same topology. In this way the
multiplicity factors $S$ are corrected to become the exact symmetry factors of the diagram with marked
links. 

In summary, we can write
\beq
\DD = \DD_0[\chi_1+\wt Q] + \DD_2[\chi_1+\wt Q, \wt Q_2^c] + \DD_3[\chi_1+\wt Q, \wt Q_2^c, \wt Q_3^c]
+ \cdots \ ,
\eeq
where the $\DD_n$ represent diagrams which are built from the original diagram by marking $n$ links
in all the possible {\it topologically inequivalent} ways and placing the interactions $Q^c$ on
the marked connected parts. Each diagram has a symmetry factor $S$ which is the number of equivalent
relabeling of the vertices, taking into account the presence of the marked links.
Clearly, summing the contribution of all the diagrams and Eq.~(\ref{gasperfetto}) we get
\beq\label{Sexpansion}
\begin{split}
\SS[\r(\bar x),&\bar\chi(\bar x,\bar y)] = N S_{harm} + \SS_0[\r,\chi_1(x-y)+\wt Q(x-y)] \\ &+ 
\SS_2[\r,\chi_1(x-y)+\wt Q(x-y),\wt Q_2^c(x,y,z)] + \cdots
\end{split}
\eeq
where $\SS_0[\r,1+f]$ is defined in the same way as the functional (\ref{Sdirho}), but
for one single copy of the system and with $f$ on the links of the diagrams, while the other
terms come from interactions involving more than two particles. 

It is worth to note at this point that the specific form of the effective potentials $\wt Q_n^c$
depends on the topology of the sub-diagram that carries them. For the three-body interactions there
is only one possible diagrams (the one in which $\ell,\ell'$ are adjacent, \ie they share one vertex)
but for $n \geq 3$ there are many possible diagrams that correspond to different interactions.

The $\wt Q^c$ functions are quite difficult to handle.
However, from their definition one
can show that, for small $A$, they are non-zero only if for all the links it holds
$|x_i-x_j| \sim D + O(\sqrt{A})$, as expected on the basis of the argument put forward 
at the beginning of this section. Moreover, for $m>0$ these functions are bounded, in the
sense that when they are different from zero, they stay finite for $A \to 0$.
This implies that, for example, the contribution coming
from diagrams containing $\wt Q^c_2$ is of order $A$, because the integrals over the two link
variables have support only on an interval of order $\sqrt{A}$. Therefore, the expression
(\ref{Sexpansion}) is an expansion in powers of $\sqrt{A}$, where the term $\SS_n$ is
proportional to $A^{n/2}$. Note that this is specific to the hard sphere potential which is always
constant except for the discontinuity in $r=D$. For smooth potentials, the corrections come from
all values of $|x_i-x_j|$, and in this case it is more convenient to integrate
over the displacement of the replicas at fixed center of mass, see \cite{MP99}.
Another remarkable property of the effective potentials is that in the limit of infinite dimension
only the two body potential should give important contributions.

\subsection{The effective potentials}

To gain physical intuition on the effective liquid described by the entropy functional
(\ref{Sexpansion}), it is convenient to compute its partition function.
First of all we define the potentials $Q$ by
\begin{widetext}
\beq\label{Qdef}
Q_n(\ell_1,\cdots,\ell_n) =
\frac{\wt Q_n(\ell_1,\cdots,\ell_n)}{\chi_1(\ell_1)\cdots\chi_1(\ell_n)} =
\int \prod_{i\in \partial L_n} d\xx_i \frac{\r(i)}\r \prod_{\ell\in L_n} 
\left[\prod_{a=2}^m \chi_a(\ell)- 1\right] 
= \left\langle \prod_{\ell\in L_n} 
\left[\prod_{a=2}^m \chi_a(\ell)- 1\right] \right\rangle_{\partial L_n}
\ ,
\eeq
and similarly for the connected potentials. The bracket indicate the average
over the $\rho(i)$ of the vertices in $\partial L_n$.

When constructing the partition function that generates the diagrams in (\ref{Sexpansion}),
we must take into account the fact that if a link is labeled it carries the function $\wt Q^c_n$
{\it instead} of the usual function $f(\ell) = \chi_1(\ell) [1+ Q(\ell)] - 1$. This leads,
using standard liquid theory, to the following grancanonical partition function:
\beq\label{Zeff}
\begin{split}
Z_{eff}[z] &= \sum_{N=0}^\io \frac{z^N}{N!} \int d^N x \ \ \prod_{\ell} \chi_1(\ell) (1 + Q(\ell)) \ \
\prod_{\ell < \ell'} \left[ 1 + \frac{Q_2^c(\ell,\ell')}{(1 + Q(\ell))(1 + Q(\ell'))} \right]
\times \\
&\times \prod_{\ell < \ell' < \ell''} 
\left[ 1 + \frac{Q_3^c(\ell,\ell',\ell'')}{(1 + Q(\ell))(1 + Q(\ell'))(1 + Q(\ell''))
\left\{\left[ 1 + \frac{Q_2^c(\ell,\ell')}{(1 + Q(\ell))(1 + Q(\ell'))} \right] \times \text{perm.}\right\}
} \right] \times
 \cdots \ ,
\end{split}
\eeq
where now the ``links'' are all possible pairs of particles. This makes it possible to identify the
$n$ body potentials of the effective liquid.
For instance, we have
\beq\label{effpot2}
\begin{split}
&e^{-\phi_{eff}(x-y)} = \chi(x-y) (1 + Q(x-y)) = \chi(x-y) \left\langle \prod_{a=2}^m \chi(x_a-y_a) \right\rangle_{x,y}
\ , \\
&e^{-\phi_{eff}^{(2)}(x-y,x-z)} =  1 + \frac{Q_2^c(x-y,x-z)}{(1 + Q(x-y))(1 + Q(x-z))} 
= \frac{\left\langle \prod_{a=2}^m \chi(x_a-y_a)
 \chi(x_a-z_a) \right\rangle_{x,y,z}
}{\left\langle \prod_{a=2}^m \chi(x_a-y_a) \right\rangle_{x,y}
\left\langle \prod_{a=2}^m \chi(x_a-z_a) \right\rangle_{x,z}}
\ , \\
\end{split}\eeq
\end{widetext}
and so on. As before the brackets denote averages over the Gaussian distribution of the molecule. 
The entropy functional (\ref{Sexpansion}) is obtained by expanding $Z_{eff}$ in a Mayer series of $z$
and then Legendre transforming to the density. Then one has
\beq
\SS[\r(\bar x),\bar\chi(\bar x,\bar y)] = N S_{harm} -\b N F_{eff} \ ,
\eeq
where $F_{eff}$ is the canonical free energy of the liquid (\ref{Zeff}).

\subsection{Correlation function of the glass}

It is interesting to compute the correlation function of the glass state. It is the correlation function
of one replica, see section~\ref{sec:correlations}.
It can be computed as follows. We choose one replica and we add to the hard core interaction
an additional potential $v(x-y)$, only to this replica. Then it is easy to show that
\beq\label{corrglassv}
\frac{\r^2}2 \gG(x-y) = - \left. \frac{\d \SS[\bar\r,\bar\chi]}{\d v(x-y)} \right|_{v=0} \ .
\eeq
This can be done by looking at the partition function $Z_m(\ee)$, Eq.~(\ref{Zstart}). If we add a potential
to replica one, we get an additional term
\beq
\exp \left\{ - \frac12 \int dx dy v(x-y) \sum_{i \neq j} \d(x-x_{1i}) \d(y - y_{1j}) \right\} \ .
\eeq
Then if we differentiate with respect to $v(x-y)$ we get
\beq\begin{split}
&\left. \frac{\d \ln Z_m(\ee)}{\d v(x-y)} \right|_{v=0}
= - \frac12 \left\langle  
 \sum_{i \neq j} \d(x-x_{1i}) \d(y - y_{1j}) \right\rangle \\ & = -\frac12 \r_{11}(x,y) = -\frac{\r^2}2 \gG(x-y) \ .
\end{split}\eeq
The derivative of $\SS$ is exactly the same because it is the Legendre transform of $\ln Z_m(\ee)$.

The expression of $\SS(m,\f;A)$ in presence of the additional interaction $v$ is very simple; indeed
the only modification is $\chi_1 \rightarrow \chi_1 e^{-v}$, that has to be substituted in the final
result (\ref{Sexpansion}).
Note that the functions $Q$, $Q_2$ etc. do not contain $\chi_1$ explicitly because they
depend only on the potentials of the replicas $2,\cdots,m$, see Eq.~(\ref{Qdef}) 
so they will not depend on $v$. Therefore the correlation of the glass is given by the correlation
of the liquid described by the partition function~(\ref{Zeff}).

\section{The two-body effective potential}
\label{app:A}

\subsection{Definitions}
\label{app:def}

The function $Q(x,y)=Q(x-y)$ is defined in Eq.~(\ref{Qdef}).
Making use of Eq.~(\ref{rhomolgauss}) we get
\beq\label{AppA1}
\begin{split}
Q&(x-y) = \\
&= \r^{-2} \int d\xx d\yy \r(x,\xx) \r(y,\yy) 
\left[\prod_{a=2}^m \chi(x_{a}-y_a) -1\right] \\ 
&= \int dX dY \g_A(x-X) \g_A(y-Y) \\
&\times \left\{ \left[ \int d\x d\h \g_A(\x)\g_A(\h) \c(X+\x-Y-\n) \right]^{m-1} - 1 \right\} \\
&=\int dX dY \g_A(x-X) \g_A(y-Y) \left\{q_A(X-Y)^{m-1} - 1 \right\}
\ .
\end{split}\eeq
where
\beq\label{F0formule}
\begin{split}
q_A(\vec r) &= \int d\x d\h \g_A(\x) \g_A(\h) \chi(\vec r+
\x- \h) 
\\ &=
\int d\vec{ r'} \g_{2A}(\vec{ r'}) \chi(\vec r-\vec{ r'}) \ .
\end{split}
\eeq
The last equality is obtained by introducing $\vec{r'}=\h-\x$ and observing that as $\x$ and
$\h$ are independent Gaussian variables with variance $A$, their difference is also
a Gaussian variable of variance $2A$.
Following a similar procedure\footnote{We start from the last line of (\ref{AppA1}),
introduce $\x=x-X$ and $\h=y-Y$, and then $\vec{r'}=\h-\x$.}, we obtain
\beq\label{QQ1}
Q(\vec r) = \int d\vec{r'} \g_{2A}(\vec{r'}) \left\{q_A(\vec r-\vec{r'})^{m-1} - 1 \right\} \ .
\eeq
We are particularly interested in the integral
\begin{equation}\label{GmAdef}\begin{split}
G_m(A) &= \frac1{V_d(D)} \int d\vec r \, \chi(\vec r) Q(\vec r) \\
&= 
\frac1{V_d(D)}\int d\vec r [ q_A(\vec r)^m - q_A(\vec r) ] \\
&=\frac1{V_d(D)}
\int d\vec r [ q_A(\vec r)^m - \c(\vec r) ] \ .
\end{split}\end{equation}
which is the
contribution to the virial of the non-trivial part of the two body effective potential 
given in (\ref{effpot2});
the different expressions in (\ref{GmAdef}) are obtained starting 
either from (\ref{QQ1}) or directly from (\ref{AppA1}) and 
performing similar manipulations to the ones in (\ref{AppA1}).

We will also be interested in the following functions:
\begin{equation}\label{FmAdef}
\begin{split}
F_m(A) &\equiv \frac{A}{1-m} \frac{\partial G_m(A)}{\partial A} \\ &= \frac{m
  A}{1-m} \frac1{V_d(D)} 
\int d\vec r q_A(\vec r)^{m-1} \frac{\partial q_A(\vec r)}{\partial A} \ , \\
H_m(A) &\equiv -m \frac{\partial G_m(A)}{\partial m}\\ & =
-\frac{m}{V_d(D)} \int d\vec r q_A(\vec r)^m \ln q_A(\vec r) \ .
\end{split}
\end{equation}
In the limit $m \to 1$ we have $G_m(A) \to 0$ while
\begin{equation}
\begin{split}
&F_m(A) \to F_1(A) \equiv -\frac{A}{V_d(D)} \int d\vec r \ln[ q_A(\vec r) ] \frac{\partial q_A(\vec r)}{\partial A} \ , \\
&H_m(A) \to H_1(A) \equiv -\frac1{V_d(D)} \int d\vec r q_A(\vec r) \ln q_A(\vec r) \ .
\end{split}
\end{equation}
In the limit $m \to 0$, $A = \a m$, we will show that $G_m(A) \to G_0(\a)$
which is a finite function of $\a$. We will also show that
\begin{equation}
\begin{split}
&F_m(A) \to F_0(\a) = \a \frac{dG_0(\a)}{d\a} \ , \\
&H_m(A) \to H_0(\a) =\a \frac{dG_0(\a)}{d\a} = F_0(\a) \ .
\end{split}
\end{equation}
Finally, we will be interested in the large $d$ limit with the following scaling:
\begin{equation}\label{calGmAdef}
\begin{split}
&\GG_m(\wh A) \equiv \lim_{d\to\io} G_m(D^2 \wh A d^{-2}) \ , \\
&\FF_m(\wh A) \equiv \lim_{d\to\io} F_m(D^2 \wh A d^{-2}) \ .
\end{split}\end{equation}

Unfortunately a full analytical evaluation of these integrals is not possible.
In the following we will discuss some particular cases in which analytical calculations
are possible.

\subsection{The function $q_A(r)$ and the virial coefficient $G_m(A)$}

The function $q_A(r)$ defined in (\ref{F0formule}) is the convolution of a Gaussian and
the theta function $\chi(r)$. This integral can be reduced to a one-dimensional integral
by mean of $d$-dimensional bipolar coordinates. This will be proved in next subsection:
the reader who is not interested in technical details can safely skip this discussion.

\subsubsection{Bipolar coordinates}

To compute the convolution (\ref{F0formule}) we introduce $d$-dimensional bipolar coordinates
in the space $\vec{r'}=(x_1,\cdots,x_d)$.
We assume, without loss of generality thanks to rotational invariance, that
$\vec r$ is
directed along the axis $x_1$ in $\RRR^d$, and we define
$u = |\vec{r'}|$, $v = |\vec r-\vec{r'}|$, and 
$R = \sqrt{x_2^2 + \cdots + x_d^2}$ the distance between $\vec{r'}$ and the axis $x_1$. 
We then look to the projection $r'_\perp$ of $\vec{r'}$ on the hyperplane 
$P_\perp = (x_2,\cdots,x_d)$
perpendicular to $x_1$; the distance of $r'_\perp$ from the origin in this plane is
just $R$, and we can introduce polar coordinates $(R,\th_{d-1})$ on this plane, thus
defining a set $\th_{d-1}$ of $d-2$ angles that specify the position of $r'_\perp$ on
the sphere of radius $R$ in $P_\perp$. We define in this way the change to bipolar
coordinates $\vec{r'} \to (u,v,\th_{d-1})$.

We wish to compute the Jacobian of such transformation. It is easy to show that
\begin{equation}
\frac{\partial (u,v,\th_{d-1})}{\partial (x_1,\cdots,x_d)} =
\begin{pmatrix}
\frac{x_1}{u} & \frac{x_2}{u}  \cdots  \frac{x_d}{u} \\
\frac{x_1-r}{v} & \frac{x_2}{v}  \cdots  \frac{x_d}{v} \\
\vec{0} & \left[ \frac{\partial \th_{d-1}}{\partial (x_2,\cdots,x_d)} \right] \\
\end{pmatrix} \ .
\end{equation}
The reason for the zeroes on the first column is that the angles $\th_{d-1}$ are
independent of $x_1$ as they describe the position of $r'_\perp$ in $P_\perp$.
We can write the determinant of this matrix as
\beq\begin{split}
J^{-1} &= \left|\frac{\partial (u,v,\th_{d-1})}{\partial  (x_1,\cdots,x_d)}\right| =
\frac{x_1 R}{u v} 
\left| \begin{matrix}
 \frac{x_2}{R}  \cdots  \frac{x_d}{R} \\
 \left[ \frac{\partial \th_{d-1}}{\partial (x_2,\cdots,x_d)} \right] \\
\end{matrix} \right| \\
&- \frac{(x_1-r) R}{u v} 
\left| \begin{matrix}
 \frac{x_2}{R}  \cdots  \frac{x_d}{R} \\
 \left[ \frac{\partial \th_{d-1}}{\partial (x_2,\cdots,x_d)} \right] \\
\end{matrix} \right| = \frac{r R}{u v} 
\left| \begin{matrix}
 \frac{x_2}{R}  \cdots  \frac{x_d}{R} \\
 \left[ \frac{\partial \th_{d-1}}{\partial (x_2,\cdots,x_d)} \right] \\
\end{matrix} \right| \ .
\end{split}\eeq
The matrix appearing in the previous equation is just the Jacobian matrix
for the change to spherical coordinates in $P_\perp$ and its determinant is
given by
\begin{equation}
\left| \begin{matrix}
 \frac{x_2}{R}  \cdots  \frac{x_d}{R} \\
 \left[ \frac{\partial \th_{d-1}}{\partial (x_2,\cdots,x_d)} \right] \\
\end{matrix} \right| = \frac{dR d\th_{d-1}}{d^{d-1}r'_\perp} = \frac1{R^{d-2} J_{d-1}(\th_{d-1})} \ .
\end{equation}
From the two previous equations we obtain the desired Jacobian
\beq
d\vec{r'} = \frac{u v}{r} R^{d-3} du dv J_{d-1}(\th_{d-1}) d\th_{d-1} = \frac{u v}{r} R^{d-3} du dv 
d\Omega_{d-1} \ ,
\eeq
where the integral of $d\Omega_{d-1}$ is just the $d-1$-dimensional solid angle $\Omega_{d-1}$.
The radius $R$ is easily expressed as a function of $u,v$ by elementary geometrical considerations:
\beq
R(u,v;r) = \frac{\sqrt{2 u^2 v^2 + 2 u^2 r^2 + 2 v^2 r^2 - u^4 - v^4 - r^4}}{2 r} \ .
\eeq
The (positive) variables $u,v,r$ have to respect the triangular inequalities, so the domain of
integration is for instance $u\in [0,\io)$, $v\in [|r-u|,r+u]$.

Then, performing the change of variable $w = v^2 - (r-u)^2$, and using 
$R(u,w;r)^2 = \frac{w (4 r u -w)}{4 r^2}$, we get
\begin{widetext}
\beq\begin{split}
q_A(r)& = \Omega_{d-1} \int_0^\io du \int_{|r-u|}^{r+u} dv \frac{u v}{r} R(u,v;r)^{d-3} \th(u-D) \g_{2A}(v) \\
& = \frac{\Omega_{d-1}}{2r} \int_D^\io du \, u \frac{e^{-\frac{(u-r)^2}{4A}}}{(4\pi A)^{d/2}} 
\int_{0}^{4 r u} dw \left(\frac{w (4 r u -w)}{4 r^2}\right)^{\frac{d-3}2} e^{-\frac{w}{4A}} \ .
\end{split}
\eeq
We now make use of the relation
\beq
\int_0^1 e^{-B x} [x(1-x)]^{\frac{d-3}2} = B^{\frac{2-d}2} e^{-\frac{B}2} \sqrt{\p} \G\left(\frac{d-1}2\right)
I_{\frac{d-2}2}(B/2) \ ,
\eeq
and of the definition $\Omega_{d-1} = \frac{2 \pi^{\frac{d-1}2}}{\G\left(\frac{d-1}2\right)}$ to obtain
the final result
\beq\label{F0unid}
q_A(r) = \int_D^\io du \left( \frac{u}{r} \right)^{\frac{d-1}{2}} 
\frac{ e^{ -\frac{(r-u)^2}{4A} }}{\sqrt{4\p A}}
\left[ e^{ -\frac{ru}{2A} } \sqrt{\pi \frac{ru}{A}} I_{ \frac{d-2}{2} } \left( \frac{ru}{2A} \right)\right] \ .
\eeq

\subsubsection{Three dimensions}

Remarkably, in $d=3$ the integrand simplifies and a full calculation of $q_A(r)$ is possible:
\beq\label{F03d}
\begin{split}
q_A(r) &= \frac{1}{r \sqrt{4 \pi A}} \int_D^\io du \, u \left[ e^{-\frac{(r-u)^2}{4A}} -
 e^{-\frac{(r+u)^2}{4A}} \right] \\
&= \frac12 \left[ \erf\left(\frac{r-D}{\sqrt{4A}}\right)-\erf\left(\frac{r+D}{\sqrt{4A}}\right)
+\frac2r \sqrt{\frac{A}\pi} \left( e^{-\frac{(r-D)^2}{4A}} -
 e^{-\frac{(r+D)^2}{4A}} \right) + 2\right] \ ,
\end{split}
\eeq
\end{widetext}
with
\beq\begin{split}
&\erf(t) \equiv \frac{2}{\sqrt{\p}} \int_0^t dx \, e^{-x^2} \ , \\
&\Th(t) = \frac{1}{2} [ 1 + \erf(t) ] = \frac{1}{\sqrt{\pi}} \int_{-\io}^t  dx \, e^{-x^2} \ .
\end{split}\eeq
Given that $q_A(r) \to 1$ for $r-D \gg \sqrt{A}$, this allows for a simple numerical evaluation
of the integral in (\ref{GmAdef}) to compute $G_m(A)$. 
A full computation of $Q(r)$ is also possible using (\ref{QQ1}).
It is also easy to check that uniformly in $r$ we 
have\footnote{Note that $\Th(t)$ is a ``smoothed'' theta function.}
\beq\label{F03dsmallA}
q_A(r) \sim \Th\left(\frac{r-D}{\sqrt{4A}}\right) +
O( \sqrt{A}) + O(e^{-\frac{D}{\sqrt{A}}}) \ .
\eeq

\subsubsection{Finite dimension: expansion in powers of $\sqrt{A}$}

In general finite dimension the integral (\ref{F0unid}) cannot be explicitly evaluated.
However, an expansion in powers of $\sqrt{A}$ similar to (\ref{F03dsmallA}) holds.
In fact, when $A$ is very small it is easy to realize that the main contribution to $q_A(r)$
comes from the integration over the component of $\vec{r'}$ which is parallel to
$\vec r$: in the
orthogonal directions $\th(|\vec r-\vec{r'}|-D)$ is essentially constant and the integration over these
components gives a correction $O(e^{-D/\sqrt{A}})$. Then we have
\beq\label{F0}
q_A(r) \sim \int_{-\io}^\io dr' \frac{e^{-\frac{r'^2}{4A}}}{(\sqrt{4\p A})} \theta(r-r'-D)
= \Th\left(\frac{r-D}{\sqrt{4A}}\right)  \ .
\eeq
The same result can be derived from (\ref{F0unid}) by observing that for $A \to 0$,
$r \sim D$, one has $z = \frac{ur}{2A} \to \io$ and for all $n$ \cite{Abramowitz}
\beq
e^{-z} \sqrt{2\pi z} I_n(z) \to 1 \ .
\eeq
Thus we have, changing variables to $t = \frac{r-D}{\sqrt{4A}}$ and $s = \frac{u-D}{\sqrt{4A}}$,
\begin{equation}\begin{split}
q_A(t) &= \frac1{\sqrt{\pi}}\int_0^\io ds \left( \frac{ D+s \sqrt{4A} }{ D+t \sqrt{4A} } \right)^{\frac{d-1}{2}}
e^{ -(t-s)^2 } \\ & \sim \frac1{\sqrt{\pi}}\int_0^\io ds \, e^{ -(t-s)^2 } = \Th(t) \ .
\end{split}\end{equation}
The next-to-leading orders in $\sqrt{A}$ can in principle be computed using the large $z$ expansion
of the Bessel functions \cite{Abramowitz}.

In this way we can derive the leading order contribution to $G_m(A)$: substituting 
$q_A(t) = \Th(t)$ in (\ref{GmAdef}) we get
\beq\label{taufirstorder}
\begin{split}
&G_m(A) = \\ &= \frac{d \sqrt{4 A}}D \int_{-D/\sqrt{4A}}^\io dt \left(\frac{D + t \sqrt{4A}}{D}\right)^{d-1}
\big[ \Th(t)^m -\th(t) \big] \\ & \sim  \frac{d \sqrt{4 A}}D \int_{-\io}^\io dt \big[ \Th(t)^m -\th(t) \big]
= \frac{d \sqrt{4 A}}D \, Q_0(m) \ ,
\end{split}
\eeq
defining
\beq
Q_0(m) = \int_{-\io}^\io dt [\Th(t)^{m}-\Th(t)] \ .
\eeq
One can check that the function $Q_0(m)$ is given by
\beq\begin{split}
&Q_0(m) = Q_0 (1-m) + O((1-m)^2) \ ,\\
&Q_0 = -\int_{-\io}^\io dt \Th(t) \ln \Th(t) \ ,
\end{split}\eeq
close to $m=1$ and that
\beq
Q_0(m) \sim \sqrt{\frac\pi{4m}} (\sqrt{m})^m 
\eeq
for $m\to 0$, by using the $t\to-\io$ expansion of the error function.

\subsubsection{Infinite dimension}

In the limit of infinite dimension, we are interested in the 
scaling $A = D^2 \wh A/d^2$. 
One could then
naively expect the small cage expansion of the previous subsection to work well. However, one has
$z = \frac{u r}{2 D^2 \wh A} d^2$, and an inspection
of the large $z$ expansion of the Bessel function $I_n(z)$ \cite{Abramowitz} shows that it is indeed an expansion
in powers of $\frac{n^2}{z}$. Then, for $n = \frac{d-2}{2}$ as in (\ref{F0unid}) one has to resum all
orders in this large $z$ expansion. 

This can be done without difficulty, either by looking at the large $z$ expansion, or by a saddle-point
evaluation of the integral representation 
\beq
I_n(z) = \frac{1}{2\pi i} \int_C dt \, t^{-n-1} e^{\frac{z}2 \left(t + \frac1t \right)} \ ,
\eeq
and one can show that
\beq
\lim_{d\to \io} e^{-d^2 z} \sqrt{2 \pi d^2 z} I_{\frac{d-2}2}(d^2 z) = e^{-\frac{1}{8 z}} \ .
\eeq
Using this result in (\ref{F0unid}), we have
\beq
q_A(r) = \int_D^\io du \left( \frac{u}{r} \right)^{\frac{d-1}{2}} 
\frac{ e^{ -\frac{(r-u)^2}{4A} }}{\sqrt{4\p A}}
e^{ -\frac{D^2 \wh A}{4 r u} }
\eeq
Changing variables again to $t = \frac{r-D}{\sqrt{4A}}$ and $s = \frac{u-D}{\sqrt{4A}}$ and
using
\beq
\left( \frac{u}{r} \right)^{\frac{d-1}{2}} =
\left( \frac{1 + \frac{s \sqrt{4\wh A}}d }{ 1 + \frac{t \sqrt{4\wh A}}d } \right)^{\frac{d-1}{2}} 
\sim e^{(s-t) \sqrt{4 \wh A}} \ ,
\eeq
we get
\beq \label{F0dinf}\begin{split}
q_A(t) &\sim e^{-\frac{\wh A}{4}} \frac1{\sqrt{\pi}} \int_0^\io ds \, e^{(s-t) \sqrt{4 \wh A} - (t-s)^2} \\
&= \Th\left( t + \frac{\sqrt{\wh A}}{2} \right) \ .
\end{split}\eeq
We can use this result to compute $\GG_m(\wh A)$ as defined in (\ref{calGmAdef}). Similarly to 
(\ref{taufirstorder}), we get
\beq\label{taudinf}
\begin{split}
\GG_m(\wh A) &= \lim_{d\to\io} \sqrt{4 \wh A} \int_{-d/\sqrt{4\wh A}}^\io dt 
\left(1 + t\frac{ \sqrt{4\wh A}}{d} \right)^{d-1} \times \\
& \hskip30pt \times \left[    \Th\left( t + \frac{\sqrt{\wh A}}{2} \right)^m -\th(t) \right] \\ & 
\sim \sqrt{4 \wh A}  \int_{-\io}^\io dt \, e^{t \sqrt{4 \wh A}} 
\left[    \Th\left( t + \frac{\sqrt{\wh A}}{2} \right)^m -\th(t) \right] \\
& = \int_{-\io}^\io dy \, e^y 
\left[    \Th\left( \frac{y + \wh A}{\sqrt{4\wh A}} \right)^m -\th(y) \right] \ .
\end{split}
\eeq
Note that if we expand the second line of (\ref{taudinf}) for small $\wh A$ 
(which amounts to setting $\wh A=0$ in the integrand) we obtain exactly the leading term
(\ref{taufirstorder}), which indicates that the two limits $d\to \io$ and $A \to 0$ can be
safely exchanged.

\subsubsection{Jamming limit}

The jamming limit corresponds to $m\to 0$ and $A = \a m$. Note that $G_m(A)$ is not
analytic in $A=m=0$, so the limits $m\to 0$ and $A \to 0$ cannot be exchanged.

However, the integral (\ref{GmAdef}) is uniformly
convergent in the jamming limit so we can exchange the limit with the integral.
As $A\to 0$, we can use the expression (\ref{F0}) for $q_A(r)$; using then the asymptotic expansion
of the error function for $|t| \to \io$, we can show that
\beq\label{thetajamming}
\lim_{m\to 0, \, A=\a m} \Th\left(\frac{r-D}{\sqrt{4A}}\right)^m = \begin{cases}
e^{-\frac{(r-D)^2}{4\a}} \hskip30pt r<D \ , \\
1 \hskip60pt r \geq D \ . \\
\end{cases}
\eeq
Then we have
\beq\label{taujamming}
G_0(\a) = \lim_{m\to 0, \, A=\a m} G_m(A) = \frac{d}{D^d} \int_0^D dr \, r^{d-1} e^{-\frac{(r-D)^2}{4\a}} \ .
\eeq
It is also possible to show that
\beq\label{dtaujamming}
\begin{split}
F_0(\a) &= \a \frac{d G_0(\a)}{d\a} 
=  \lim_{m\to 0, \, A=\a m} \frac{A}{1-m} \frac{\partial G_m(A)}{\partial A} \\
&=  -\lim_{m\to 0, \, A=\a m} m \frac{\partial G_m(A)}{\partial m} \\ & =
\frac{d}{4\a D^d}  \int_0^D dr \, r^{d-1} (r-D)^2 e^{-\frac{(r-D)^2}{4\a}} \ .
\end{split}
\eeq
In the limit $d\to\io$, $\a = D^2 \wh \a / d^2$, Eq.~(\ref{taujamming}) becomes, by
changing variable to $y=d(D-r)/D$,
\beq\begin{split}\label{G0adef}
\GG_0(\wh \a) &= \lim_{d\to\io} G_0(D^2 \wh \a / d^2) = \lim_{m\to 0, \, \wh A=\wh \a m} \GG_m(\wh A) \\
& =
\int_0^\io dy \, e^{-y -\frac{y^2}{4 \wh \a}} \ ,
\end{split}\eeq
which can be also directly derived from (\ref{taudinf}) and (\ref{thetajamming}),
confirming that the limits $d\to\io$ and $A\to 0$ can be exchanged without problems.

\subsection{The function $Q(r)$ in the jamming limit}
\label{app:Qasy}

\begin{figure}[t] \centering
\includegraphics[width=8cm]{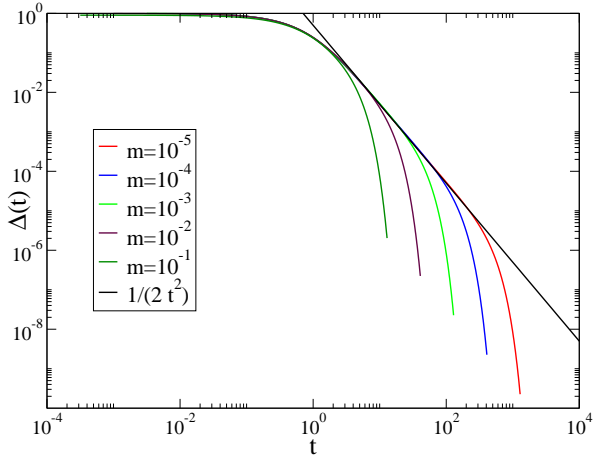}
\caption{
The function $\D_m(t)$ for different values of $m$ compared with the
asymptotic behavior for $m\to 0$ and large $t$.
}
\label{fig2}
\end{figure}

Even if $q_A(r)$ can be fully computed,
the function $Q(r)$, defined in (\ref{QQ1}), cannot be computed analytically in finite
dimension. 
Therefore we limit ourselves here to the computation at first order in $\sqrt{A}$.
In this case we have $q_A(r) \sim \Th\left(\frac{r-D}{\sqrt{4A}}\right)$; the integration
over the components of $\vec{r'}$ orthogonal to $\vec r$ gives 1 neglecting subleading corrections and we get
\beq\begin{split}
Q(r)&= \int_{-\io}^\io dr' \frac{e^{-\frac{(r')^2}{4A}}}{\sqrt{4\p A}} 
\left[ \Th\left(\frac{r-r'-D}{\sqrt{4A}}\right)^{m-1} - 1 \right] \\
&\equiv \frac1m \D_m\left(\frac{r-D}{\sqrt{4mA}}\right) \ ,
\end{split}\eeq
where the function $\D_m(t)$ is
\beq\label{Dmt}
\D_m(t) = m \int_{-\io}^\io \frac{ds}{\sqrt{\p}} e^{-(t \sqrt{m}-s)^2} [ \Th(s)^{m-1}-1 ] \ .
\eeq
It is easy to show that
\beq\label{Dintegrale}\begin{split}
\int_0^\io & dt \D_m(t) = \sqrt{m} \int_{-\io}^\io ds [\Th(s)^m - \Th(s)] \\ &=\sqrt{m} Q_0(m)
\rightarrow_{m\to 0} \sqrt{\pi/4}  \ .
\end{split}\eeq
Using, for $s \to -\io$,
$\Th(s) \sim \frac{e^{-s^2}}{2\sqrt{\p}|s|}$,
it is easy to show that for $m \to 0$,
$\D_m(0) \to 1$,
for $1 \ll t \ll 1/\sqrt{m}$,
$\D_0(t) \sim \frac{1}{2t^2}$, and for very large $t \gg 1/\sqrt{m}$,
$\D_m(t) \sim e^{-m t^2}$.
Indeed we have for any finite $t$
\beq\label{Dlimite}\begin{split}
\D_0(t)& \equiv \lim_{m\to 0} \D_m(t) = 2 \int_0^\io dy \, y \, e^{-y^2 - 2 t y} \\
&= 1 - \sqrt{\pi} t e^{t^2} (1 - \erf(t)) \sim_{t \gg 1} \frac{1}{2t^2} -
\frac{3}{4t^4} + \cdots
\ .
\end{split}\eeq
The (one-dimensional) Fourier transform of $\D_0(t)$ is given by
\beq\begin{split}
\wh \D_0(q) &= \int_{-\io}^\io dt \, e^{i q t} \D_0(|t|) \\ &= \sqrt{\pi} 
\left[ 1 -e^{\frac{q^2}4} \frac{\sqrt{\pi}}2 |q| \left( 1 -\erf\left(\frac{|q|}2\right)
\right)\right]\\ & = \sqrt{\pi} \D_0\left(\frac{|q|}2\right) \ .
\end{split}\eeq
The function $\D_m(t)$ is reported for different values of $m$
in Fig.~\ref{fig2}.

\section{Derivatives of the correlation functions}
\label{app:deriva}

In this appendix we will show how to compute the derivative
${d g(u,v)}/{d \ln\c(x,y)}$ that is needed to compute the correlation
function in the first order small cage approximation, Eq.~(\ref{gQg}).
We start from the partition function as a function of the activity
\beq\begin{split}
&Z[z(x),\c(x,y)] = \sum_{N=0}^\io \int \frac{d^N x}{N!} \prod_i z(x_i) \prod_{i<j}
\chi(x_i,x_j)
\\ &= \int \exp\left({\int dx \hat\rho(x) \ln z(x)
+ \frac{1}{2} \int dx dy \hat \rho_2(x,y) \ln\chi(x,y) }\right) \ .
\end{split}\eeq
where the $\int$ sign is a shorthand for $\sum_{N=0}^\io \int \frac{d^N
  x}{N!}$, and $\hat\rho(x) = \sum_i \d(x-x_i)$, 
$\hat\rho_2(x,y)=\sum_{i\neq j}\d(x-x_i)\d(y-x_j)$.
From the last expression it is straightforward to show that
\beq\begin{split}
&\frac{\partial \ln Z}{\partial \ln z(x)} = \la \hat\rho(x) \ra = \r(x) \ , \\
&\frac{\partial \ln Z}{\partial \ln \c(x,y)} = \frac12 \la \hat\rho_2(x,y) \ra =
\frac{\r^2}2 g(x,y) \ .
\end{split}
\eeq
We define the entropy functional, the Legendre transform of $Z$:
\beq\begin{split}
\SS &[\rho(x),\chi(x,y)] = \\ &= \max_{z(x)} \left[ \ln Z[z(x),\c(x,y)] - \int dx
  \r(x) \ln z(x) \right] \ .
\end{split}\eeq
From this it follows that
\beq\begin{split}
&\frac{\partial \SS}{ \partial\r(x)} = -\ln z(x) \ , \\
&\frac{\partial \SS}{\partial \ln \c(x,y)} = \frac{ \partial\ln Z}{ \partial\ln \c(x,y)} =
\frac{\r^2}2 g(x,y) \ ,
\end{split}
\eeq
because the explicit derivative with respect to $z(x)$ vanishes due to the 
maximum condition.

Therefore 
we need to compute $\frac{\partial^2 \SS}{\partial\ln\c(x,y)\partial \ln\c(u,v)}$. To simplify the notation
let us write $\ln\c(x,y)=J_1$, $\ln\c(u,v)=J_2$ and $\ln z(x)=j(x)$.
Then
\beq\label{Sder}
\begin{split}
&\frac{\partial^2 \SS}{\partial J_1 \partial J_2} =
\frac{\partial^2 \ln Z}{\partial J_1 \partial J_2} + \int dt \frac{\partial^2
  \ln Z}{\partial J_1 \partial j(t)}
\frac{d j(t)}{d J_2} \\ &=
\frac{\partial^2 \ln Z}{\partial J_1 \partial J_2} - \int dt ds
\frac{\partial^2 \ln Z}{\partial J_1 \partial j(t)}
\left[ \frac{\partial^2 \ln Z}{\partial j(t) \partial j(s)} \right]^{-1}
\frac{\partial^2 \ln Z}{\partial J_2 \partial j(s)}
\end{split}
\eeq
where we computed $\frac{d j(t)}{d J_2}$ as follows. Recalling that the
derivatives in the last equation are done at constant $\rho$
we can write
\beq\begin{split}
0 &= \frac{d}{d J_2} \r(t) = \frac{d}{dJ_2} \frac{\partial \ln Z}{\partial j(t)} \\
&= \frac{\partial^2 \ln Z}{\partial j(t) \partial J_2} + \int ds
\frac{\partial^2 \ln Z}{\partial j(t) \partial j(s)}
\frac{d j(s)}{d J_2}
\end{split}\eeq
from which
\beq
\frac{d j(t)}{d J_2} = - \int ds 
\left[ \frac{\partial^2 \ln Z}{\partial j(t) \partial j(s)} \right]^{-1}
\frac{\partial^2 \ln Z}{\partial j(s) \partial J_2}
\eeq

The explicit derivatives of $Z$ that appear in Eq.~(\ref{Sder}) can be
explicitly related to correlation functions of the liquid as follows
\cite{Hansen}:
\begin{widetext}
\beq\begin{split}
\frac{\partial^2 \ln Z}{\partial \ln\c(x,y) \partial \ln\c(u,v)} & =
\frac14 \left[ \la \hat\rho_2(x,y)\hat\rho_2(u,v)\ra- \la
  \hat\rho_2(x,y)\ra\la\hat\rho_2(u,v)\ra\right] \\
& = \frac{\r^2}4 [\d(x-u)\d(y-v)+\d(x-v)\d(y-u)] g(x,y) \\
& + \frac{\r^4}4 \Big\{ g_4(x,y,u,v) + \r^{-1} \big[ \d(u-x) g_3(y,u,v) +
\text{3 perm.} \big] - g(x,y)
g(u,v) \Big\} \\
\frac{\partial^2 \ln Z}{\partial \ln \c(x,y) \partial \ln z(t)} &=
\frac12 \left[ \la \hat\rho_2(x,y)\hat\rho(t)\ra- \la
  \hat\rho_2(x,y)\ra\la\hat\rho(t)\ra\right] \\
&= \frac{\r^3}2 \Big\{ g_3(x,y,t) + [\r^{-1} (\d(x-t)+\d(y-t)) -1 ] g(x,y)
\Big\} \\
\frac{\partial^2 \ln Z}{\partial \ln z(t) \partial \ln z(s)} &=
 \la \hat\rho(t)\hat\rho(s)\ra- \la
  \hat\rho(t)\ra\la\hat\rho(s)\ra
 = \r \Big\{ \d(t-s) + \r h(t,s) \Big\} \\
\left[\frac{\partial^2 \ln Z}{\partial \ln z(t) \partial \ln
    z(s)}\right]^{-1} &=
\frac1\r\d(t-s) - c(t,s)
\end{split}\eeq
where the last equation follows from the Ornstein-Zernicke relation
\cite{Hansen}.

Finally we have
\begin{equation}\label{dgdchi}
\begin{split}
\frac{d g(u,v) }{d \ln\c(x,y)} &= \frac2{\r^2}  
\frac{\partial^2 \SS}{\partial\ln\c(x,y)\partial \ln\c(u,v)} =
\frac{1}2 [\d(x-u)\d(y-v)+\d(x-v)\d(y-u)] g(x,y) \\
&+ \frac{\r^2}2 \Big\{ g_4(x,y,u,v) + \r^{-1} \big[ \d(u-x) g_3(y,u,v) +
\text{3 perm.} \big] - g(x,y)
g(u,v) \Big\} \\
&- \frac{\r^4}2\int dt ds
\Big\{ g_3(x,y,t) + [\r^{-1} (\d(x-t)+\d(y-t)) -1 ] g(x,y)
\Big\}
\left[\frac1\r\d(t-s) - c(t,s)\right] \times \\
& \hskip2cm \times \Big\{ g_3(u,v,s) + [\r^{-1} (\d(u-s)+\d(v-s)) -1 ] g(u,v)
\Big\}
\end{split}\end{equation}
\end{widetext}
By inspection of these contribution to Eq.~(\ref{gQg})  one can show that only the first one is
relevant to describe the delta peak of $\wt g(x,y)$, which gives Eq.~(\ref{gfirst}).

\section{Scaling close to jamming}
\label{app:scaling}

We will show here how to compute the behavior of $m(\f,\f_j)$ for $\f \to \f_j$.
For this we will need the asymptotic behavior of $Q_0(m)$ for $m\to 0$; one can show
that
\beq
Q_0(m) \sim \sqrt{\frac{\pi}{4m}} e^{\frac{m}2 \ln m} R_m + S_m
\eeq
where $R_m$ and $S_m$ are $C^\io$ functions of $m$ with $R_m = 1 + R_1 m + \cdots$ and
$S_m = S_1 m + S_2 m^2 + \cdots$, so that
\beq
\ln Q_0(m) \sim \frac12 \ln \frac\pi4 - \frac12 \ln m + \frac{m}2 \ln m +
\ln R_m + O(m^{3/2}) \ .
\eeq
where $O(x^\a)$ represents a quantity which is bounded by $C x^\a$ for $x \to 0$.
From this relation it follows that
\beq\label{QpQ}\begin{split}
&\frac{Q'_0(m)}{Q_0(m)} = -\frac1{2m} + \frac{\ln m}{2} + \frac12 +
\frac{R'_m}{R_m} + O(m^{1/2}) \\
&\frac{d}{dm} \frac{Q'_0(m)}{Q_0(m)} = \frac{1}{2m^2} + \frac{1}{2m} + O(m^{-1/2})
\end{split}\eeq
We have then from (\ref{siSi}):
\beq\label{sipm}\begin{split}
\partial_m \Si(m,\f) & = \frac{d}{2m} + \frac{d}{1-m} + 2 d m \frac{Q'_0(m)}{Q_0(m)}
- d \\ &-d m (1-m) \frac{d}{dm} \frac{Q'_0(m)}{Q_0(m)} \\
= -d +O(\sqrt{m})
\end{split}\eeq
On the other hand
\beq\begin{split}
\partial_\f \Si(m,\f) & = S'(\f) + d \frac{\partial_\f (A^*(m))^{-1/2}}{(A^*(m))^{-1/2}} \\
&= S'(\f) + d \frac{\partial_\f[ \f Y(\f)]}{\f Y(\f)} \\
&= S'(\f) + \frac{d}{\f} + d \frac{Y'(\f)}{Y(\f)}
 \ ,
\end{split}\eeq
and we obtain, close to $\f_j$,
\beq
m(\f,\f_j) \sim -\frac{1}{d} \left[
S'(\f) 
+ \frac{d}{\f} + d \frac{Y'(\f)}{Y(\f)}
\right] (\f_j - \f) \ .
\eeq
\ie Eq.~(\ref{mJ}).

\bibliography{HS}

\bibliographystyle{miormp}

\end{document}